\documentclass[preprint, 12pt]{elsarticle}

\journal{Transportation Research Part C}

\usepackage[margin=1in]{geometry}
\newcommand{\R}{\mathbb{R}}

\usepackage[colorlinks=true,linkcolor=black, citecolor=blue, urlcolor=blue]{hyperref}
\usepackage{url}
\usepackage{array}
\usepackage{graphicx}
\usepackage{subcaption} 
\usepackage{amssymb, dsfont} 
\usepackage{amsmath} 
\usepackage{commath} 
\usepackage{hyperref}
\usepackage{enumerate}
\usepackage{listings}
\usepackage{booktabs}

\usepackage[toc,page]{appendix}
\usepackage{bbm}
\usepackage{upgreek}

\usepackage{multirow}
\usepackage{mwe,tikz}
\usepackage[percent]{overpic}
\usepackage{float}
\usepackage{caption}

\makeatletter
\def\ps@pprintTitle{%
	\let\@oddhead\@empty
	\let\@evenhead\@empty
	\def\@oddfoot{\reset@font\hfil\thepage\hfil}
	\let\@evenfoot\@oddfoot
}
\makeatother

\begin{document}
	
	\begin{frontmatter}

\title{Evaluation of Ride-Sourcing Search Frictions and Driver Productivity: A Spatial Denoising Approach}

\author[mymainaddress,mysecondaryaddress]{Natalia Zuniga-Garcia\corref{mycorrespondingauthor}}
\ead{nzuniga@utexas.edu}
\author[mysecondaryaddress]{Mauricio Tec}
\ead{mauriciogtec@utexas.edu}
\author[mysecondaryaddress,mythirdaddress]{James G. Scott}
\ead{james.scott@mccombs.utexas.edu}
\author[myforthaddress]{Natalia Ruiz-Juri} \ead{nruizjuri@mail.utexas.edu}
\author[mymainaddress]{Randy B. Machemehl}
\ead{rbm@mail.utexas.edu}

\cortext[mycorrespondingauthor]{Corresponding author}
\address[mymainaddress]{The University of Texas at Austin, Department of Civil, Architectural and Environmental Engineering, 301 E. Dean Keeton St. Stop C1761, Austin, TX 78712, United States}
\address[mysecondaryaddress]{The University of Texas at Austin, Department of Statistics and Data Sciences, 2317 Speedway  Stop D9800, Austin, TX 78712, United States}
\address[mythirdaddress]{The University of Texas at Austin, Department of Information, Risk, and Operations Management,  2110 Speedway Stop B6000, Austin, TX 78705, United States}
\address[myforthaddress]{The University of Texas at Austin, Network Modeling Center, Center for Transportation Research,  3925 West Braker Lane Austin, TX 78759, United States}
		
\begin{abstract} 
\noindent
This paper considers the problem of measuring spatial and temporal variation in driver productivity on ride-sourcing trips.  This variation is especially important from a driver's perspective: if a platform's drivers experience systematic disparities in earnings because of variation in their riders' destinations, they may perceive the pricing model as inequitable.  This perception can exacerbate search frictions if it leads drivers to avoid locations where they believe they may be assigned ``unlucky'' fares. To characterize any such systematic disparities in productivity, we develop an analytic framework with three key components.  First, we propose a productivity metric that looks two consecutive trips ahead, thus capturing the effect on expected earnings of market conditions at drivers' drop-off locations.  Second, we develop a natural experiment by analyzing trips with a common origin but varying destinations, thus isolating purely spatial effects on productivity.   Third, we apply a spatial denoising method that allows us to work with raw spatial information exhibiting high levels of noise and sparsity, without having to aggregate data into large, low-resolution spatial zones.  By applying our framework to data on more than 1.4 million rides in Austin, Texas, we find significant spatial variation in ride-sourcing driver productivity and search frictions.  Drivers at the same location experienced disparities in productivity after being dispatched on trips with different destinations, with origin-based surge pricing increasing these earnings disparities.  Our results show that trip distance is the dominant factor in driver productivity: short trips yielded lower productivity, even when ending in areas with high demand. These findings suggest that new pricing strategies are required to minimize random disparities in driver earnings.
\end{abstract}


\end{frontmatter}


	
\section{Introduction}
\noindent
Ride-sourcing companies, also known as transportation network companies (TNCs), provide pre-arranged or on-demand transportation service for compensation \cite{shaheen2016}.  They operate as a two-sided market that connects drivers of personal vehicles with passengers.
TNCs have been controversial in cities around the world due to multiple factors, including lack regulation of their pricing system, driver selection, and the perception that they are unfair competition to taxi services. Pricing strategies for TNCs have also been criticized due to concerns for the welfare of providers and consumers \cite{surge12}.

In this paper, we consider the problem of spatial and temporal mispricing of ride-sourcing trips from a driver's perspective.  For drivers, a desirable characteristic of a ride-sourcing platform is \textit{destination invariance}: the principle that two drivers dispatched on different trips from the same location at the same time do not envy each other's expected future income \cite{spatial1}. It is possible, however, for some trip opportunities to yield higher continuation payoffs relative to others, leading to inefficiencies on a network level.  Such incorrect pricing results in a loss of service reliability if it leads drivers to select a specific kind of trip or decline to accept trips from particular locations, destinations, or time frames. Moreover, excessive hourly or daily volatility or perceived arbitrariness in driver earnings can limit long-term driver participation, reducing the service's supply through limited driver availability.  In principle, a pricing scheme that accurately accounted for this variation in driver productivity could at least partially mitigate this volatility in earnings.  

\subsection{Background}
\noindent
Spatial and temporal mispricing can lead to several market failures \cite{spatial1}, such as drivers ``chasing the surge'' or avoiding short trips thus leaving riders in other areas without access to the service.\footnote{These examples explain why platforms do not show the trip destination before the driver accepts the ride (e.g., \cite{ignorance, short}).} For example, Uber and Lyft tried to provide drivers with more flexibility by adding destination filters, where drivers can select the desired drop-off location that would allow them to relocate themselves \cite{filter1, lyft}. However, this feature caused a negative impact on the platform by increasing riders waiting time and other drivers' pick-up time, as strategic drivers used the filter to select trips with better earning potential\footnote{Uber decided to restrict this feature to two times per day.}\cite{filter2}. Strategic and experienced drivers can learn how to improve their earnings by predicting profitable times and locations, which exacerbates disparities in driver earnings and satisfaction (as in \citet{gender}). 

Recent research efforts have addressed ride-sourcing's spatial mispricing problem by proposing different pricing strategies and driver-passenger matching functions. The primary goal of this line of work is to reduce search frictions, i.e.~imbalances between driver supply and passenger demand across geographic areas that cause long matching and reaching times.\footnote{Market frictions are those factors that prevent the market clearing, leaving some buyers and sellers unable to immediate trade as both may need to invest in a costly search process to locate matching partners. Search frictions are present in ride-sourcing markets as a consequence of the spatial mis-allocation of drivers and passengers; often drivers are in one place and passengers in another.} Some examples include incorporating spatial surge pricing models \cite{spatial3, spatial2}, spatio-temporal pricing mechanisms \cite{spatial1}, search and matching models \cite{friction8, spatial4, spatial5, spatial15}, and non-linear pricing models \cite{friction9}. However, the majority of the methods focus on optimization of platform revenue and do not evaluate the pricing problem from a driver perspective. There is also very limited evidence on the driver's opportunity costs associated with different trip destinations. Research on spatial pricing suggests that accounting for prices based on both origin and destination does not provide a substantial gain when optimizing for platform revenue \cite{spatial2}. Yet other authors have demonstrated with simulation how a spatio-temporal pricing mechanism can result in higher social (consumer and provider) welfare when using a model based on origin and destination prices that preserves driver equity \cite{spatial1}. Moreover, there is a general lack of understanding of the spatial structure of driver productivity\footnote{We define the driver productivity in terms of profit per unit time.} and limited empirical evaluations of possible methods for characterizing this structure.


\subsection{Objective and Contributions}
\noindent
The principal objective of this research is to analyze the spatial structure of ride-sourcing search frictions and driver performance variables, with the ultimate goal of helping to provide insight on how new, more equitable pricing strategies might be developed.  To that end, we propose a three-pronged analytical framework for understanding spatial variation in driver continuation payoffs.
\begin{enumerate}
\item We propose a productivity metric that, conditional on a trip's origin time and location, looks at a driver's hourly earnings across two consecutive trips ahead.\footnote{Ideally we would be able to look further ahead to get at true continuation payoffs, but further look-ahead windows result in radically smaller data sets.  See Section \ref{sec:sec4}.}  This metric allows us to capture the effect on expected earnings of market conditions at drivers' drop-off locations, and allows us to understand how different kinds of trips affect productivity---e.g.~a short, low-revenue trip to a dense urban area with high subsequent demand for a second trip, versus a long, high-revenue trip to a suburban destination with lower subsequent demand.
\item We isolate exogenous destination-based variability in driver productivity via a natural experiment that examines trips with a common origin but varying destinations.  This setup allows us to find purely destination-based effects on productivity that cannot be explained by unobserved differences in driver preferences of when and where to work.
\item Finally, we apply a spatial denoising method that addresses the significant data-analysis challenges created by our approach.  As space is discretized more finely, our ability to detect high-resolution spatial variation improves, but the amount of data in each discrete sub-area decreases, leading to high levels of noise as well as sparsity (where some areas have no data at all).  In the literature related to the pricing issue, some studies do not account for this spatial heterogeneity in the variables under study \cite{surge6,surge3,surge12}, while others incorporate it in a highly spatially aggregate manner \cite{spatial3, spatial4, friction8}.  In contrast, we employ a spatial denoising technique based on the graph-fused lasso \cite{tibshirani2011}, a well established method in the image-processing literature for analyzing high-definition spatial maps. This approach compensates for high levels of noise in small spatial areas, addresses the data-sparsity problem, and allows for a very fine-resolution analysis yielding highly interpretable summaries.  With this form of spatial smoothing, a very fine discretization of space can be used: in our case, we analyze data at the traffic analysis zone (TAZ)\footnote{TAZs are geographic areas dividing a planning region into relatively similar areas of land use and land activity.} level, and we smooth the data values associated with all TAZs jointly via a highly efficient convex optimization routine.
\end{enumerate}

We then apply this analytical framework using data made available by RideAustin, a TNC company in Austin, Texas, including trips made during the period that Uber and Lyft, the leading national TNC companies, were temporarily out of the city.\footnote{Uber and Lyft left the city from May 2016 to May 2017 after the Austin City Council passed an ordinance requiring ride-hailing companies to perform fingerprint background checks on drivers, a stipulation that already applies to Austin taxi companies \cite{samuels_2017}.} During this period, RideAustin was responsible for one-third of Austin's ride-sourcing market share, suggesting high representativeness of the data \cite{Felipe}. 

On the methodological side, the contributions of this work include
(i) development of performance metrics that capture the effects of the trip destination on driver earnings,
(ii) design of a study to isolate destination-based spatial effects on driver earnings, using a natural experiment; and
(iii) implementation of a spatial-denoising methodology to analyze high-resolution spatial variables at the TAZ level.

Our principal empirical findings suggest that current dispatching and pricing schemes do not adequately ensure driver equity. Drivers at the same origin enjoyed different continuation payoffs after getting dispatched on trips to different locations.  Origin-based surge pricing actually increased these disparities in drivers' earnings. Furthermore, we find that trip distance is a major predictive factor on driver productivity, with short-distance trips resulting in lower expected earnings, even when those trips end in areas with high demand.  We conclude that new pricing strategies---ones that account for spatio-temporal variability in continuation payoffs---are required to guarantee driver equity.


\subsection{Outline}
\noindent
Subsequent sections of the paper are organized as follows: 
Section \ref{sec:sec2} provides a literature review of the principal aspects of pricing strategies in ride-sourcing markets; 
Section \ref{sec:sec3} describes the dataset; 
Section \ref{sec:sec4} presents the methodology for implementing driver performance measures;
Section \ref{sec:sec5} presents the selected smoothing method; 
Section \ref{sec:sec6} includes the empirical analysis and provides results and discussion; 
and finally, Section \ref{sec:sec7} contains conclusions and final remarks.

\section{Literature Review}
\label{sec:sec2}
\noindent
This section summarizes prior related research on pricing strategies, spatial pricing, and spatial aggregation in ride-sourcing systems. 

The popularity of ride-sourcing platforms relies not only on the advanced technology of connecting users and providers through cell phone applications but also on pricing strategies. An essential tool used by TNCs is dynamic (or surge) pricing that helps in managing both supply and demand. Surge pricing consists of raising the cost of a trip when demand outstrips supply within a fixed geographic area. The existing literature on pricing mainly addresses the problem of dealing with temporal demand fluctuations at a given location. Recent research in price strategies has focused on comparing the impact of static versus dynamic prices and analyzing benefits, as in \citet{surge15, surge6, surge3, surge12}. Authors agree on the benefits of dynamic pricing to both users and providers \cite{surge15, surge12} and the need for regulation \cite{surge1, surge6}. Additional research has been dedicated to studying pricing policies and effects on the labor supply (or drivers' choice of hours), such as by \citet{surge1, surge4, surge5}. In the context of ride-sourcing trips, labor supply elasticities have substantial implications on the effectiveness of surge pricing because the temporary increase in wages can have an immediate effect on whether or not drivers continue to work. 

Research that incorporates the spatial distribution of demand and supply in ride-sourcing system pricing schemes is recent. Authors have attempted to balance spatial supply-demand using different mechanisms. For example, \citet{spatial3} studied pricing and penalty strategies for platform revenue maximization and social welfare optimization in a hybrid (street-hailed taxi and ride-sourcing) market with variable demand across space. 
\citet{friction8} and \citet{spatial4} modeled the passenger search and driver matching process using a hybrid market \cite{friction8} and only taxis \cite{spatial4}, and evaluated their models using empirical information.
\citet{spatial5} proposed a geometric matching method for ride-sourcing systems based on market equilibrium and assuming a revenue maximization platform. They suggest the use of a rate cap regulation to avoid excessively high pricing and provide an empirical study using data from a Chinese TNC. 
Also assuming platform revenue maximization, \citet{spatial15, spatial13, spatial2} considered the problem of matching ride-sourcing costumers to ``strategic'' drivers in a geometric area by examining how the platform should respond to a short-term \cite{spatial15}  and long-term \cite{spatial13, spatial2} supply-demand imbalance. Drivers are considered strategic because they move in equilibrium in a simultaneous move game, choosing where to reposition based on prices, supply levels, and driving costs. 

For this study, the work by \citet{spatial1} is especially relevant. They addressed spatial and temporal variations using origin-destination prices that warrant driver equity. The authors proposed a spatio-temporal pricing (STP) mechanism that considers multiple locations and periods along with rider demand, willingness to pay, and driver supply varying over space and time. They show with simulation that this mechanism provides higher social welfare than the myopic pricing scheme.  \citet{spatial2} also analyze pricing models using origin-destination prices. The authors found that if the demand pattern is not balanced, the platform can benefit substantially from pricing rides differently depending on the origin location. They discovered that in comparison to optimal origin-based pricing, the extra gain from origin-destination based price is not as significant as the gain from origin-based pricing in comparison to uniform pricing. 

Research evaluating driver and taxi fleet performance metrics includes the work by \citet{friction9}. The authors defined an ``expected profit" measure to estimate the profit that a taxi driver expects to receive from picking up a customer in a particular zone and used this metric to calculate the probability that a vacant taxi in a drop-off zone will seek customers in any other zone. They proposed a non-linear fare structure with a continuously declining charge per unit distance to address problems with drivers that offer illegal discounts for long-distance trips (known as``taxi discount gangs") and long waiting queues at problematic locations such as an airport. 
Similarly, productivity has been measured by accounting for idle and reach time \cite{idle}, average speed \cite{can}, capacity utilization rate \cite{can, capacity},
average earning per unit time \cite{learning}, and fuel cost \cite{fuel}.

Incorporating spatial information is a complex challenge. First, the availability of empirical ride-sourcing data is limited. Thus, some authors rely on simulations \cite{spatial1, spatial2} or limit their research to taxi-only data \cite{spatial3, spatial4}. Second, when available, spatial information is subject to noise and high sparsity, so researchers tend to aggregate the data in large areas, losing valuable high-resolution insights. Further, data processing does not include spatial denoising steps. Examples include \citet{spatial3} who summarized the demand pattern of a taxi network in Beijing using eighty-one squared-area zones of 5.4 $km^2$ to model pricing and penalty strategies. With the objective of modeling the search and matching process, \citet{spatial4} aggregated NYC taxi data in thirty-nine zones conformed by uniting census tracks, and \citet{friction8} summarized Uber and taxi information in NYC using forty geographic locations. In our work, we summarized ride-sourcing information in more than 1,500 TAZs in Austin, Texas. TAZ areas vary from 0.01 $km^2$ in the Central Business District (CBD) to 30 $km^2$ in the rural area, with an average of 2 $km^2$. 

Our original contributions to current literature encompass the implementation of driver productivity measures that incorporate the market conditions on the destination area, and a natural experiment to provide empirical proof of space and time heterogeneity of driver performance and search frictions. 
We make use of a spatial smoothing technique to evaluate high-definition denoised spatial information. To date, pricing mechanisms are mainly focused on maximizing platform revenue, minimizing the evaluation of driver productivity. But drivers are an essential part of the success of ride-sourcing systems, and the provision of fair working conditions can result in more reliable services.


\section{Ride-Sourcing Data}
\label{sec:sec3}
\noindent
We use a dataset that an Austin-based TNC made available in early 2017 \cite{data}. It consists of 1,494,125 rides between June 2, 2016, and April 13, 2017. The dataset provides a description of the trip, rider, and driver (anonymized), payment, cost, location, among other trip information. Because demand during the first few months was limited, we focus our analysis on data from September 1, 2016, to April 13, 2017, corresponding to 1,417,282 trips. 
A total of 261,632 (18.5 percent) trips started in the CBD, 58,558 (4.1 percent)  started at the Austin-Bergstrom International Airport (ABIA), and 85,072 (6.0 percent) ended there. During this period, there was a monthly average of 177,160 trips and a daily average of 6,271 trips. Figure \ref{fig:des}, in \ref{app:A}, provides a broader description of the trips. A majority of trips occurred during the Friday to Sunday timeframe as opposed to during typical weekdays, and the hourly ridership pattern indicates that a large number of the trips happened between 6 PM and 6 AM, suggesting that ride-sourcing services in Austin are used more for social and leisure activities than work-related activities. Other studies using this dataset found the same pattern (as in \citet{Felipe} and \citet{cstrb}).    
 
\begin{figure}[H]
	\begin{center}
		\begin{subfigure}[h]{0.495\linewidth}
			\begin{overpic}[width=\linewidth]{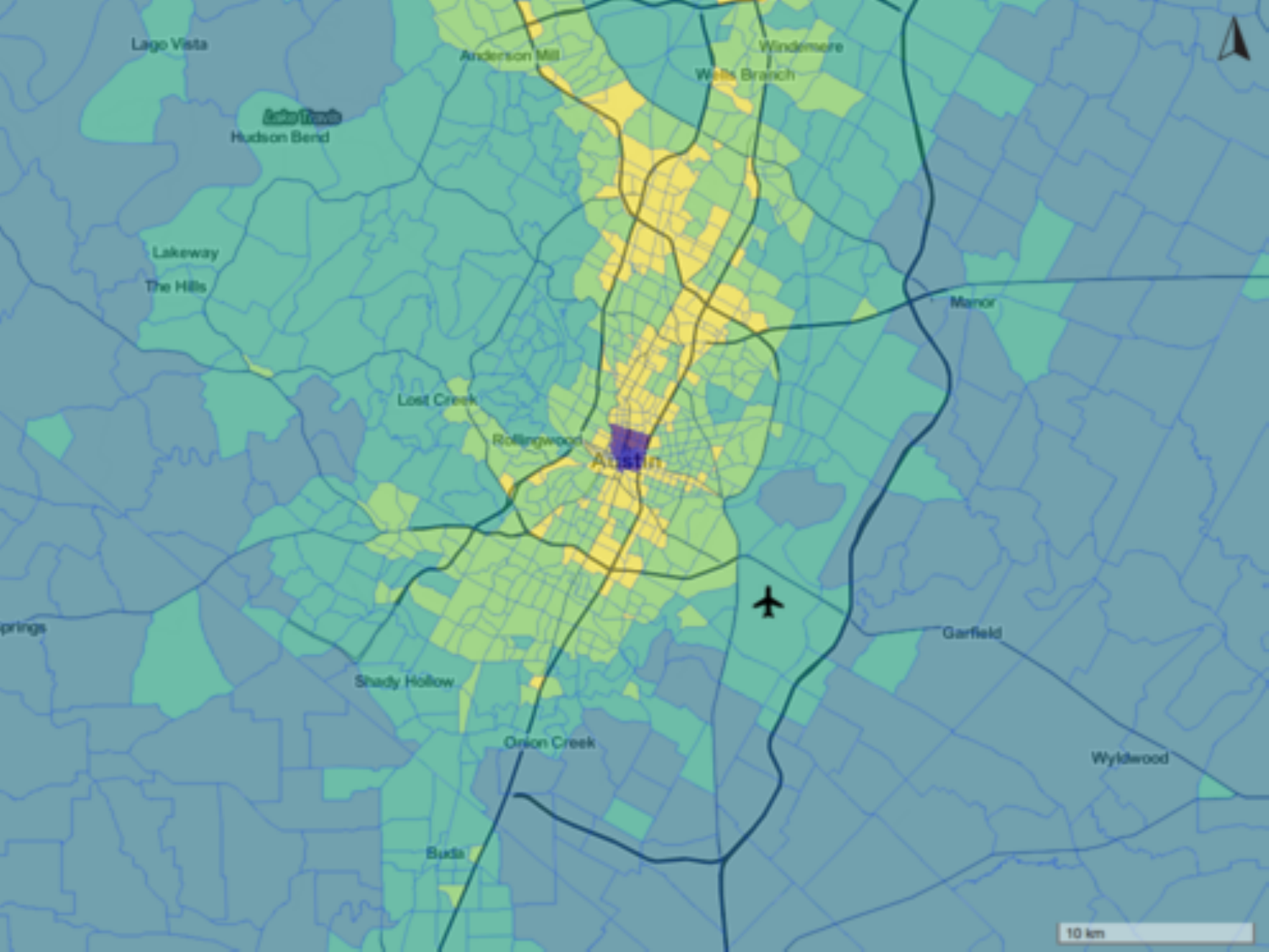}
				\put(0,0){\includegraphics[width=.22\linewidth]{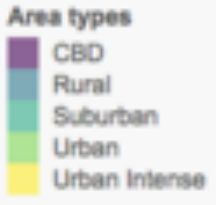}}
			\end{overpic}
			\caption{Area types by TAZ, 2015}
		\end{subfigure}
		\begin{subfigure}[h]{0.495\linewidth}
			\begin{overpic}[width=1\linewidth]{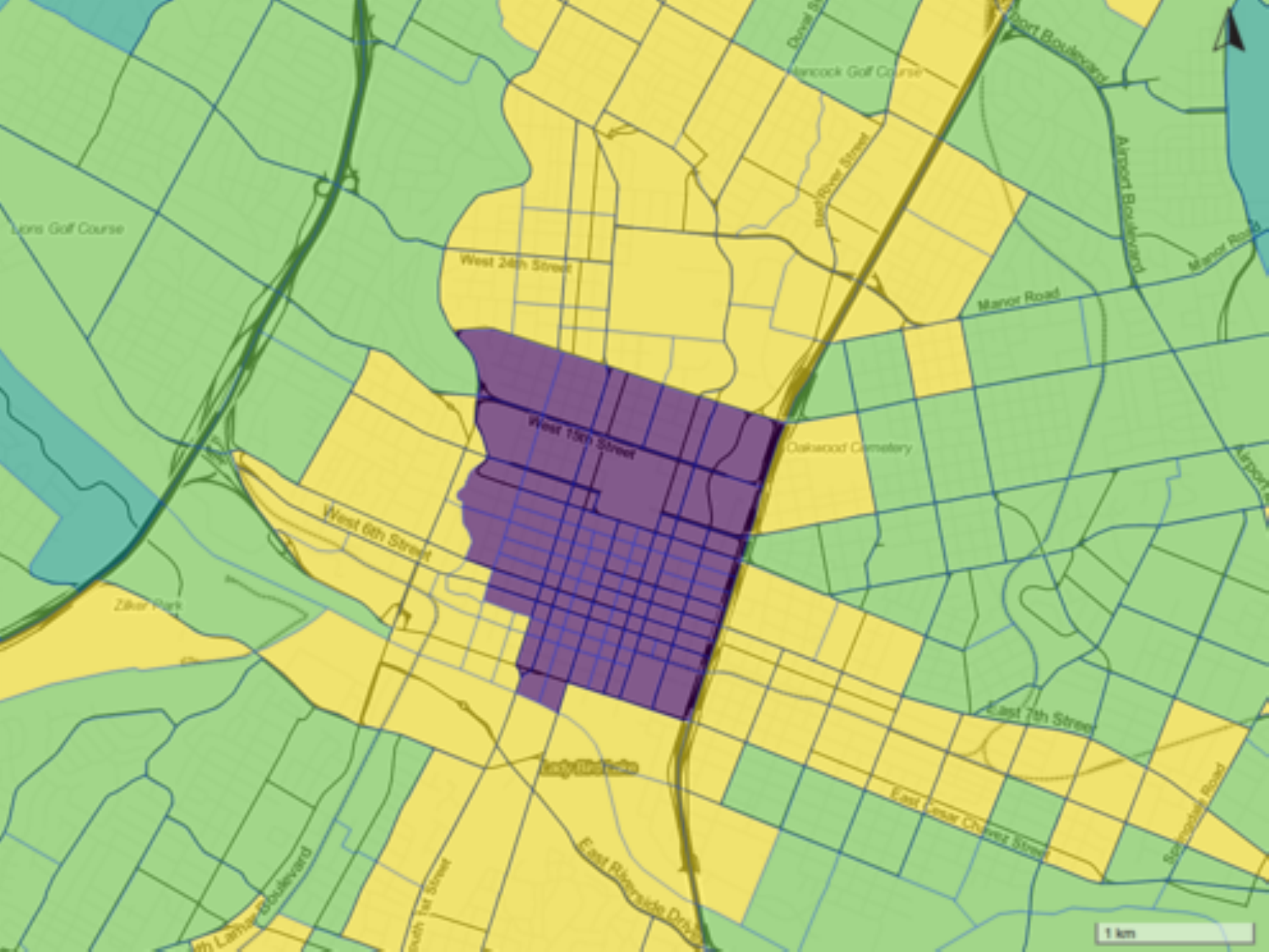}
				\put(0,0){\includegraphics[width=.22\linewidth]{fig/area/type/Simb.pdf}}
			\end{overpic}
			\caption{CBD area}
		\end{subfigure}
		\begin{subfigure}[h]{0.495\linewidth}
			\begin{overpic}[width=1\linewidth]{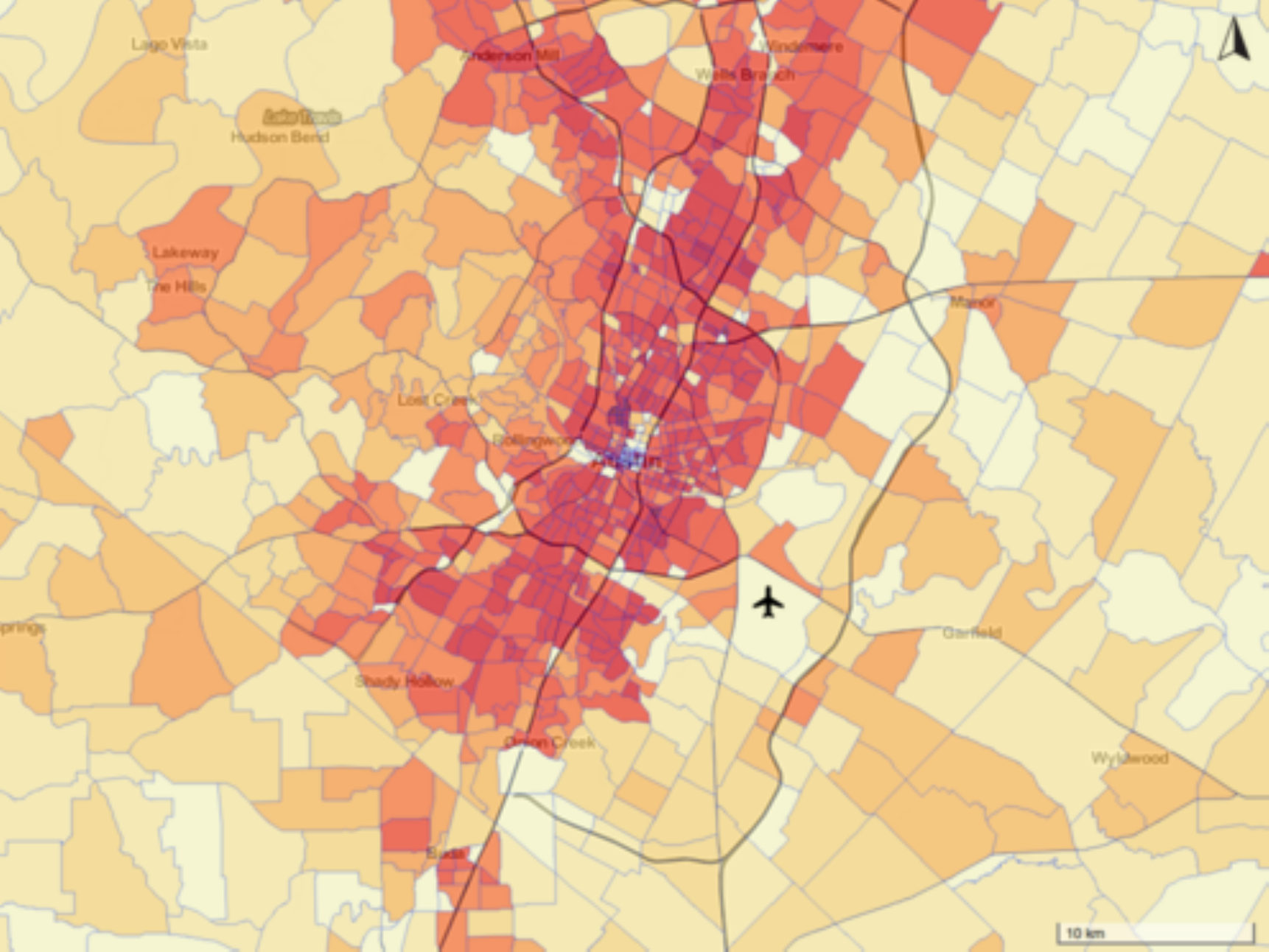}
				\put(0,0){\includegraphics[width=.22\linewidth]{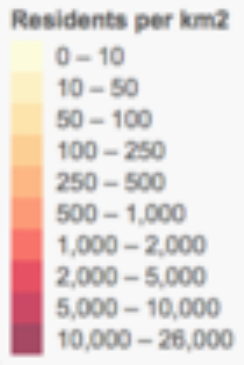}}
			\end{overpic}
			\caption{Population density by TAZ, 2015}
		\end{subfigure}
		\begin{subfigure}[h]{0.495\linewidth}
			\begin{overpic}[width=\linewidth]{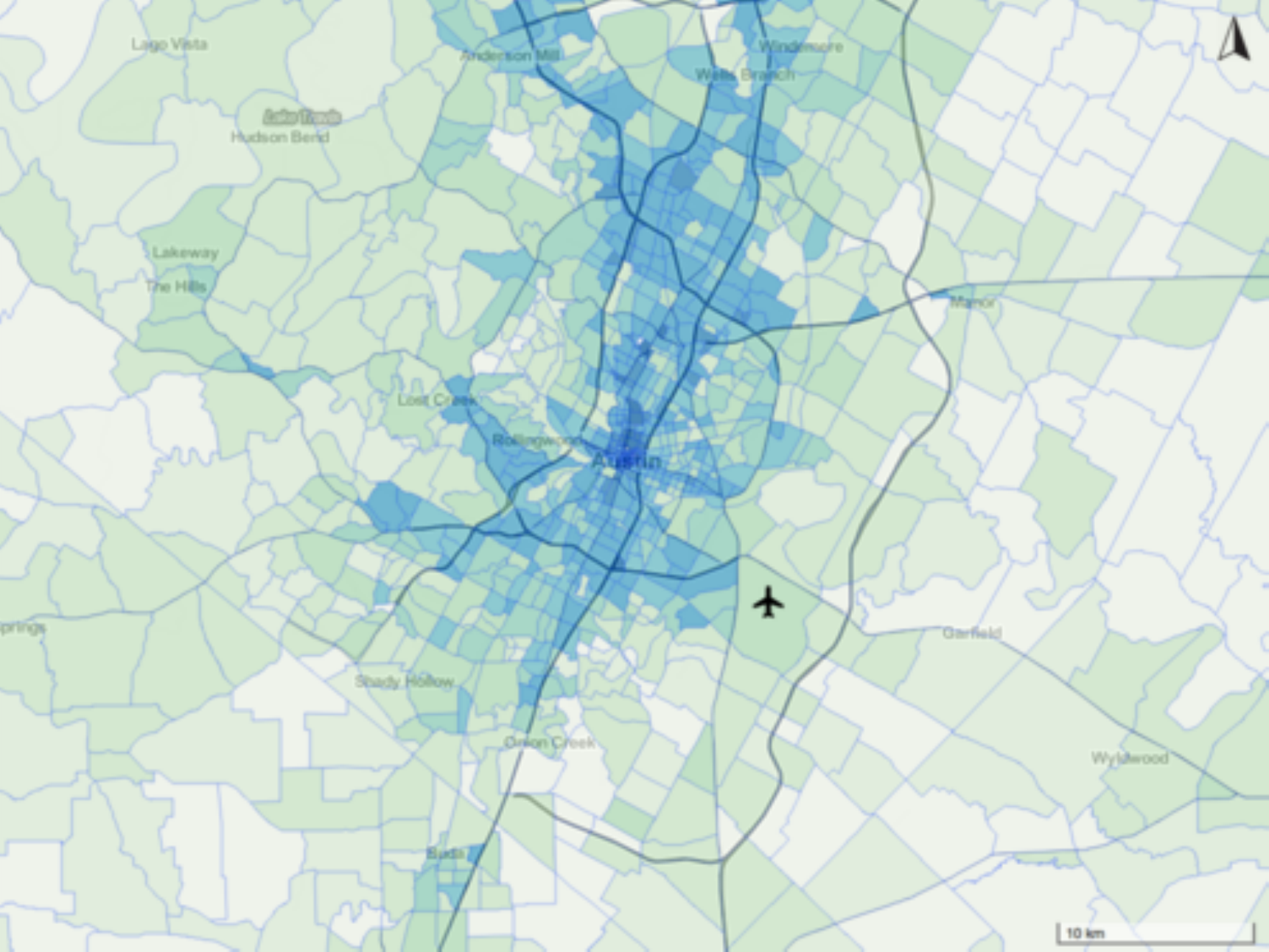}
				\put(0,0){\includegraphics[width=.22\linewidth]{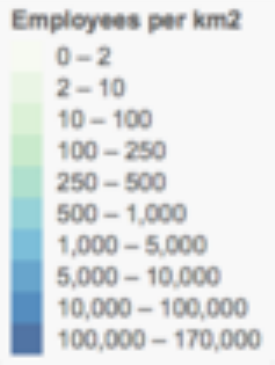}}
			\end{overpic}
			\caption{Employment density by TAZ, 2015}
		\end{subfigure}
		\caption{Description of the study area}
		\label{fig:area}
	\end{center}
\end{figure}

We select rides with origin and destination coordinates within the city of Austin and surrounding areas, with an approximate population of 916,906 residents (during 2017) \cite{censo}. 
We use TAZ-level demographic data obtained from the Capital Area Metropolitan Planning Organization (CAMPO) website\footnote{The CAMPO website can be accessed at \url{https://www.campotexas.org/}.} to characterize the study zone. Figure \ref{fig:area}(a) provides a spatial description of Austin area types during the year 2015, Figure \ref{fig:area}(b) presents a detailed view of the downtown area, and Figures \ref{fig:area}(c) and (d) present the population and employment densities estimated using CAMPO information for the year 2015.

\section{Driver Productivity }
\label{sec:sec4}
\noindent
We propose metrics to describe destination-based driver productivity that are sensitive to the market conditions at the drop-off location. These measures are defined as profit-per-unit-time and take into account the occupied trip time and the unproductive time, defined as the idle (or dead-heading) time and the reach (or pick-up) time. This section presents the methodology used to estimate these metrics and the implementation of a natural experiment to evaluate destination-based variation in productivity for a fixed trip origin. We also provide a description of the spatial aggregation approach and the variables used in the computation of our metrics.

\subsection{Trip Fare}
\noindent
To approximate driver profit per trip, we use the trip fare information provided in the dataset. RideAustin has four types of vehicle classes (standard, premium, luxury, and sport utility vehicle [SUV]), and each class has a different driver rate. A total of five percent of the trips are made using a non-standard vehicle class. In some cases, drivers provide standard services even if their vehicle category is different. For the analysis, we use the information on trips, including all vehicle categories, but we scaled the driver rates to the standard class to ensure that the different rates do not influence the results.

The trip fare we use in the analysis represents the actual driver earnings, which includes mileage, time, and base fare.
This fare corresponds to the total passenger cost excluding tip, roundup amount\footnote{RideAustin allows riders to round up the total fare to the nearest dollar and designate it to a local charity.}, and other operational fees like booking and airport fees.
We estimate this fare using the information provided in the dataset (trip distance and duration) and RideAustin driver rates and fare structure. 
For the standard car category, the fare consists of the sum of a base fare (\$1.50), time rate (\$0.25/minute), and distance rate (\$0.99/mile). The fare value increases linearly with trip distance and duration, except for short trips (approximately less than 0.5 miles [0.8 km]), where a minimum fare is charged (\$4.00) in the standard car category. 
For the fare estimation, consider a trip from the pick-up location $r \in R$ to the drop-off location $s \in S$, where $R$ is the set of riders origin locations and $S$ is the set of rider destination locations. 
The fare for this trip $F_{rs}$ is calculated from Expression \ref{eq:fare} using the trip duration $t_{rs}$, distance $d_{rs}$, and minimum fare.

\begin{equation}\label{eq:fare}
\begin{aligned}
F_{rs} := \max{\bigg\{1.5 + 0.25\ t_{rs} + 0.99\ d_{rs}, \textbf{ } 4\bigg\}} 
\end{aligned}
\end{equation}

RideAustin also provides details about the applied surge factor.\footnote{We do not have details of the methodology employed by RideAustin to determine surge multipliers.} During the analysis period, a total of 17.5 percent of system-wide trips had an applied surge factor, while for CBD-origin trips, 28.5 percent trips had a surge factor applied. Figure \ref{fig:surge}, in \ref{app:A} provides a spatial description of the percentage of trips with surge price during weekdays and weekends. 
Our driver-performance evaluation includes two analyses. We estimate results using the ``flat'' fare, corresponding to the fare estimated using \eqref{eq:fare}, and we estimate results using the surge price. Expression \ref{eq:surge} presents the estimation of surge fare $\check{F}_{rs}$ using the surge factor multiplier $\alpha$ and the flat fare $F_{rs}$. 

\begin{equation}\label{eq:surge}
\begin{aligned}
\check{F}_{rs} := \alpha F_{rs}
\end{aligned}
\end{equation}

\subsection{Idle and Reach Time}
\noindent
Idle time, also known as dead-heading or waiting time, is obtained from the driver's unique identification (ID) information. First, we filter trips by drivers' unique ID and order them based on date and time. Second, for every two consecutive trips, we calculate the difference between drop-off time of the first trip and trip-assignation time of the second trip (corresponding to the moment a ride is assigned to the driver).
This estimation does not account for drivers working for other platforms\footnote{During the analysis period, other companies also started their operations in Austin (e.g., Fasten and Fare).} or taking breaks between trips. To reduce the effect that these uncertain driver activities may have on the idle time variable, we only analyze consecutive trips where idle time values are below sixty minutes. Reach or pick-up time is calculated from the dataset as the difference between trip-assignment time and pick-up time for a specific trip. 

We now describe the variables used in our methodology. Consider a driver serving a rider during the first trip ($Trip\ 1$), originated from the pick-up location $r \in R$ to the destination drop-off location $s \in S$, during the ride duration $t_{rs}$. After finishing this trip, the driver must wait for the system to assign the second trip ($Trip\ 2$) at the pick-up location $r^* \in R$. Waiting time will depend on the market conditions of the location $s$, i.e., the balance between trip demand and driver availability. We call this waiting period the driver-idle time $w_{sr^*}$. The reach time  $p_{r^*}$ for $Trip\ 2$ corresponds to the time between the trip assignment and the rider pick-up. Finally, the analysis period ends after the driver finishes the second trip, with duration $t_{r^*s^*}$, at the drop-off location $s^* \in S$. The unproductive time $u_{s,r^*}$  between $Trip\ 1$ and $Trip\ 2$ is defined as the sum of idle and reach time:
\begin{equation}\label{eq:unp}
\begin{aligned}
u_{sr^*} := w_{sr^*} + p_{r^*}
\end{aligned}
\end{equation}
Figure \ref{fig:diag} exhibits a visual summary of the variables just defined. Figure \ref{fig:fric} in the appendix contains histograms with variable descriptions. 
The average idle time value is 12.8 minutes (standard deviation of 14.5 minutes), and the average reach time value is 6.4 minutes (standard deviation of 4.0 minutes). On average, the trip duration is 13.2 minutes, with a standard deviation of 8.0 minutes.


\begin{figure}[H]
	\begin{center}
		\includegraphics[width=0.7\linewidth]{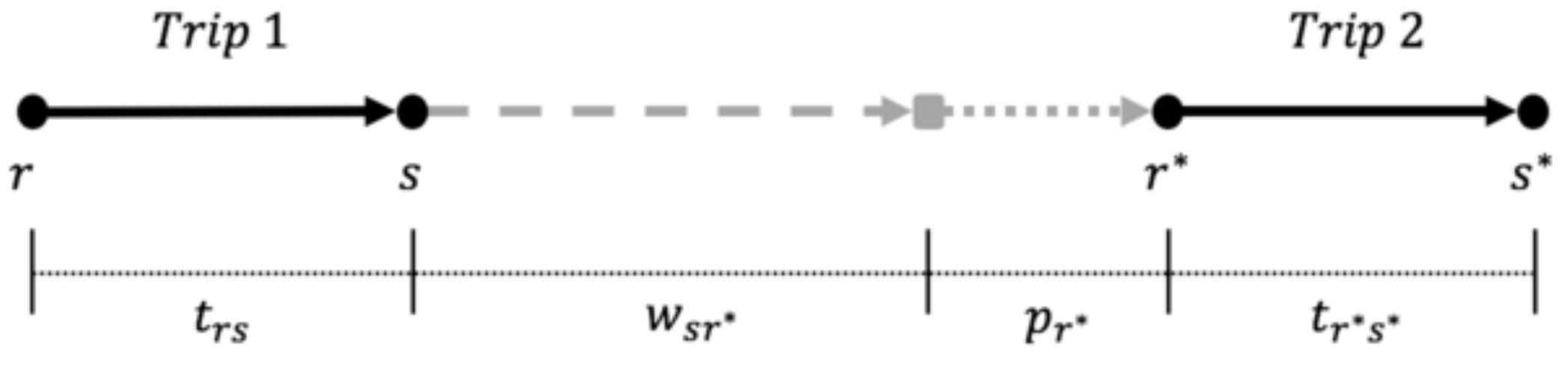}
		\caption{Description of the driver time used for the analysis}
		\label{fig:diag}
	\end{center}
\end{figure}



\subsection{Productivity Metrics}
\noindent
We develop performance metrics that evaluate the effect of market conditions of the trip destination and further spatial dynamics for ending in a specific region.
First, we propose a measure to assess the value or productivity of a region, inherent to characteristics of an area, such as expected trip revenue and unproductive time. 
This metric is an indirect measure of a driver expected continuation payoff and intends to capture the market conditions of the trip destination. 
To define this metric, consider a driver providing a service in $Trip\ 1$ described in Figure \ref{fig:diag}. 
We estimate the productivity as the expected revenue from the following ride, $Trip\ 2$, given the market conditions of the destination of $Trip\ 1$ in region $s \in S$.  We refer to this measure $\pi_s$ as the \textit{destination continuation payoff} or the \textit{destination productivity}, and we calculate it using Expression \ref{eq:prodD}.

\begin{equation}\label{eq:prodD}
\begin{aligned}
\pi_{s} := \frac{F_{r^*s^*}}{u_{sr^*}+t_{r^*s^*}}
\end{aligned}
\end{equation}
We acknowledge that this definition deliberately ignores the characteristics of the trip that led to that position. It does not account for the first trip ($Trip\ 1$) distance and duration because we want to isolate the effects of the destination location on further driver earnings. 

We further extend our analysis and consider the effects of the first trip characteristics (e.g., distance and duration). Moreover, we want to know if specific kinds of trips have a strong effect on the expected payoff of a driver after taking into account both the features of the first trip and the continuation payoff of the destination.

We propose a second metric, denoted $\Pi_{s \mid r}$, for this goal, and we use it to compare trips with common origin $r \in R$ but varying destinations $s\in S$. We refer to it as \textit{driver productivity} and it is calculated using Expression \ref{eq:prodC}, below.  It takes into account the search frictions on the destination, the penalty suffered by drivers for traveling to low-demand zones, as well as the immediate earnings benefits of long trips.
\begin{equation}\label{eq:prodC}
\Pi_{s \mid r} := \frac{F_{rs} + F_{r^*s^*}}{t_{rs} + u_{sr^*}+t_{r^*s^*}}
\end{equation}
In the following section, we detail a study design based on a natural experiment, used to estimate the value of this metric for our data. We obtain the following identity expressing $\Pi_{s \mid r}$ as a weighted average of the value of the first trip and the continuation payoff of the destination:
\begin{equation}\label{eq:procIdentity}
\Pi_{s \mid r} = \left(\frac{t_{rs}}{t_{rs} + u_{sr^*}+t_{r^*s^*}}\right)\cdot \frac{F_{rs}}{t_{rs}} + \left(\frac{u_{sr^*}+t_{r^*s^*}}{t_{rs} + u_{sr^*}+t_{r^*s^*}}\right) \cdot \pi_s
\end{equation}
The first factor $F_{rs} / t_{rs}$ is the hourly revenue of the first trip, and it depends only on the tariff per minute, surge price, and minimum fare. The second factor $\pi_s$ is the continuation payoff defined in Equation \ref{eq:prodD}, and it depends on the market conditions at the destination of the trip. The weights assigned to each factor depend proportionally on the duration of the first trip relative to the the unproductive time and duration of the subsequent trip.
 
\subsection{Design of the Natural Experiment}
\noindent
We developed a study design to estimate the driver productivity $\Pi_{s\mid r}$ using a natural experiment that allows us to consider a common initial location. We select trips with the same origin $r \in R$, the CBD area, and analyze the destination zones, $s \in S$, using the performance measures proposed in the previous section. The CBD accounts for 18.5 percent of the trips and presents 28.5 percent of trips with surge price during the analysis period. This design is a useful setup for causal reasoning since all trips considered are undifferentiated at the beginning (they all depart from downtown) and we may assume that the treatments (destinations) are assigned completely at random.

\subsection{Spatial and Temporal Aggregation}
\noindent
Our objective is to evaluate the spatial structure of the ride-sourcing operational and driver performance measure. We aggregate the variables in space and time to assess spatio-temporal changes.

\subsubsection{Space Discretization}
\begin{figure}[H]
	\begin{center}
		\begin{subfigure}[h]{0.495\linewidth}
			\begin{overpic}[width=1\linewidth]{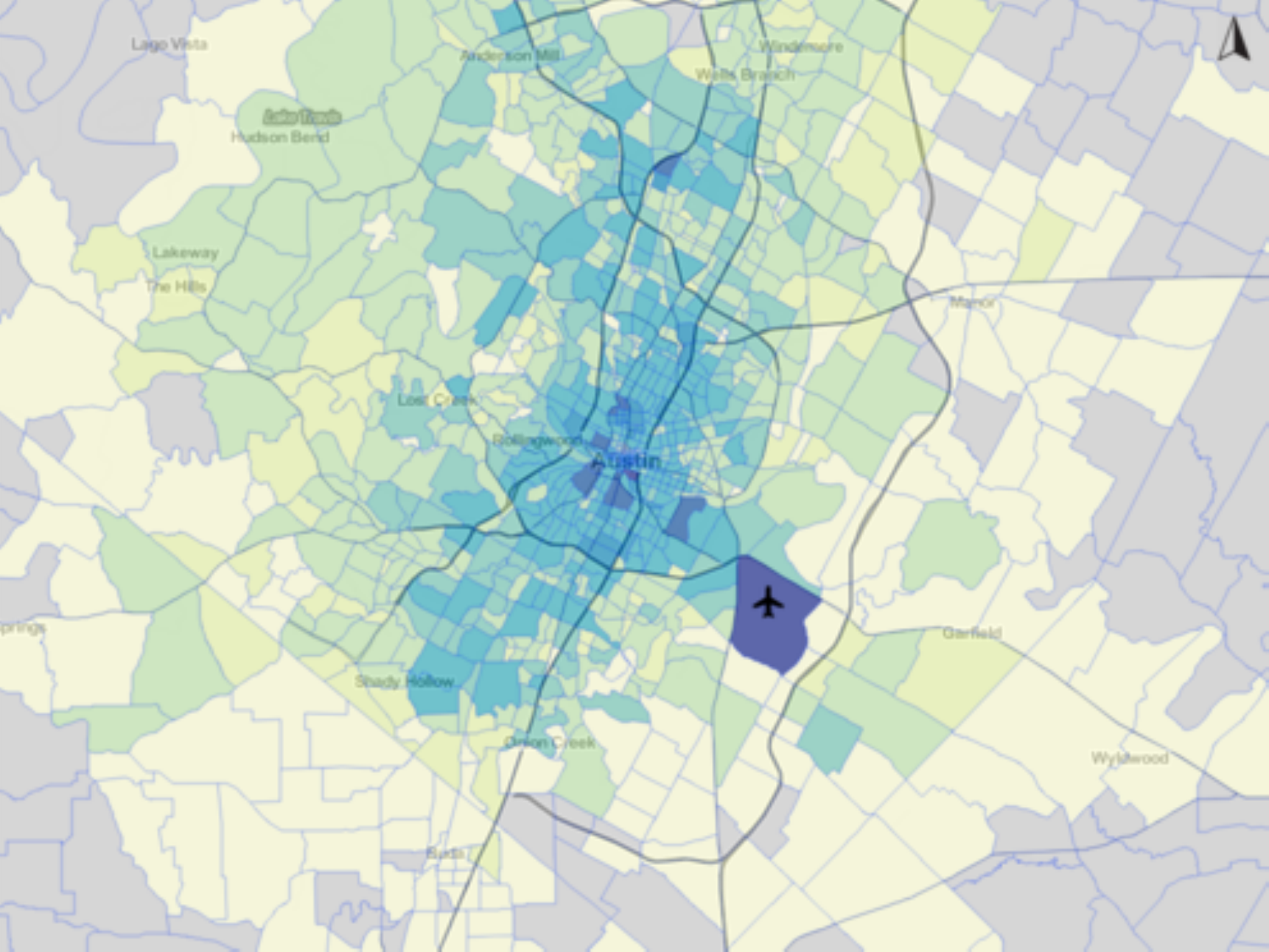}
				\put(0,0){\includegraphics[width=.22\linewidth]{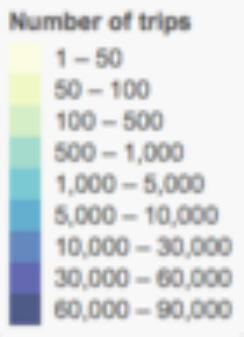}}
			\end{overpic}
			\caption{Trip origins}
		\end{subfigure}
		\begin{subfigure}[h]{0.495\linewidth}
			\begin{overpic}[width=1\linewidth]{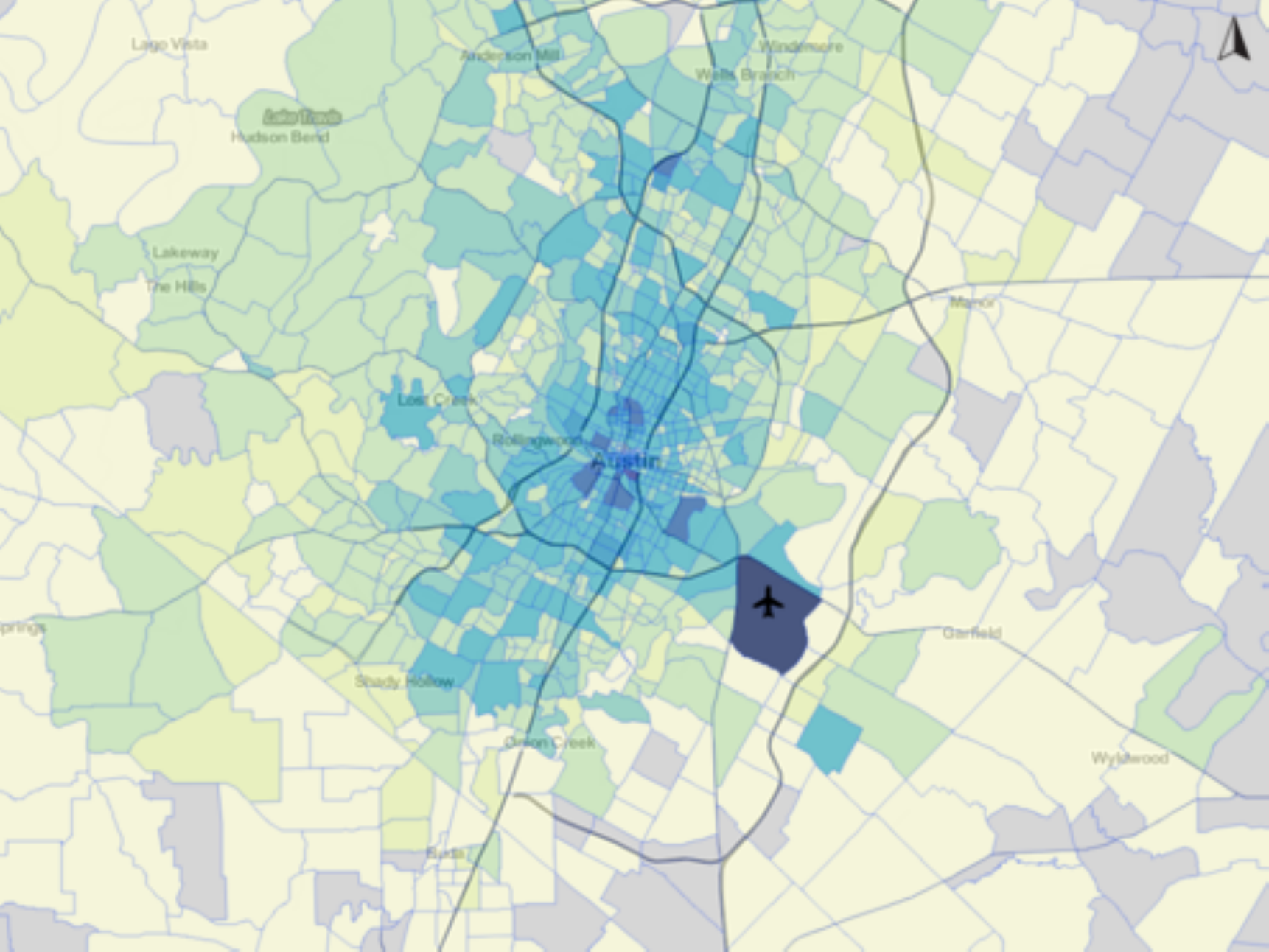}
				\put(0,0){\includegraphics[width=.22\linewidth]{fig/area/count/Simb.pdf}}
			\end{overpic}
			\caption{Trip destinations}
		\end{subfigure}
		\caption{Count of trips by TAZ-origin and TAZ-destination}
		\label{fig:maps}
	\end{center}
\end{figure}

\noindent
Space discretization involves aggregating origin/destination-based trip properties at the TAZ level, a common spatial unit used in transportation planning. We match the trip pick-up (origin) and drop-off (destination) longitude and latitude coordinates with the corresponding TAZ location. Table \ref{tab:TAZ} describes trip counts per TAZ, and Figure \ref{fig:maps} presents the maps including origin and destination counts. The mean count of trips by origin is 1,079 and by destination is 921. The ABIA airport TAZ presents the highest demand with 58,558 origin
trips and 85,072 destination trips.

\begin{table}[H]
	\centering
	\caption{Description of trip counts by TAZ}
	\label{tab:TAZ}
	\begin{tabular}{lrrrrrrr}
		\hline
		\multicolumn{1}{c}{TAZ} & \multicolumn{1}{c}{Mean} & \multicolumn{1}{c}{Std. Dev.} & \multicolumn{1}{c}{Min.} & \multicolumn{1}{c}{1$^{st}$ Qu.} & \multicolumn{1}{c}{Median} & \multicolumn{1}{c}{3$^{rd}$ Qu.} & \multicolumn{1}{c}{Max.}  \\ \hline
		Origin                  & 1,079.0                  & 2,901.7                       & 1.0                      & 32.0                         & 236.0                      & 931.0                        & 58,558.0                  \\ 
		Destination             & 921.0                    & 3,157.6                       & 1.0                      & 12.0                         & 153.0                      & 691.0                        & 85,072.0              \\ \hline
	\end{tabular}
\end{table}

\subsubsection{Analysis periods}
\noindent
We select four analysis periods based on typical assumptions regarding trip purpose and demand intensity. During the Monday to Friday period, we separated peak hours trips (from 6 to 9 AM and from 4 to 7 PM) corresponding to work-related trips, mid-day (from 10 AM to 3 PM) which represents the period with lower demand, and overnight (from 8 PM to 5 AM) corresponding to a majority of recreational trips. We also use weekends (Saturday and Sunday) corresponding to recreational trips. Table \ref{tab:count} provides a description of the trip counts per period for the system-wide and CBD-origin trips. Weekend encompasses nearly 50 percent of the trips, while the mid-day period is less than 15 percent.

\begin{table}[H]
	\centering
	\caption{Description of trip counts by period}
	\label{tab:period}
	\begin{tabular}{llrrrr}
		\hline
		\multicolumn{2}{c}{\multirow{2}{*}{Period}} & \multicolumn{2}{c}{System-wide}                            & \multicolumn{2}{c}{CBD-origin}                             \\ \cline{3-6} 
		\multicolumn{2}{c}{}                        & \multicolumn{1}{c}{Total} & \multicolumn{1}{c}{Percentage} & \multicolumn{1}{c}{Total} & \multicolumn{1}{c}{Percentage} \\ \hline
		\multirow{3}{*}{Weekday}    & Peak Hours    &  290,572                   & 21\%                           &  36,901                    & 14\%                           \\ 
		& Mid-Day       &  209,005                    & 15\%                            &  27,845                     & 11\%                            \\
		& Overnight     &  265,093                   & 19\%                           &  69,968                    & 27\%                           \\ 
		\multicolumn{2}{l}{Weekend}                 &  652,612                    & 46\%                           &  126,918                    & 49\%                           \\ \hline
		\multicolumn{2}{l}{\textit{Total}}          & \textit{1,417,282}        & \textit{100\%}                 & \textit{261,632}          & \textit{100\%}                 \\ \hline
	\end{tabular}
	\label{tab:count}
\end{table}


\section{Spatial Denoising Approach}
\label{sec:sec5}
\noindent
Spatial smoothing techniques are typically used for a wide range of applications. For example, in the image processing field smoothing approaches are used for image denoising \cite{chambolle2004}; in computational geometry and object modeling they are used to reconstruct surfaces \cite{yu2013, tasdizen2002}; and, in machine learning, they are used to impute missing values \cite{compton2014}. 
Other applications include spatial statistical analysis wherein, for instance, they are used to predict crime hotspots by smoothing incident report locations \cite{mclafferty2000}, detect crash hotspots using historical crash data \cite{thakali2015}, or for event detection using taxi trips \cite{wang2015}. 
In this research effort, we make use of the GFL spatial smoothing technique to provide an empirical analysis of the spatial structure of ride-sourcing variables. This section presents the smoothing approach used. For an extended background on smoothing techniques, refer to \ref{app:B} or \citet{zuniga2018}. 

\subsection{Graph-Fused Lasso}
\noindent
Assume that we have observations $y_i$ that are noisy versions of a true signal $x_i$ associated to the vertices of an undirected graph $\mathcal{G}= (\mathcal{V}, \mathcal{E})$ with vertex set $\mathcal{V}$ and edge set $\mathcal{E}$. The edge set determines which sites are neighbors on the graph. The goal of the smoothing techniques is to estimate the unknown true signal $x_i$ in a way that leverages the assumption of spatial smoothness over $\mathcal{E}$. The underlying statistical model is described by Equation \ref{eq:31}.
\begin{equation}\label{eq:31}
y_i = x_i + \epsilon_i \, , \quad i = 1, \ldots, n, 
\end{equation}
where the $\epsilon_i$ are mean-zero errors. The GFL smoothing optimization problem shown in Expression \ref{eq:37} seeks to minimize the differences between the estimated true signal and the observations with a penalization for drastic changes in neighboring values of the true signal. The first term $\ell(  \textbf{y},  \textbf{x}) $ corresponds to a smooth convex loss function arising from the negative loglikelihood of the observation process; and the second term $\sum_{(r,s) \in \mathcal{E}} |x_r - x_s | $ is the $\ell_1$ penalty that rewards a solution for having small absolute first differences across the edges of the graph. The penalty hyperparameter $\lambda > 0$ is used to calibrated the trade-off between fitting the observed data and having smooth values.

\begin{equation}\label{eq:37}
\begin{aligned}
& \underset{x \in \R^n}{\text{minimize}}
& & 
\ell(  \textbf{y},  \textbf{x})  + \lambda \sum_{(r,s) \in \mathcal{E}} |x_r - x_s |.
\end{aligned}
\end{equation}
Equation \ref{eq:37} does not have a closed-form solution. Therefore, convex optimization approaches such as the alternating direction method of multipliers (ADMM)\footnote{The ADMM is an algorithm that solves convex optimization problems by breaking them into smaller pieces, each of which is then easier to handle.} \cite{boyd2011} are required. 
Many efficient, specialized procedures using ADMM have been developed (c.f., \citet{wahlberg2012}, \citet{barbero2014}, and \citet{tansey2015}). We implemented the  method developed by \citet{tansey2015}, which leads to an efficient approach that presents a fast solution and is also scalable. \ref{app:C} provides more details of the method. 

\subsubsection{Loss Function}
\noindent
In this study, we selected a penalized weighted least squared-error loss function to take into account the differences in the number of observations within each zone, so that each trip in the data is assigned the same weight. 
Let $y_{i1}...,y_{i\eta_i}$ be the observations from trips ending within the $i$-th TAZ and denote by $y_i$ the mean of those observations. The loss function is then given by
\begin{equation}\label{eq:weightedLoss}
\ell(  \textbf{y},  \textbf{x}) = \frac{1}{2}\sum_{i=1}^{n} \eta_i (y_i-x_i)^2,\quad\quad y_i := \frac{1}{\eta_i}\sum_{j=1}^{\eta_i}y_{ij}
\end{equation}
This loss function corresponds to independent Gaussian errors with variances proportional to $1/\eta_i$ (c.f. Expression \ref{eq:31}). To see why it assigns the same weight to each observation, we leverage the fact that
$$
\sum_{i=1}^{n} \sum_{j=1}^{\eta_i} (y_{ij}-x_i)^2 = \sum_{i=1}^{n} \eta_i (y_i-x_i)^2 + C
$$
where $C$ is a factor that does not depend on the unknowns $x_i$, and thus plays no role in the optimization. The resulting objective function of our graph-fused lasso problem is
\begin{equation}\label{eq:weightedObj}
\begin{aligned}
& \underset{\mathbf{x} \in \R^n}{\text{minimize}}
& & 
\frac{1}{2} \sum_{i=1}^{n} \eta_i (y_i-x_i)^2  + \lambda \sum_{(r,s) \in \mathcal{E}} |x_r - x_s |.
\end{aligned}
\end{equation}

\subsubsection{Choosing the Regularization Parameter}
\noindent
The regularization parameter $\lambda$ controls the amount of smoothing. With $\lambda = 0$, no smoothing is done, and as $\lambda \to \infty$, the estimated values become the same in every location. 
To select the optimal $\lambda$ for each variable, we first randomly split the data into a training and a test set of respective sizes 90 percent and 10 percent. Then, we used the training set to estimate the smoothed value of each region using different values of $\lambda$ ranging from 0.001 to 100. For a new data point outside the training set, the prediction would be the smoothed value of the region where it was observed. The predictions of the test data are used to estimate the out-of-sample prediction error using the root mean square error (RMSE) criterion, described by Expression \ref{eq:RMSE}. 
\begin{equation}\label{eq:RMSE}
\begin{aligned}
\widehat{RMSE} = \left(\frac{1}{N}\sum_{i=1}^{n} \sum_{j = 1}^{\eta_i}(y_{ij} - \hat{x}_i)^2 \right)^{1/2},
\end{aligned}
\end{equation}
where $n$ is the total number of TAZ regions and $N=\sum_{i = 1}^n \eta_i$ the total number of observations in the test set.

The optimal $\lambda$ corresponds to the minimum RMSE from the possible $\lambda$ values tested and it is the amount of smoothing that best filters the noise and represents what we would expect from a newly observed point. After selecting the best $\lambda$, we obtained the final estimates from the model with the optimal $\lambda$ using both the training and test data.

As pointed out in the previous section, minimizing the RMSE is equivalent to minimizing the loss $\ell(  \mathbf{y}_\mathrm{test},  \mathbf{\hat{x}}) $ on the test data $ \mathbf{y}_\mathrm{test}$ using $\mathbf{\hat{x}}$ inferred from the training data.

\subsubsection{Graph Definition}
\noindent
The edges for joining the TAZ nodes were chosen according to a $k$-nearest neighbors principle. The location of a TAZ was computed as the mean longitude and latitude of all the points observed in that region. Once the node locations were calculated, an edge $(r,s)$ was added for all $s$ within the $k$-nearest neighbors of each node $r$. We used $k=4$ so that the graph represented spatial adjacency \cite{zuniga2018, zuniga2019}. We note that there was little variation in the final results for other close values of $k$.

\subsubsection{GFL Denoising Example}
\begin{figure}[H]
	\centering
	\captionsetup{justification=centering}
	\begin{center}
		
		\begin{subfigure}[h]{0.325\linewidth}
			\begin{overpic}[width=1\linewidth]{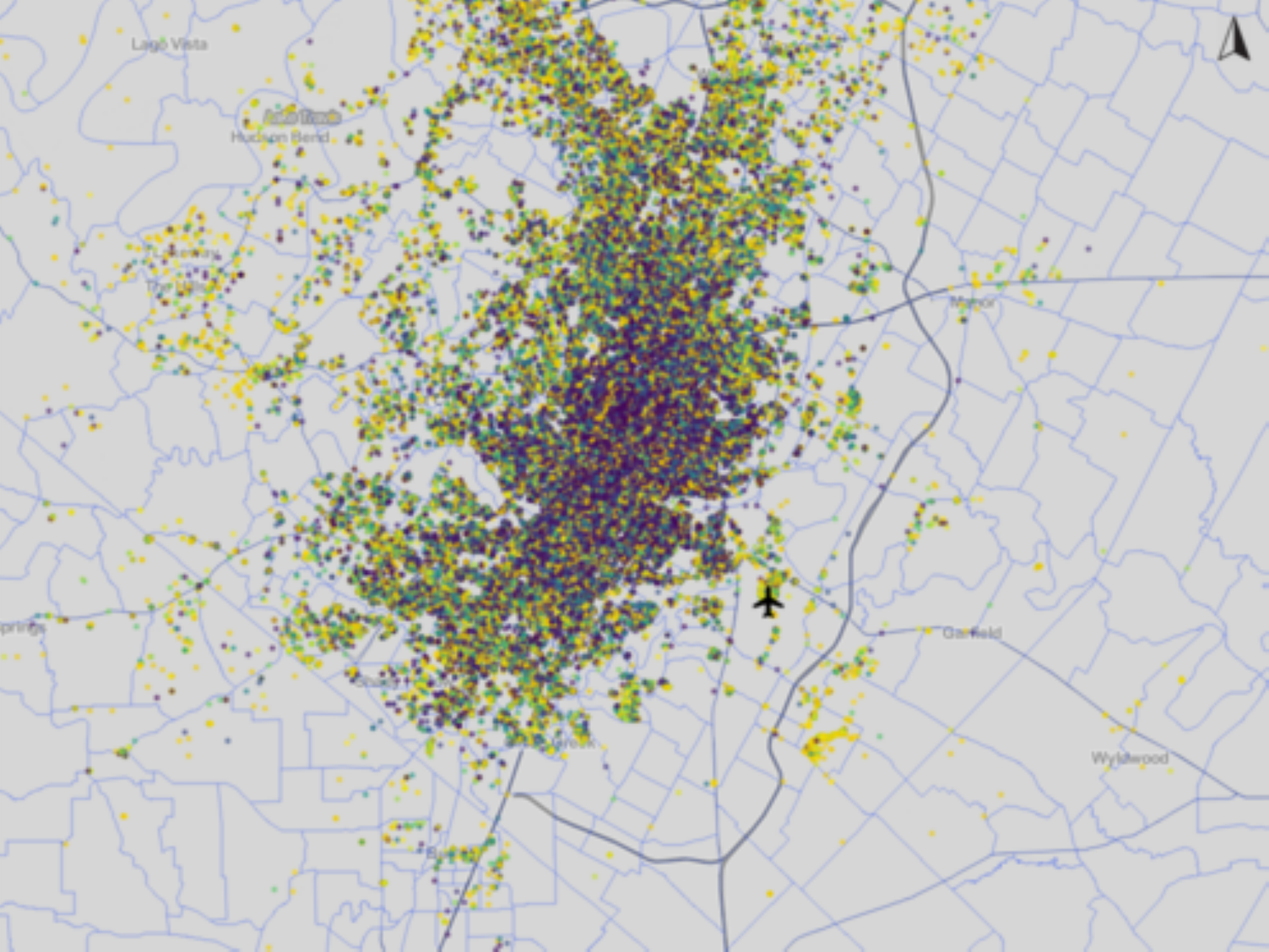}
				\put(0,0){\includegraphics[width=.09\linewidth]{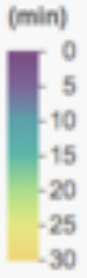}}
			\end{overpic}
			\caption{Idle time data points}
		\end{subfigure}
		\begin{subfigure}[h]{0.325\linewidth}
			\begin{overpic}[width=1\linewidth]{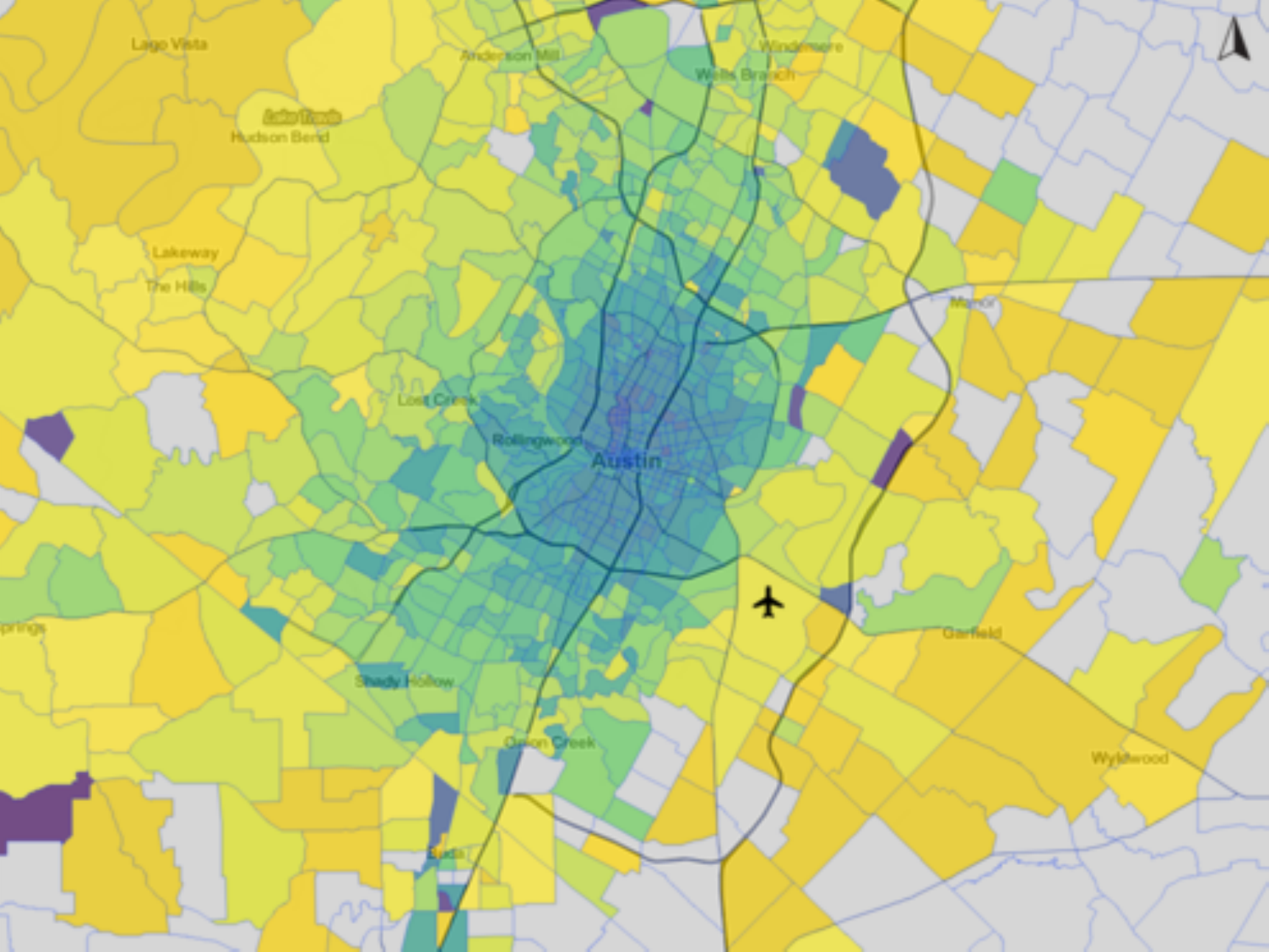}
				\put(0,0){\includegraphics[width=.09\linewidth]{fig/idle-aft/Simb.pdf}}
			\end{overpic}
			\caption{Idle time in TAZs}
		\end{subfigure}
		\begin{subfigure}[h]{0.325\linewidth}
			\begin{overpic}[width=1\linewidth]{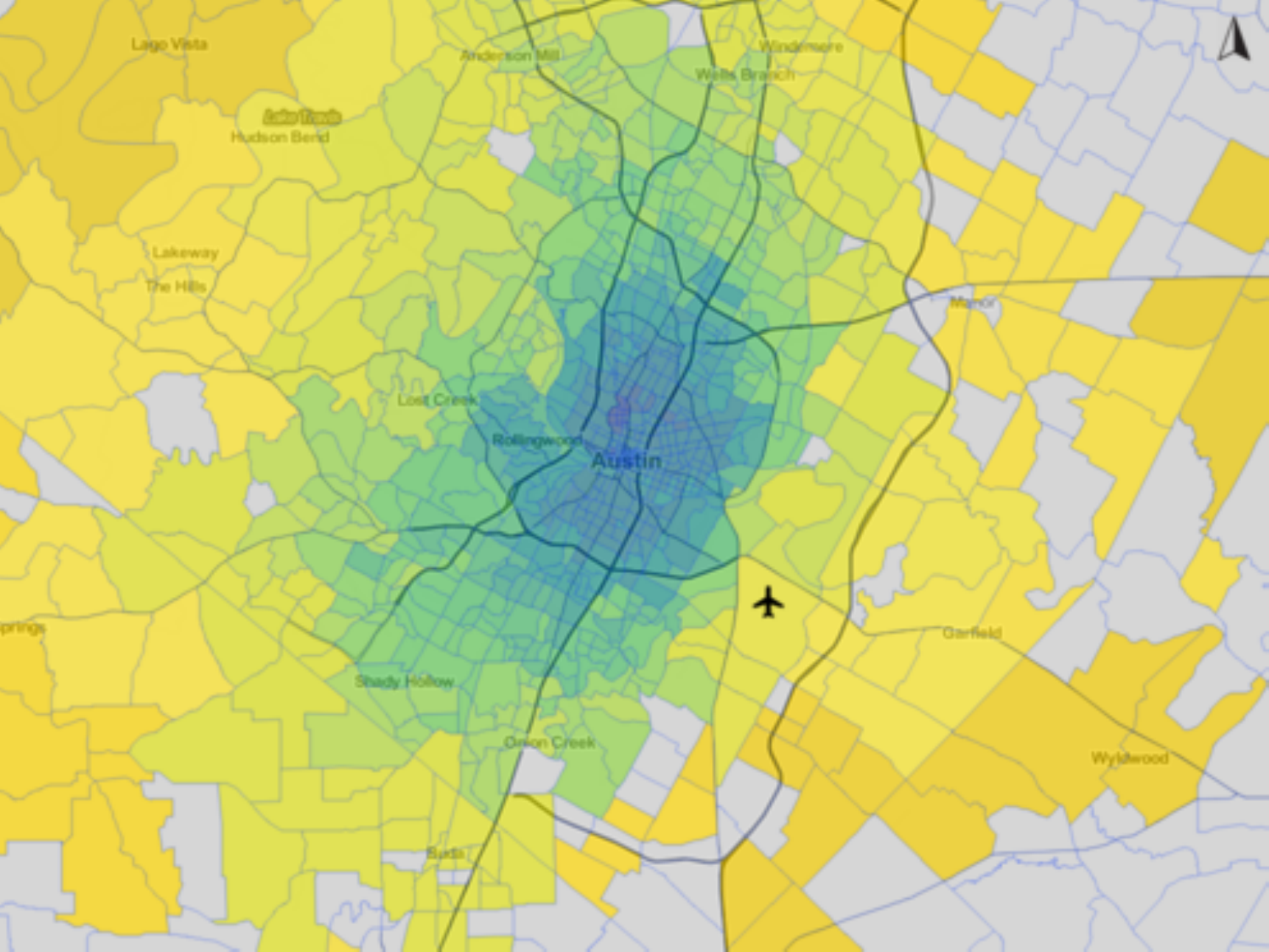}
				\put(0,0){\includegraphics[width=.09\linewidth]{fig/idle-aft/Simb.pdf}}
			\end{overpic}
			\caption{Idle time GFL denoised}
		\end{subfigure}
		\begin{subfigure}[h]{0.325\linewidth}
			\begin{overpic}[width=1\linewidth]{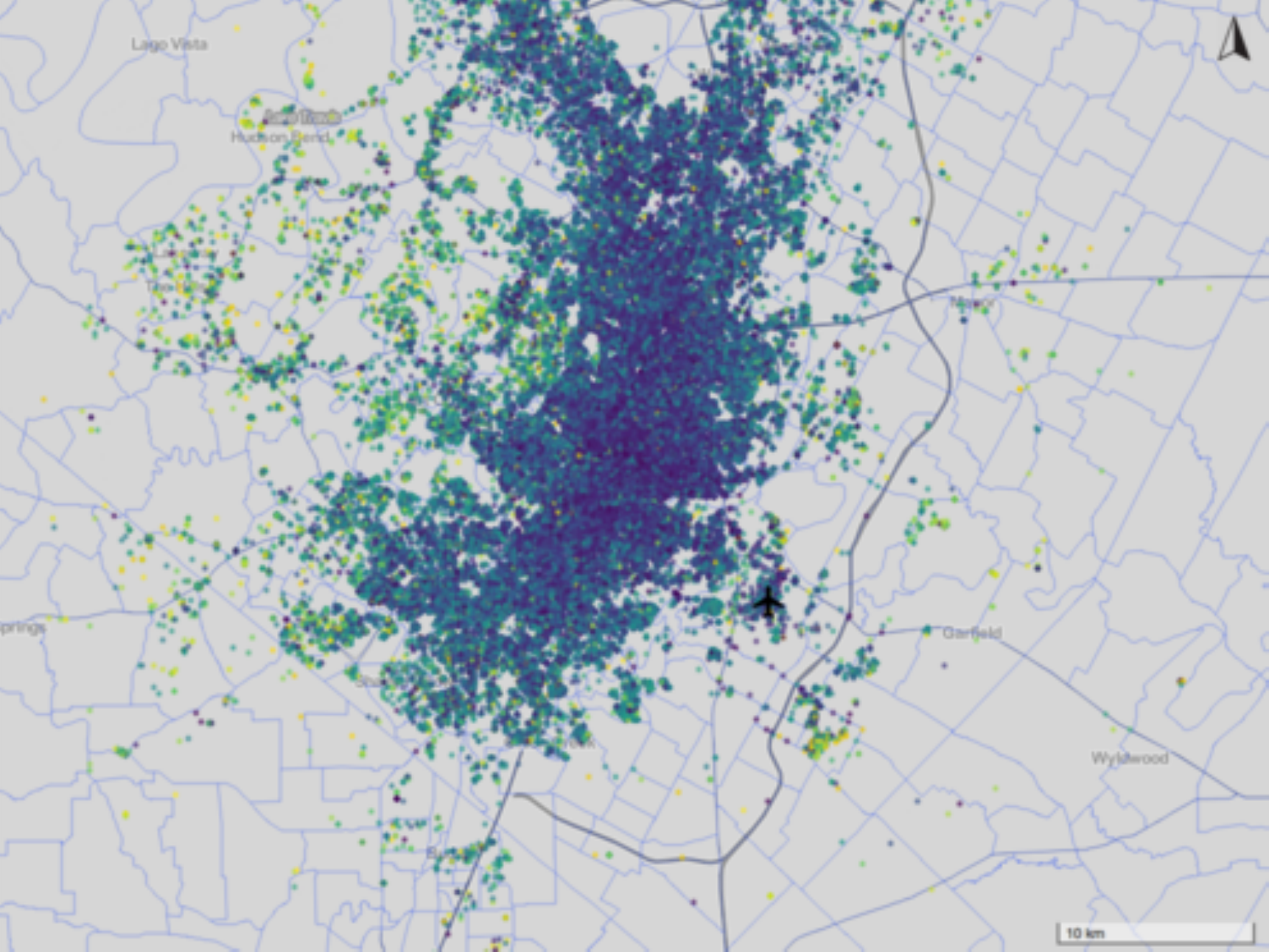}
				\put(0,0){\includegraphics[width=.09\linewidth]{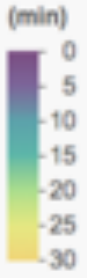}}
			\end{overpic}
			\caption{Reach time data points}
		\end{subfigure}
		\begin{subfigure}[h]{0.325\linewidth}
			\begin{overpic}[width=1\linewidth]{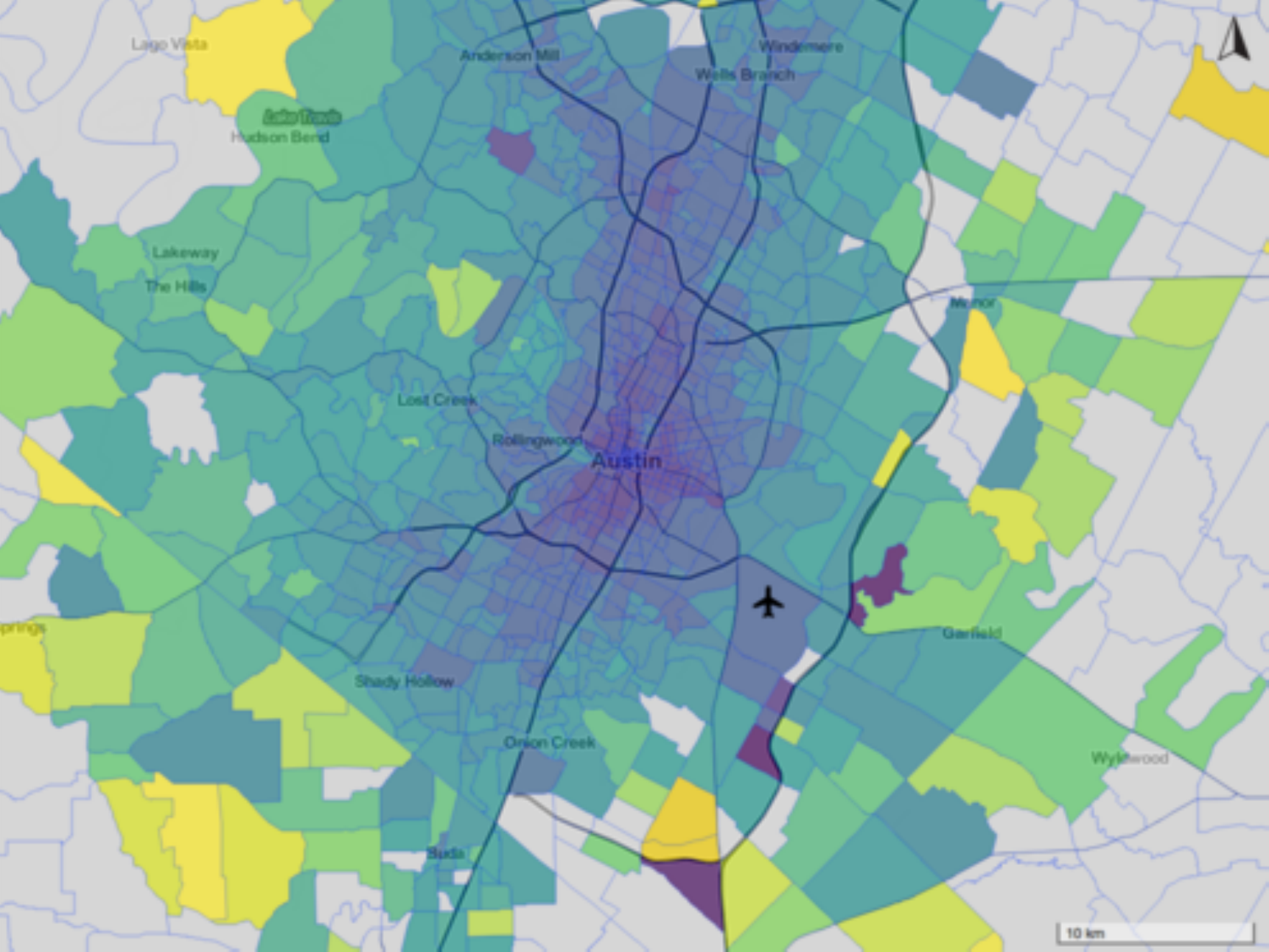}
				\put(0,0){\includegraphics[width=.09\linewidth]{fig/reach/Simb.pdf}}
			\end{overpic}
			\caption{Reach time data in TAZs}
		\end{subfigure}
		\begin{subfigure}[h]{0.325\linewidth}
			\begin{overpic}[width=1\linewidth]{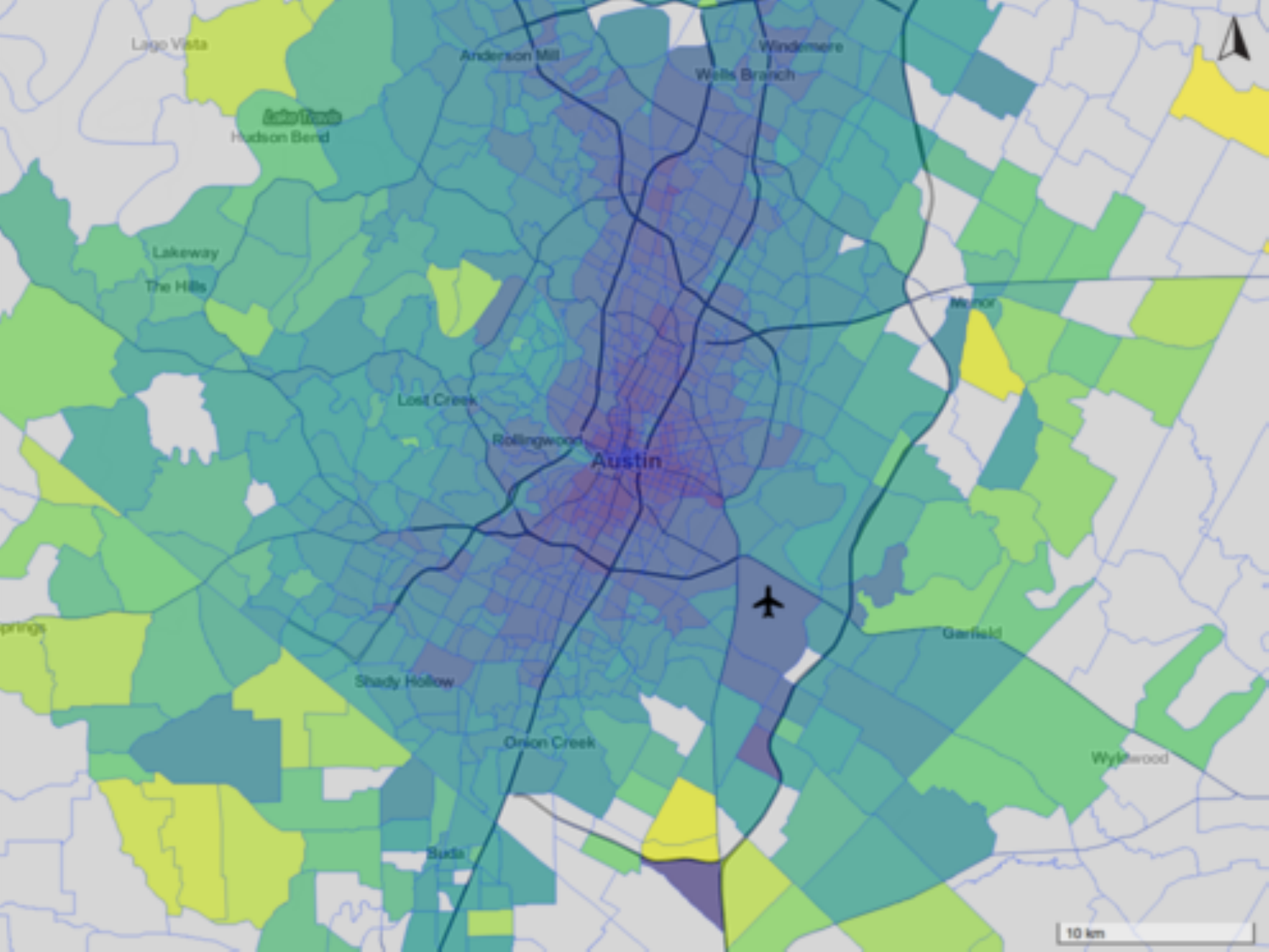}
				\put(0,0){\includegraphics[width=.09\linewidth]{fig/reach/Simb.pdf}}
			\end{overpic}
			\caption{Reach time GFL denoised}
		\end{subfigure}
		\caption{GFL denoising examples (system-wide weekend trips). The left column shows all the observations in the data. The middle column shows averages by TAZ, providing a first level of smoothing, but with noisy data on regions with low counts. The right column shows the GFL output with reduced noise and preserved definition.}
		\label{fig:GLFexamples}
	\end{center}
\end{figure}

\noindent
Figure \ref{fig:GLFexamples} shows examples of the application of the GFL smoothing to the variables of interest. The first image presents the raw data points, where each dot in the map represents the origin of a trip. The next image provides the information summarized per TAZ. Finally, the third image presents the denoised graph. 
The denoised image allows a better interpretation of the spatial distribution of the variables.

\section{Results and Discussion}
\label{sec:sec6}
\noindent
This section presents the principal results and a discussion of the main findings. First, we present an analysis of the system-wide search frictions to provide a general characterization of the conditions on the Austin area. Second, we analyze the results of the proposed performance metrics.  

\subsection{Search Frictions}
\begin{figure}[H]
	\centering
	\captionsetup{justification=centering}
	\begin{center}
		\begin{subfigure}[h]{0.325\linewidth}
			\begin{overpic}[width=1\linewidth]{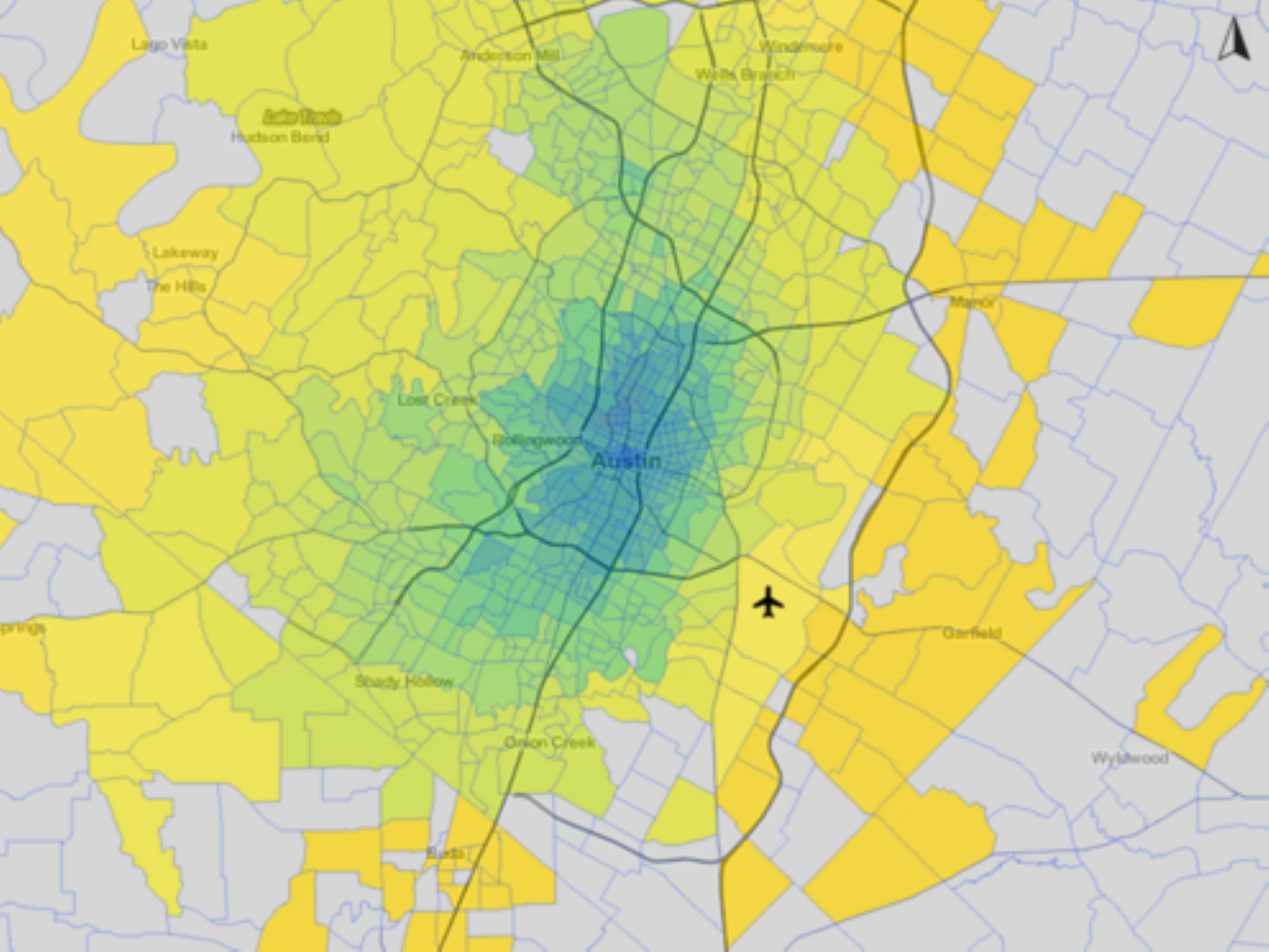}
				\put(0,0){\includegraphics[width=.09\linewidth]{fig/idle-aft/Simb.pdf}}
			\end{overpic}
			\caption{Idle time, peak hours \\ (by destination)}
		\end{subfigure}
		\begin{subfigure}[h]{0.325\linewidth}
			\begin{overpic}[width=1\linewidth]{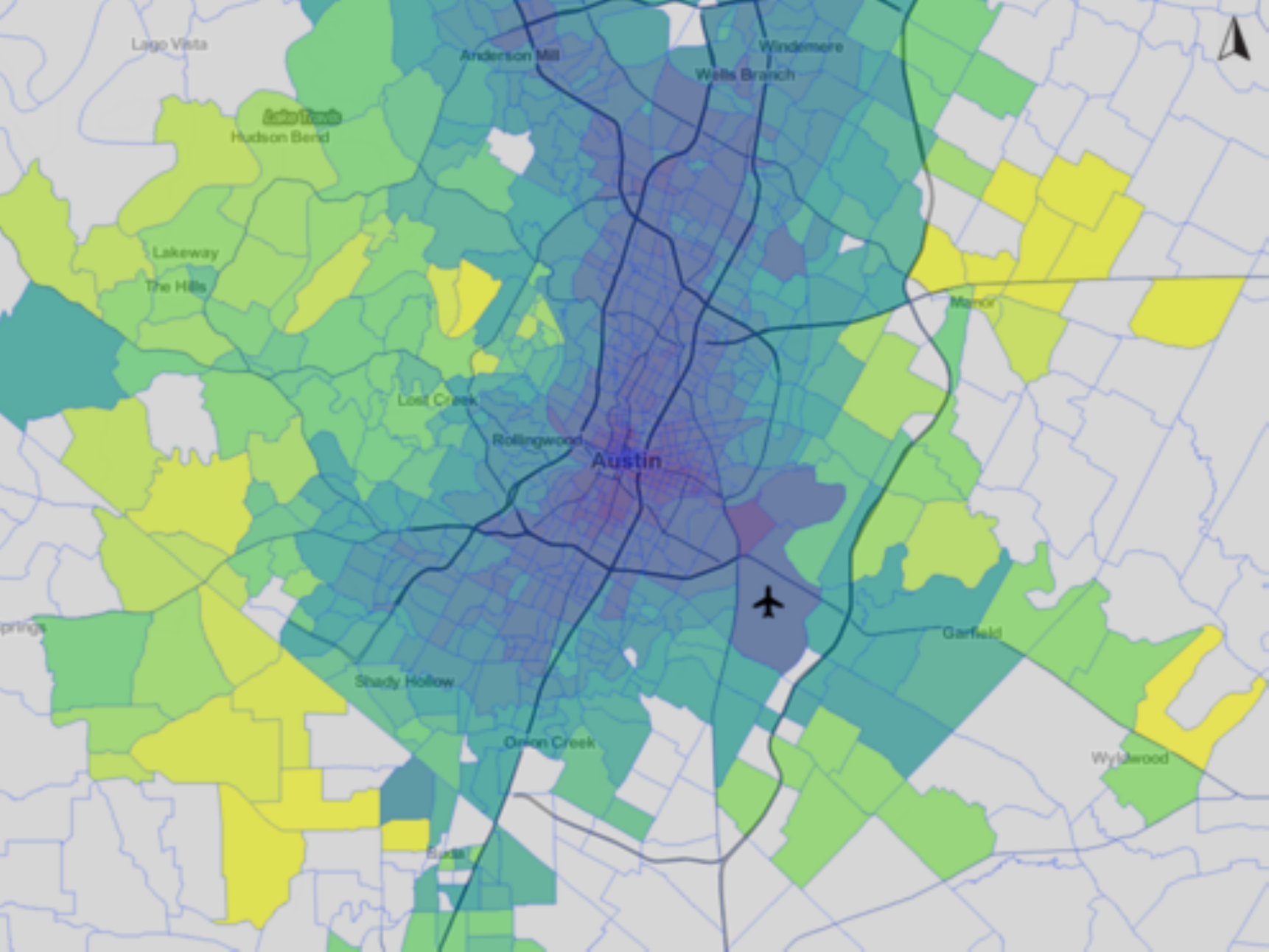}
				\put(0,0){\includegraphics[width=.09\linewidth]{fig/reach/Simb.pdf}}
			\end{overpic}
			\caption{Reach time, peak hours \\ (by origin)}
		\end{subfigure}
	
		\begin{subfigure}[h]{0.325\linewidth}
			\begin{overpic}[width=1\linewidth]{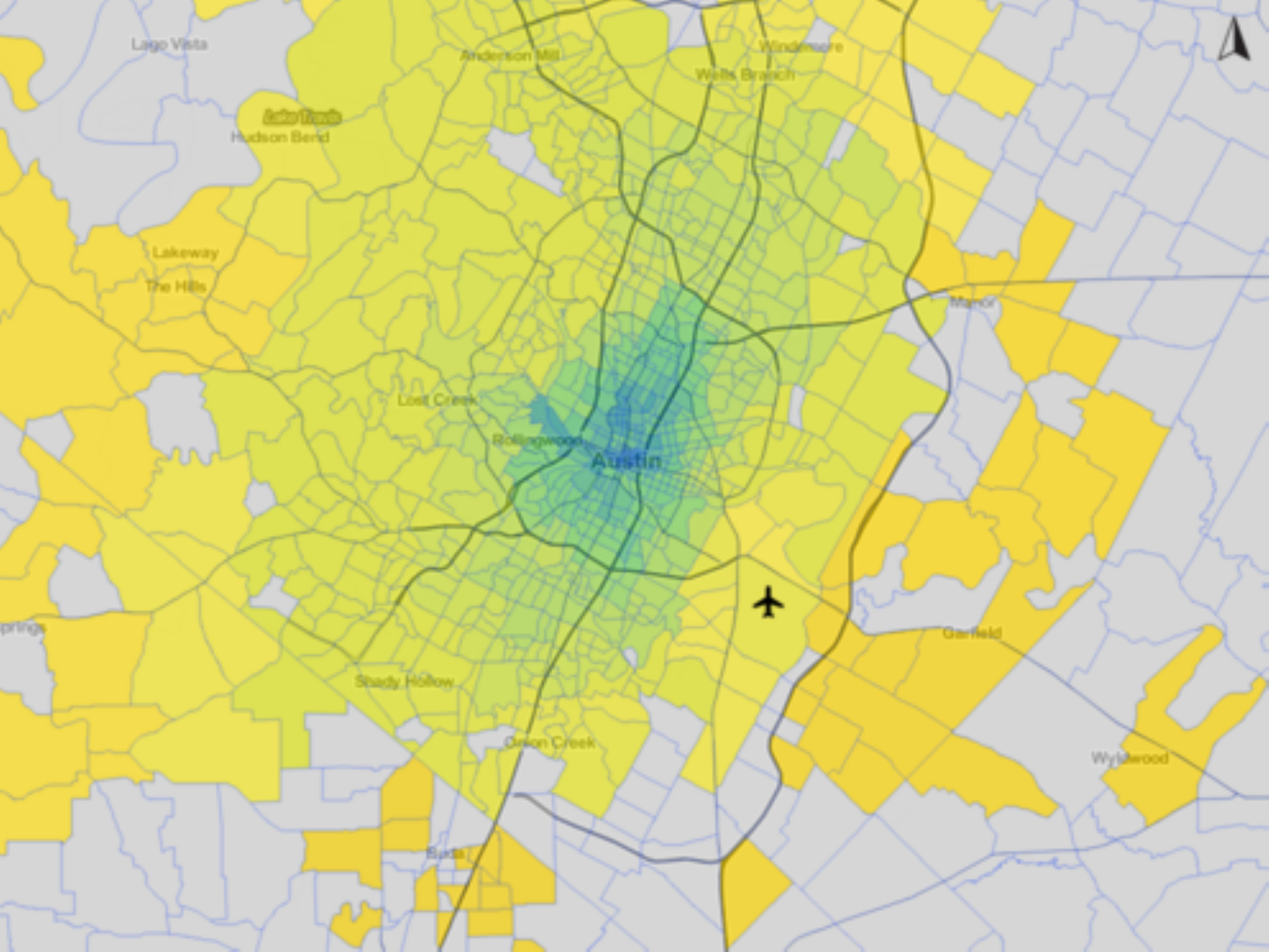}
				\put(0,0){\includegraphics[width=.09\linewidth]{fig/idle-aft/Simb.pdf}}
			\end{overpic}
			\caption{Idle time, mid-day \\ (by destination)}
		\end{subfigure}
	\begin{subfigure}[h]{0.325\linewidth}
			\begin{overpic}[width=1\linewidth]{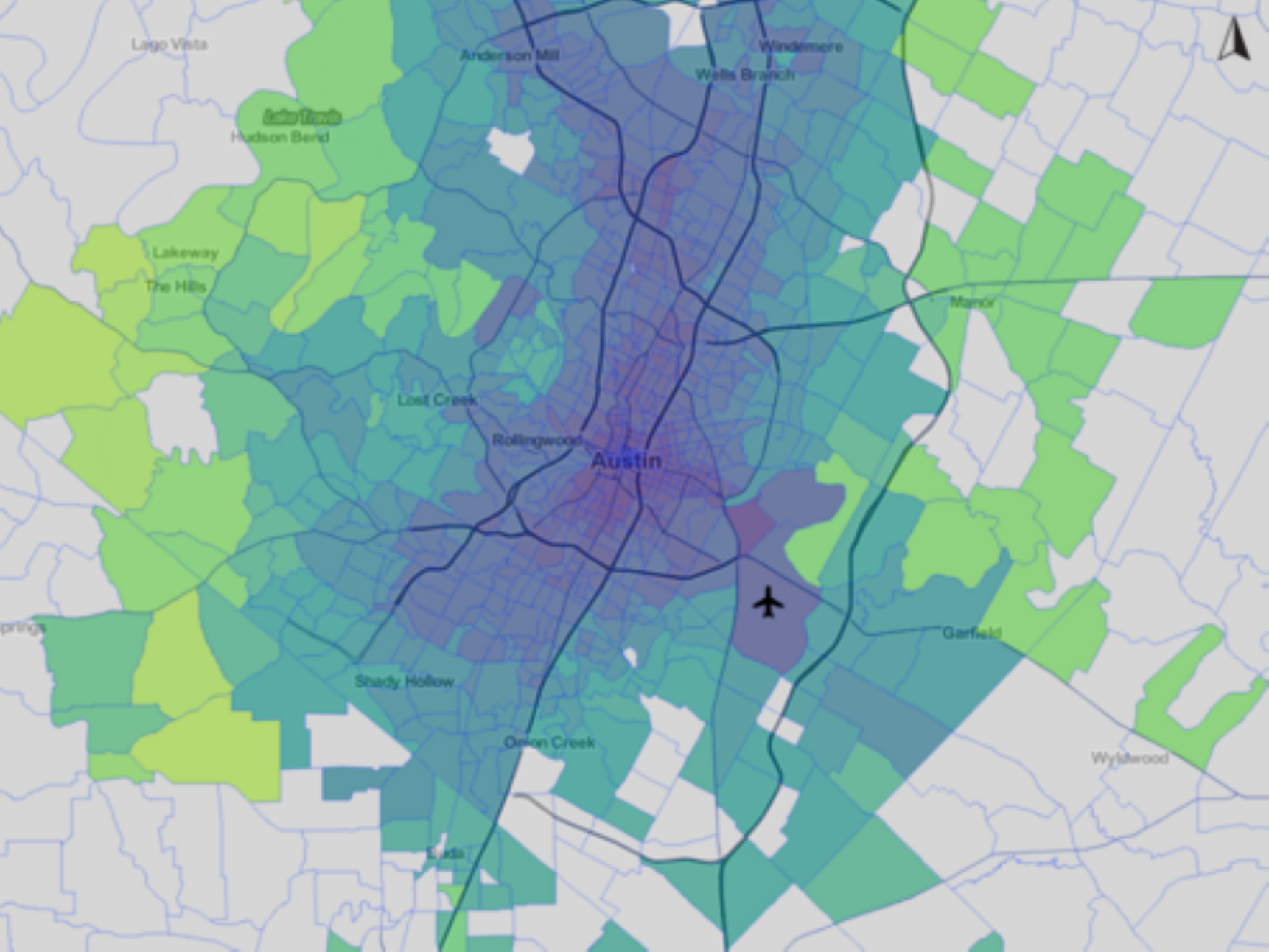}
				\put(0,0){\includegraphics[width=.09\linewidth]{fig/reach/Simb.pdf}}
			\end{overpic}
			\caption{Reach time, mid-day \\ (by origin)}
		\end{subfigure}
		
		\begin{subfigure}[h]{0.325\linewidth}
			\begin{overpic}[width=1\linewidth]{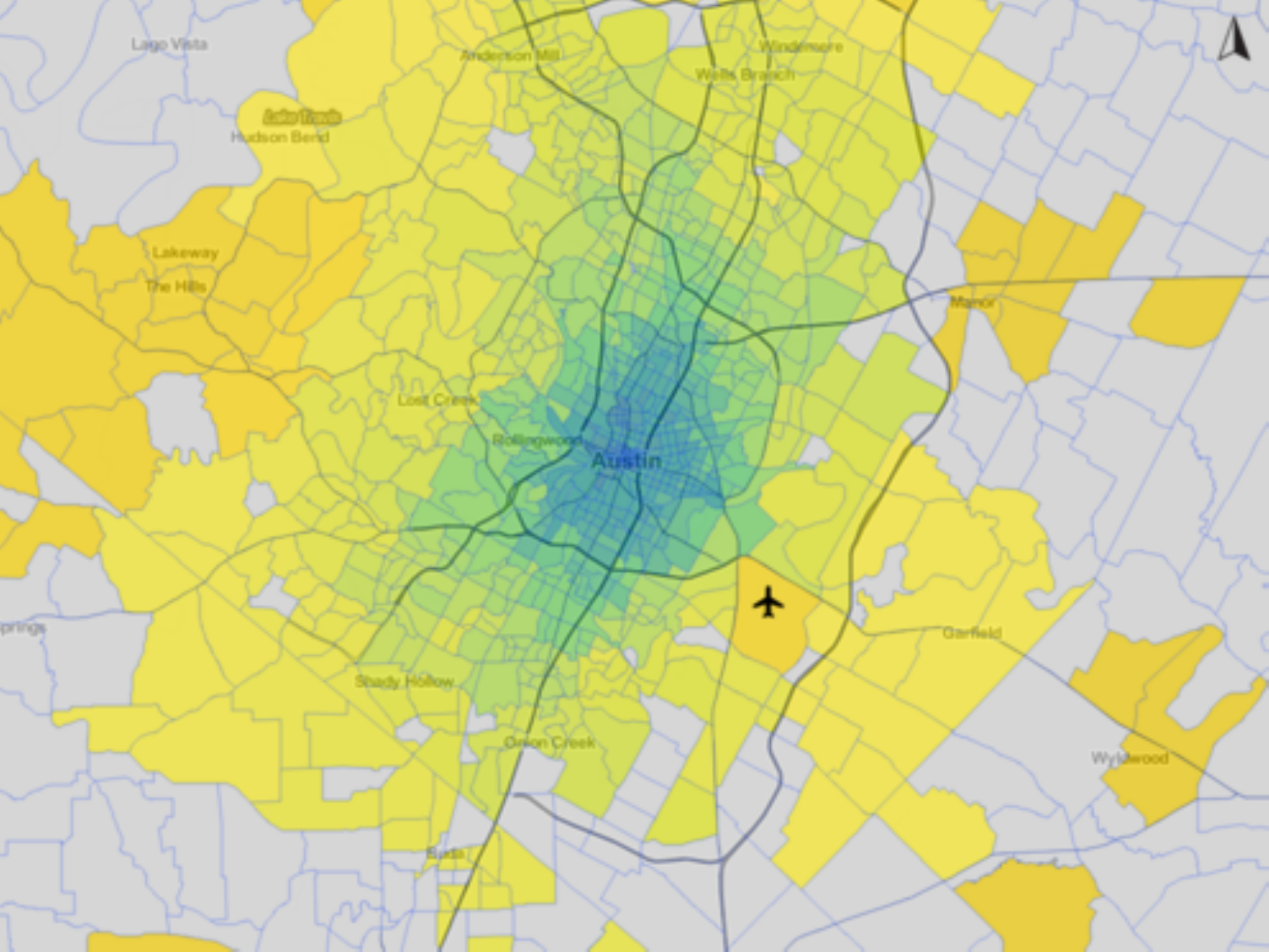}
				\put(0,0){\includegraphics[width=.09\linewidth]{fig/idle-aft/Simb.pdf}}
			\end{overpic}
			\caption{Idle time, overnight \\ (by destination)}
		\end{subfigure}
	\begin{subfigure}[h]{0.325\linewidth}
			\begin{overpic}[width=1\linewidth]{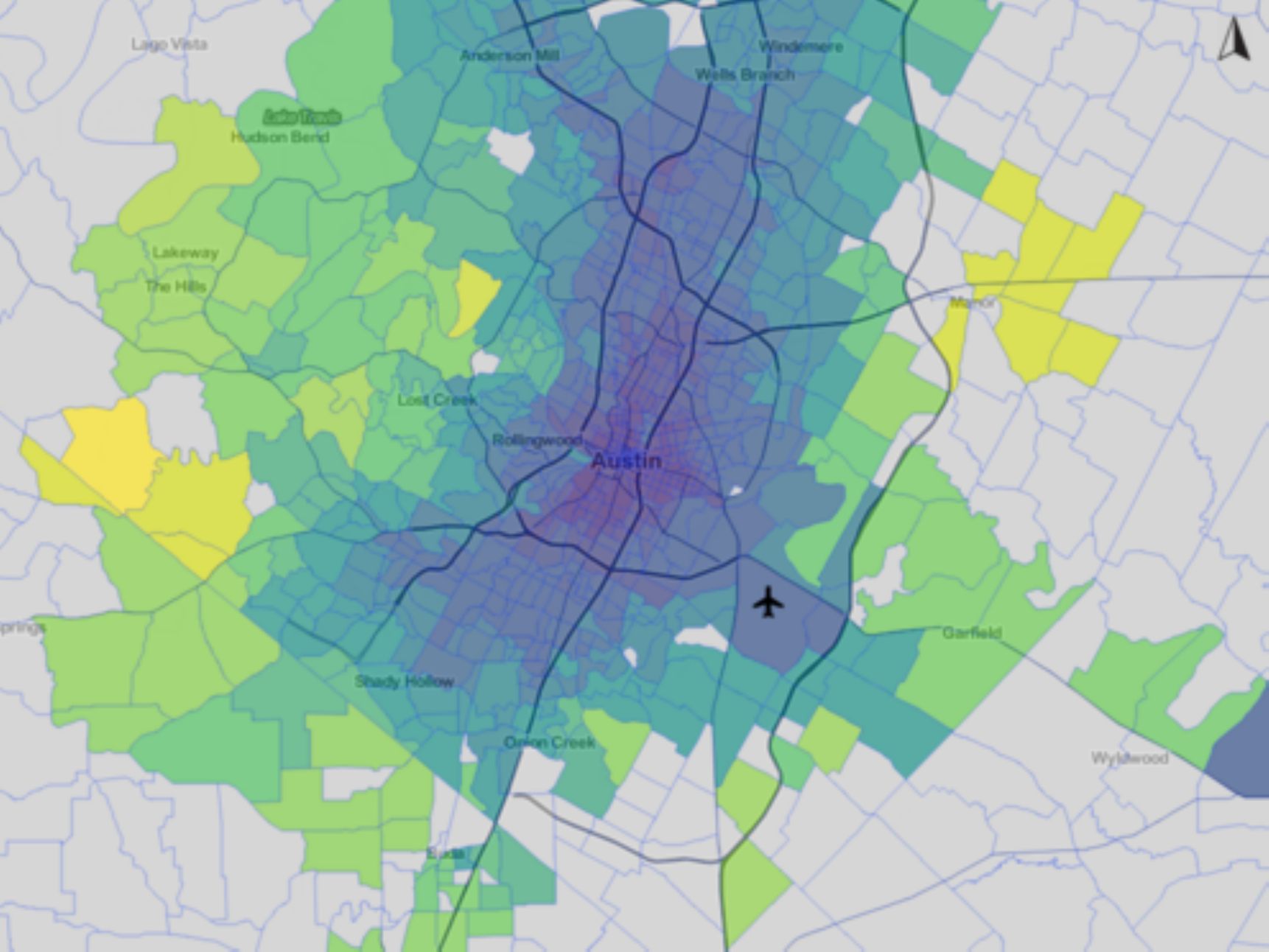}
				\put(0,0){\includegraphics[width=.09\linewidth]{fig/reach/Simb.pdf}}
			\end{overpic}
			\caption{Reach time, overnight \\ (by origin)}
		\end{subfigure}
	
	 	\begin{subfigure}[h]{0.325\linewidth}
	 		\begin{overpic}[width=1\linewidth]{fig/idle-aft/Weekend-smooth.pdf}
	 			\put(0,0){\includegraphics[width=.09\linewidth]{fig/idle-aft/Simb.pdf}}
	 		\end{overpic}
	 		\caption{Idle time, weekend \\ (by destination)}
	 	\end{subfigure}
 		\begin{subfigure}[h]{0.325\linewidth}
 			\begin{overpic}[width=1\linewidth]{fig/reach/Weekend-smooth.pdf}
 				\put(0,0){\includegraphics[width=.09\linewidth]{fig/reach/Simb.pdf}}
 			\end{overpic}
 			\caption{Reach time, weekend \\ (by origin)}
 		\end{subfigure}
		\caption{Search frictions comparison for system-wide trips}
		\label{fig:opersys}
	\end{center}
\end{figure}

\noindent
Using the spatial and time aggregation described in Section \ref{sec:sec4}, we make use of the GFL to estimate the smoothed TAZ-level average values. This section presents results from idle and reach time variables. Idle time is analyzed based on the driver destination, i.e., trip drop-off location. Reach time is analyzed based on the trip origin,  i.e., pick-up location. The results, shown in Figure \ref{fig:opersys}, consist of Austin-area maps for each period (peak hours, mid-day, overnight, and weekend). To facilitate the maps' reading, Table \ref{tab:res1} presents the average value for the total TAZs (or system-wide location), CBD TAZs, and ABIA airport TAZ, for the four analysis periods.


\begin{table}[H]
	\centering
	\caption{Summary of search friction results by location and periods}
	\label{tab:res1}
	\begin{tabular}{lllrrr}
		\hline
		\multicolumn{1}{c}{\multirow{2}{*}{Variable}}       & \multicolumn{2}{c}{\multirow{2}{*}{Period}} & \multicolumn{3}{c}{Location}                                                            \\ \cline{4-6} 
		\multicolumn{1}{c}{}                                & \multicolumn{2}{c}{}                        & \multicolumn{1}{c}{System-wide} & \multicolumn{1}{c}{CBD} & \multicolumn{1}{c}{Airport} \\ \hline
		\multirow{5}{4 cm}{Average idle time, (by destination)
			(min)} & \multirow{3}{*}{Weekday}     & Peak Hours    & 18.4                            & 11.9                    & 24.7                       \\  
		&                             & Mid-Day       & 19.5                            & 13.3                    & 22.5                       \\  
	&                             & Overnight     & 18.6                            & 10.9                    & 27.4                       \\ 
& \multicolumn{2}{l}{Weekend}                 & 17.9                            & 10.0                    & 22.8                                 \\ \hline
		\multirow{5}{4 cm}{Average reach time, (by origin) (min)}     & \multirow{3}{*}{Weekday}    & Peak Hours    & 10.0                            & 5.1                     & 7.2                         \\ 
		&                             & Mid-Day       & 8.9                           & 4.6                     & 6.6                         \\  
		&                             & Overnight     & 9.3                             & 4.7                     & 7.2                         \\ 
		& \multicolumn{2}{l}{Weekend}                 & 10.0                             & 5.1                     & 7.3                         \\ \hline
	\end{tabular}
\end{table}

\subsubsection{Idle Time}
\noindent
The estimated average idle time, summarized by TAZ, varies from 5 to 30 minutes. There is a clear concentration of short times (under 15 minutes) in the central zone, corresponding to dense areas with high population and employment density and with a high concentration of ridership (defined as the number of origin-based trips). For example, the CBD shows idle times between 10 and 13 minutes. Idle time greater than 20 minutes corresponds to suburban areas. For example, a driver ending a trip in the city of Buda (south of Austin) has an expected waiting time of approximately 20-30 minutes until the next trip, similar to areas such as Manor (east of Austin) and the city of Lakeway (west of Austin).
The differences across periods indicates that weekends have the lowest idle time, while mid-day period has the longest average values. During weekends, the majority of urban areas have idle times lower than 15 minutes. During mid-days, corresponding to the lower demand period, the idle time can be up to 30 minutes in the cities of Buda and Garfield. 

The airport zone presents a relatively high idle time, varying from 22 to 27 minutes across the analysis periods. This result can be attributed to the high imbalance between trips into and out of the airport. The total airport-origin trips are 58,558, while the total airport-destination trips are 85,072, a difference of 45 percent with respect to the origin trips.
The lowest idle time corresponds to weekends and mid-day hours. During weekdays, the highest idle time is in the overnight period, while the lowest is during mid-day hours. ABIA airport hourly departure and arriving variation \cite{abia} indicates that there is a high peak of departures in the early morning (before 6 AM), while for midday period the number of departures and arrivals are balanced. Thus, it is possible that the overnight period is capturing this imbalance and, as a consequence, drivers must wait longer for the next trip.

\begin{figure}[H]
	\centering
	\captionsetup{justification=centering}
	\begin{center}
		\begin{subfigure}[h]{0.495\linewidth}
			\includegraphics[width=1\linewidth]{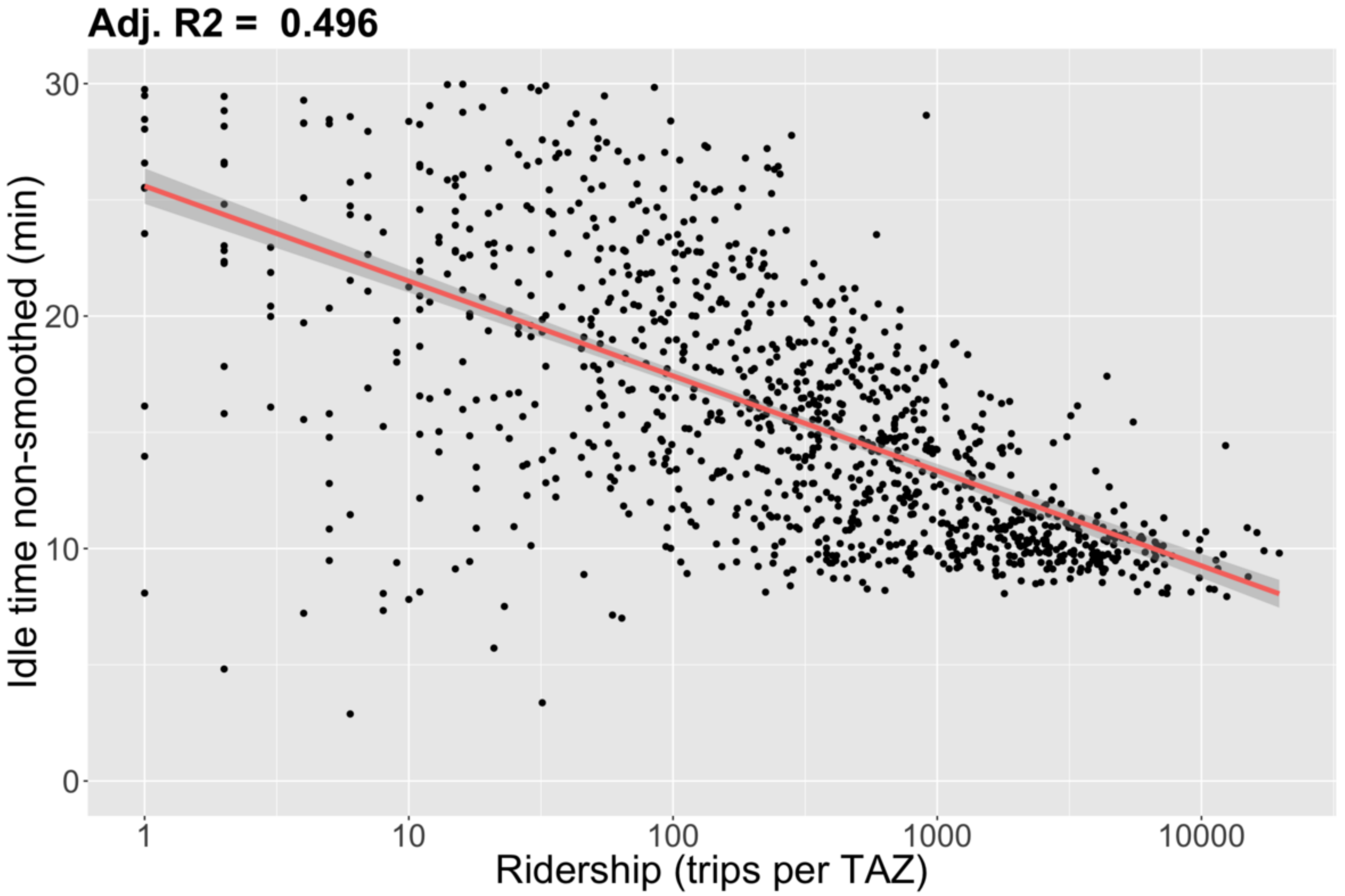}
			\caption{Idle time non-smoothed \\ example (weekend)}
		\end{subfigure}
		\begin{subfigure}[h]{0.495\linewidth}
			\includegraphics[width=1\linewidth]{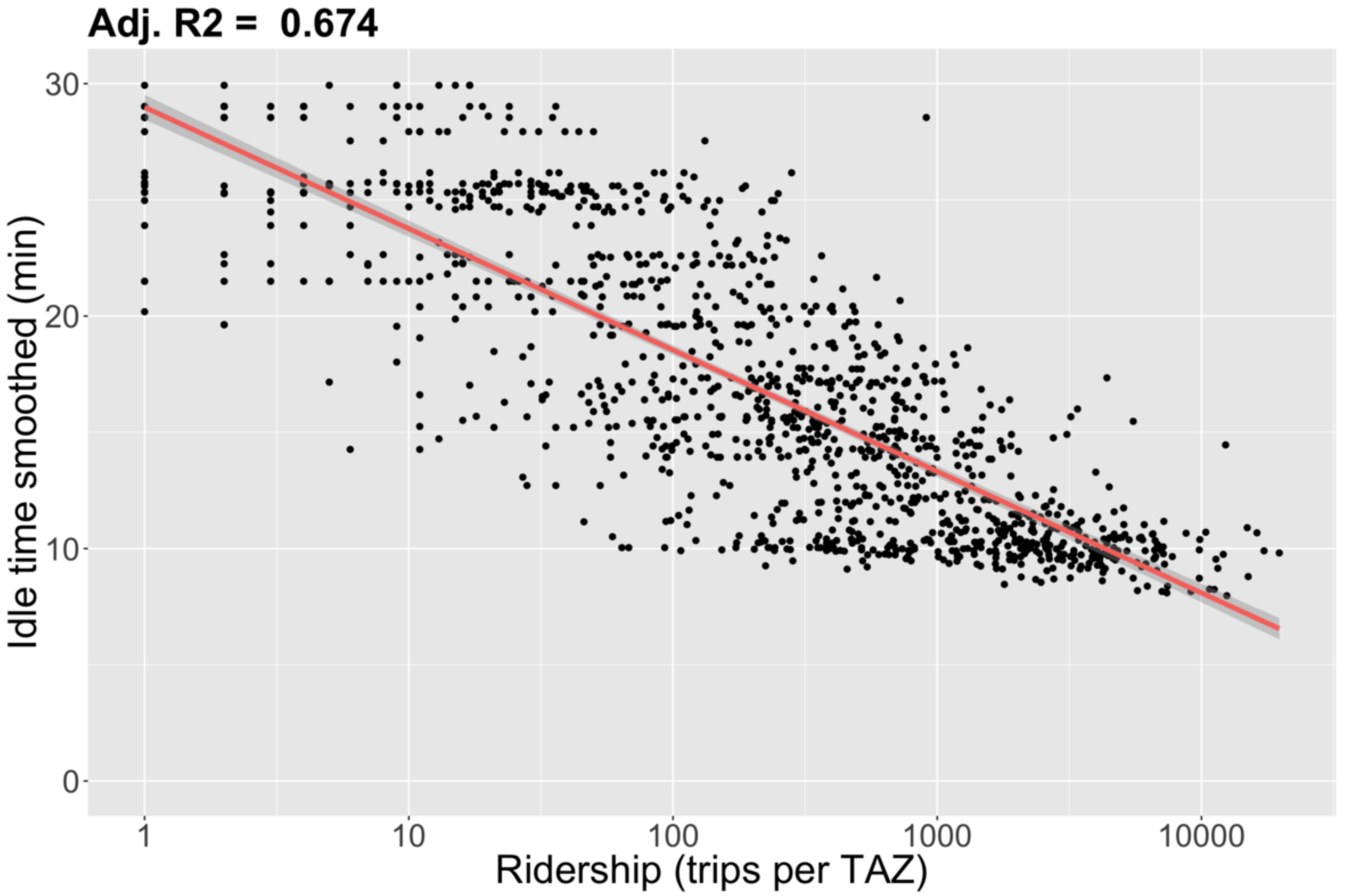}
			\caption{Idle time smoothed \\ example (weekend)}
		\end{subfigure}
		\caption{Relationship between idle time and ridership before and after the GFL smoothing}
		\label{fig:fricresults}
	\end{center}
\end{figure}

\newpage
Driver idle time is related to the demand for the service. In areas where the demand is high, it is expected that the driver idle time is lower than in areas with low demand \cite{idle2019}. We analyze the relationship of idle time and RideAustin ridership, estimated as the trip count, as a surrogate for demand in the area.\footnote{We do not include the airport TAZ in this analysis because of its extremely high ridership values.} We model idle time and the logarithm of ridership and found a high linear correlation between these variables. Figure \ref{fig:fricresults} shows the relationship before and after the smoothing application, for an example of the weekend period. Figure \ref{fig:fricresults}(a) includes the non-smoothed idle time, while Figure \ref{fig:fricresults}(b) shows smoothed idle time. The coefficient of determination or R-squared value (shown at the top of the figure) indicates an improvement in the relationship after the GFL smoothing process, suggesting that the denoising process helped to improve the model.


\begin{figure}[H]
	\centering
	\captionsetup{justification=centering}
	\begin{center}
	\begin{subfigure}[h]{0.325\linewidth}
		\begin{overpic}[width=1\linewidth]{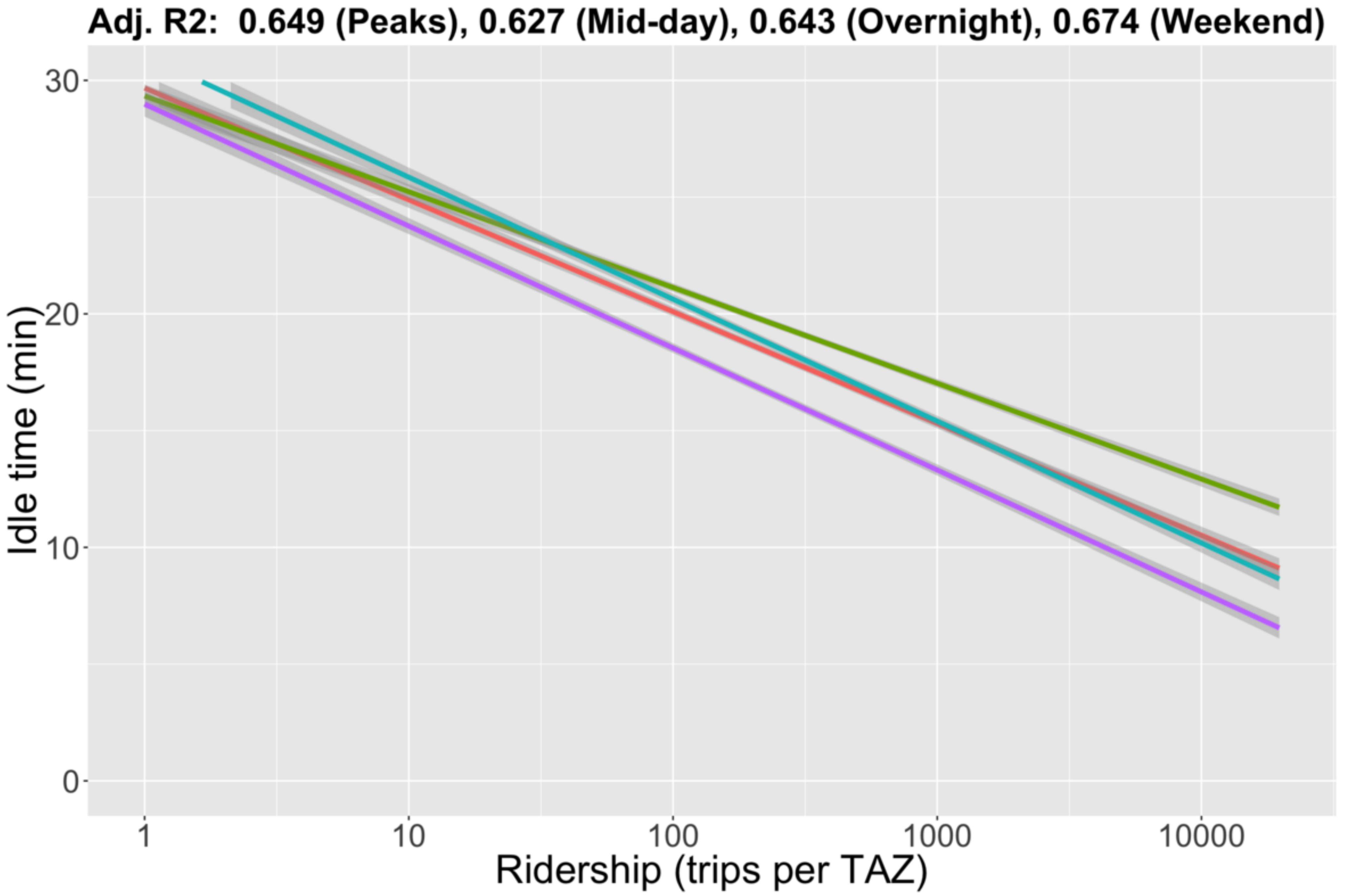}
			\put(40,57){\includegraphics[width=0.55\linewidth]{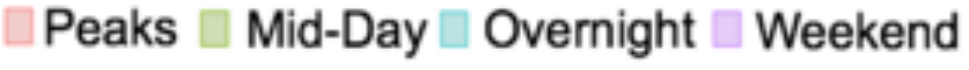}}
		\end{overpic}
		\caption{Idle time and ridership}
	\end{subfigure}
		\begin{subfigure}[h]{0.325\linewidth}
			\begin{overpic}[width=1\linewidth]{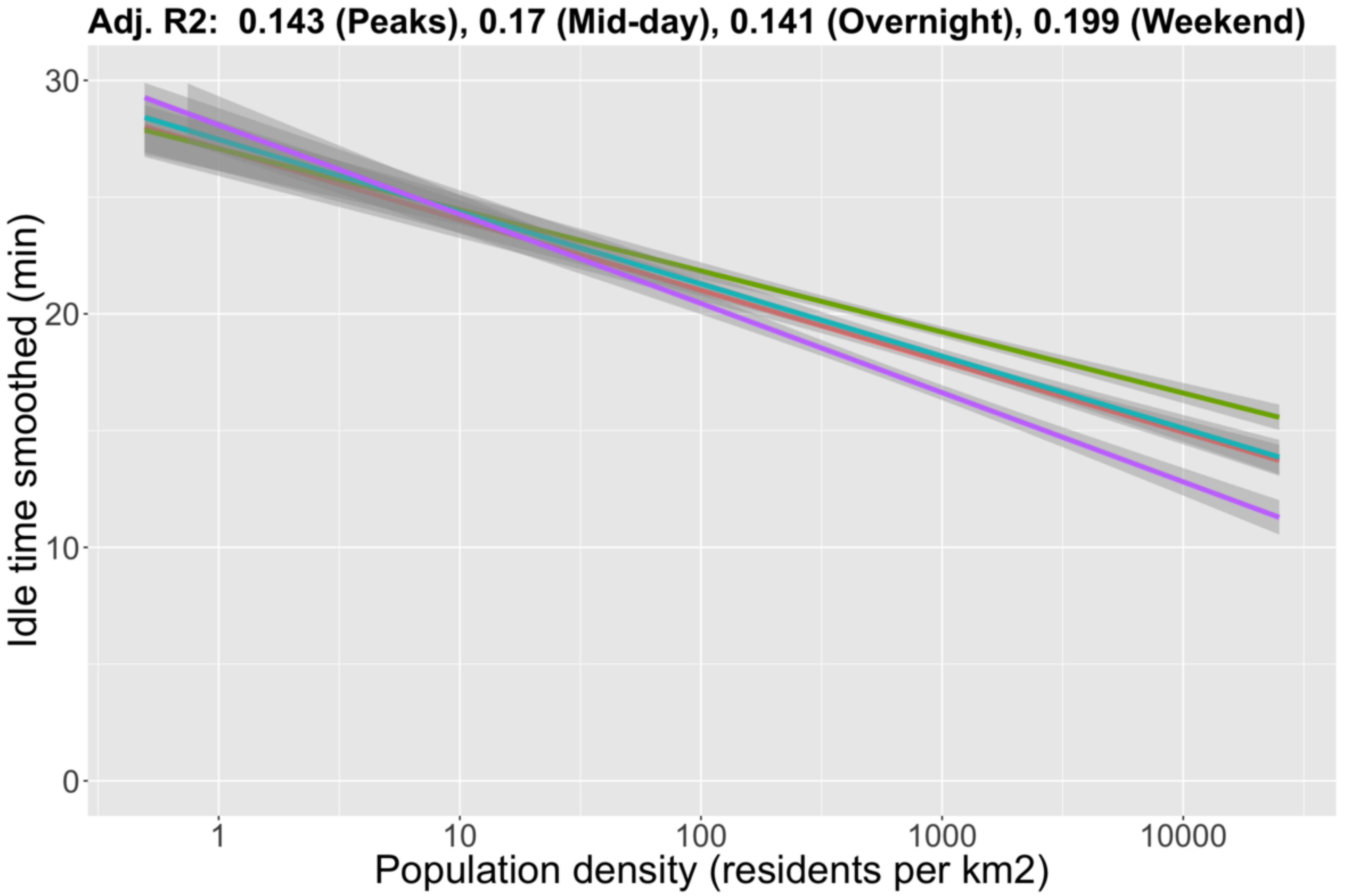}
				\put(40,57){\includegraphics[width=0.55\linewidth]{fig/analysis/Simb2.pdf}}
			\end{overpic}
			\caption{Idle time and population density}
		\end{subfigure}
		\begin{subfigure}[h]{0.325\linewidth}
			\begin{overpic}[width=1\linewidth]{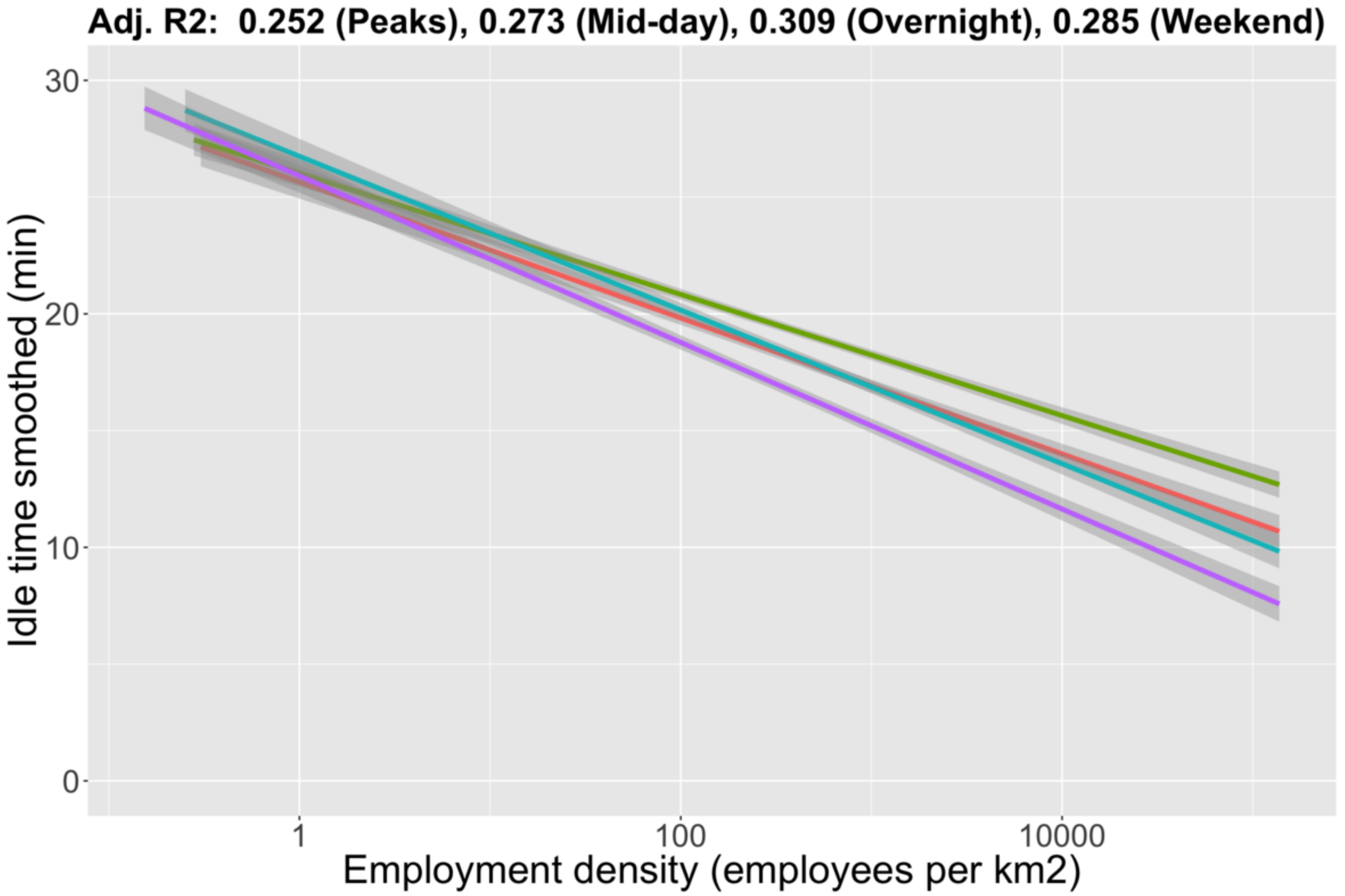}
				\put(40,57){\includegraphics[width=0.55\linewidth]{fig/analysis/Simb2.pdf}}
			\end{overpic}
			\caption{Idle time and employment density}
		\end{subfigure}
		\caption{Idle time for different periods (smoothed data)}
		\label{fig:fricex}
	\end{center}
\end{figure}

Population and employment density are also presumably related to more frequent trip requests and thus to a lower driver waiting times \cite{idle2019}. Using the smoothed values, we analyze the relationships between idle time and the logarithm of ridership, population density, and employment density, during the four analysis periods. 
The results are shown in Figure \ref{fig:fricex}, the R-squared values are shown at the top of the figures.  Figure \ref{fig:fricex}(a) indicates that there is a strong relationship between idle time and ridership, with R-squared value from 0.627 (mid-day) to 0.674 (weekend). Areas with higher ridership are related to low driver idle time and the effect of ridership on idle time is smaller during the midday period, which is the period with lowest trip requests.
The weekend period presents significantly lower idle time compared to the other periods. 

Figures \ref{fig:fricex}(b) and (c) indicate that the relationship between idle time and population/employment density is not as significant as the relationship found with the ridership variable. R-squared values for population density vary from 0.141 (overnight) to 0.199 (weekend), while for employment density it varies from 0.252 (peak hours) to 0.309 (overnight). Results suggest that higher population density and higher employment density are associated with shorter waiting times and that the effect of population and employment density is weakest for the mid-day period. 
Similarly, an evaluation of ride-sourcing waiting times in Seattle, Washington, using data from Uber \cite{idle2019}, found that the effect of population density is weakest shortly after the morning peak hour. However, for
employment density, the relationship found is weakest around the evening peak hour. The authors' explanation for these results hypothesizes that the pool of available drivers in high-density residential areas has been depleted following the morning peak, while it is depleted in high-density employment centers following the afternoon rush. 
In our case, we found that the system peak hours are not the platform peak hours, thus, it
 could be that the relationship is not strong because the majority of trips are concentrated
on non-peak hours and are more related to recreational trips (e.g., weekend rides).

\subsubsection{Reach Time}
\noindent
Reach time results, shown in Figure \ref{fig:opersys} and Table \ref{tab:res1}, indicate average reach times lower than 10 minutes in the majority of the area. The CBD presents driver reach time of approximately 5 minutes across all the evaluated periods. 
Analysis of the different periods suggests that the lowest reach time period corresponds to the mid-day hours, which can be related to the low demand (only 15 percent of the total trips). Weekday peak hours and weekend present the highest reach time values.

For the ABIA airport, the lowest value occurs during weekday mid-day (6.6 minutes) and the highest during weekends (7.3 minutes) and weekday overnight (7.2 minutes). This result can be related to the unbalanced airport demand during weekday overnight, where the majority of trips are concentrated in hours before 6 AM. A greater airport inbound demand can mean longer queues to access the rider pick-up area.
Additionally, Figure \ref{fig:opersys} indicates that there is shorter reach time in northern and southern areas and longer reach time in eastern and western areas that can be attributed to differences in road density in different regions.


\subsection{Productivity}
\noindent
The proposed productivity variables are analyzed based on the destination (drop-off) TAZ of the trips, which allows us to investigate the impact of the trip destination on driver productivity. Using the spatial and time aggregation described in Section \ref{sec:sec4}, we make use of the GFL to estimate the smoothed TAZ-level average values. This section describes the results for destination and driver productivity.

\begin{figure}[H]
	\centering
	\captionsetup{justification=centering}
	\begin{center}
		\begin{subfigure}[h]{0.325\linewidth}
			\begin{overpic}[width=0.95\linewidth]{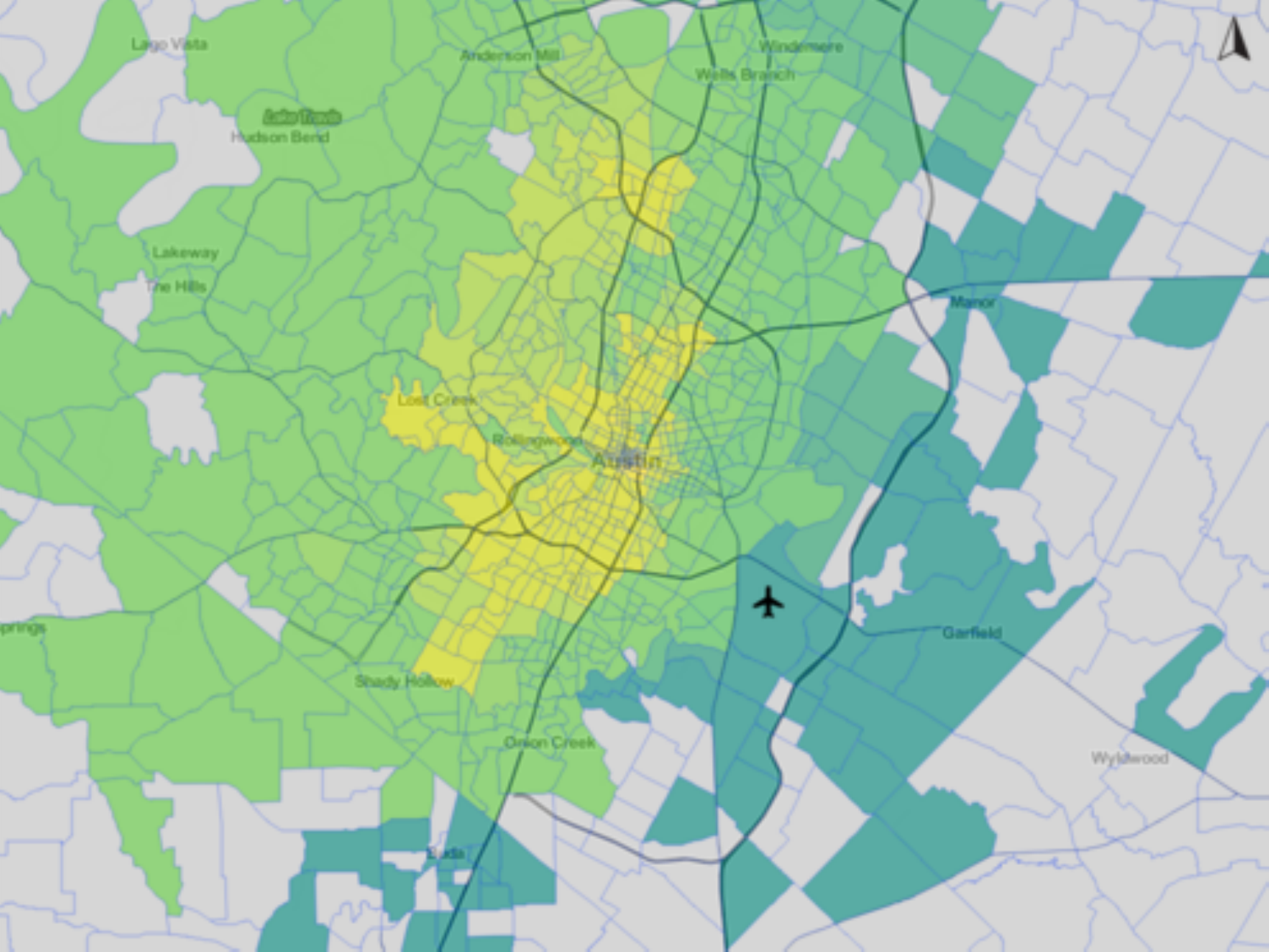}
				\put(0,0){\includegraphics[width=.09\linewidth]{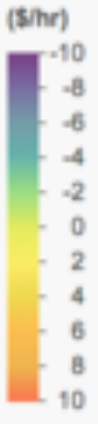}}
			\end{overpic}
			\caption{Peak hours}
		\end{subfigure}
		\begin{subfigure}[h]{0.325\linewidth}
			\begin{overpic}[width=0.95\linewidth]{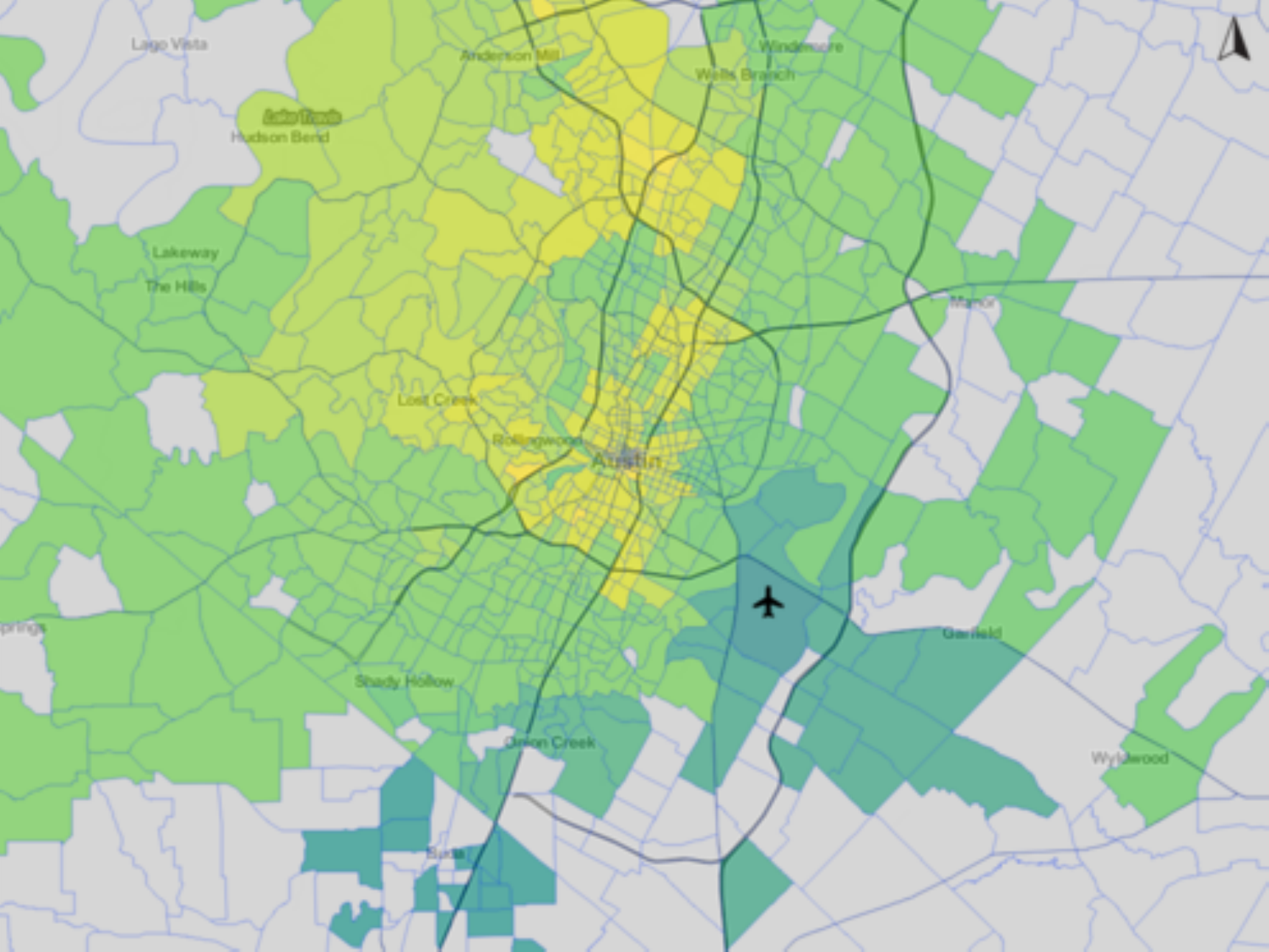}
				\put(0,0){\includegraphics[width=.09\linewidth]{fig/prodDs/Simb.pdf}}
			\end{overpic}
			\caption{Mid-day}
		\end{subfigure}
		
		\begin{subfigure}[h]{0.325\linewidth}
			\begin{overpic}[width=0.95\linewidth]{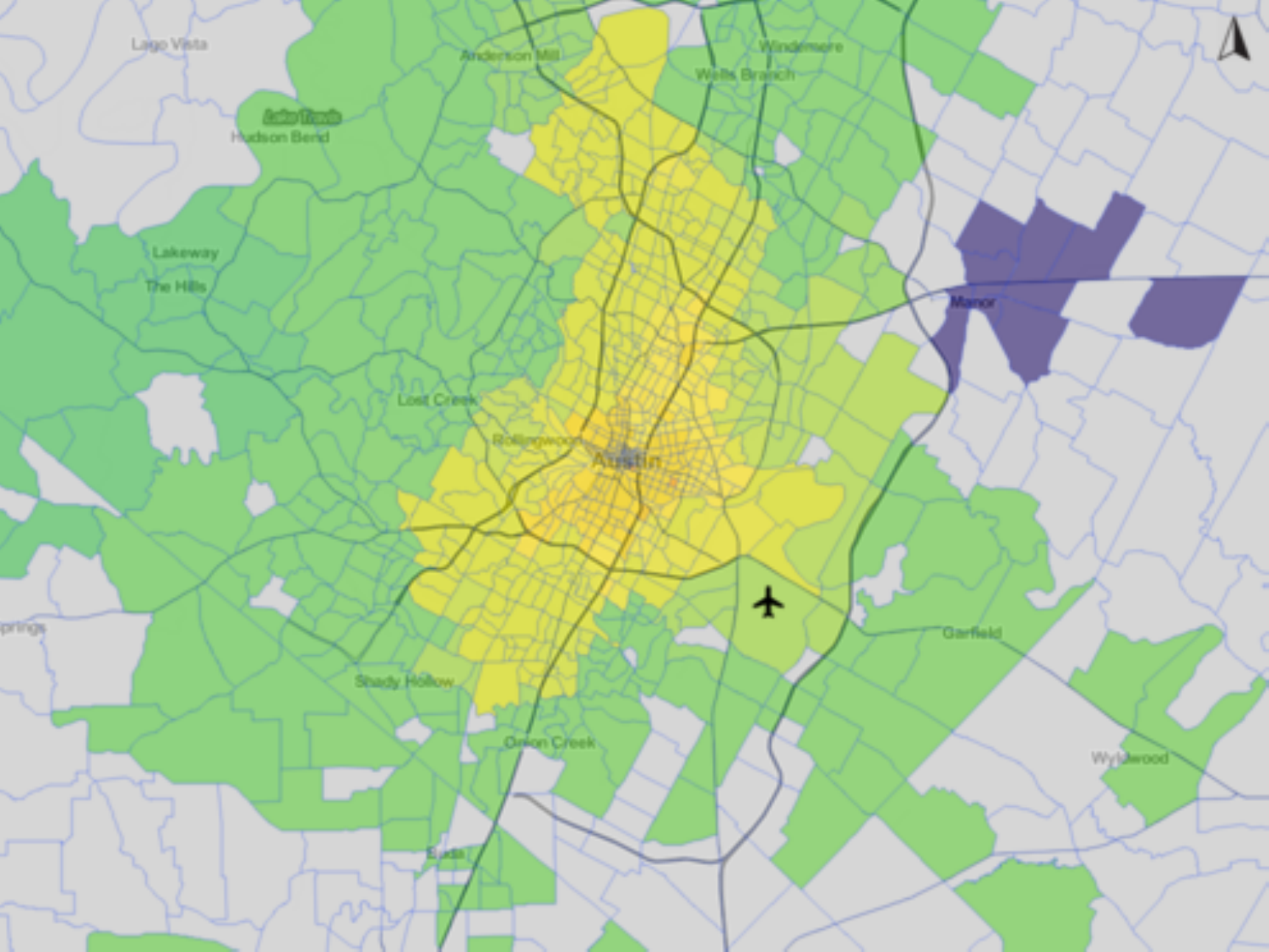}
				\put(0,0){\includegraphics[width=.09\linewidth]{fig/prodDs/Simb.pdf}}
			\end{overpic}
			\caption{Overnight}
		\end{subfigure}
		\begin{subfigure}[h]{0.325\linewidth}
			\begin{overpic}[width=0.95\linewidth]{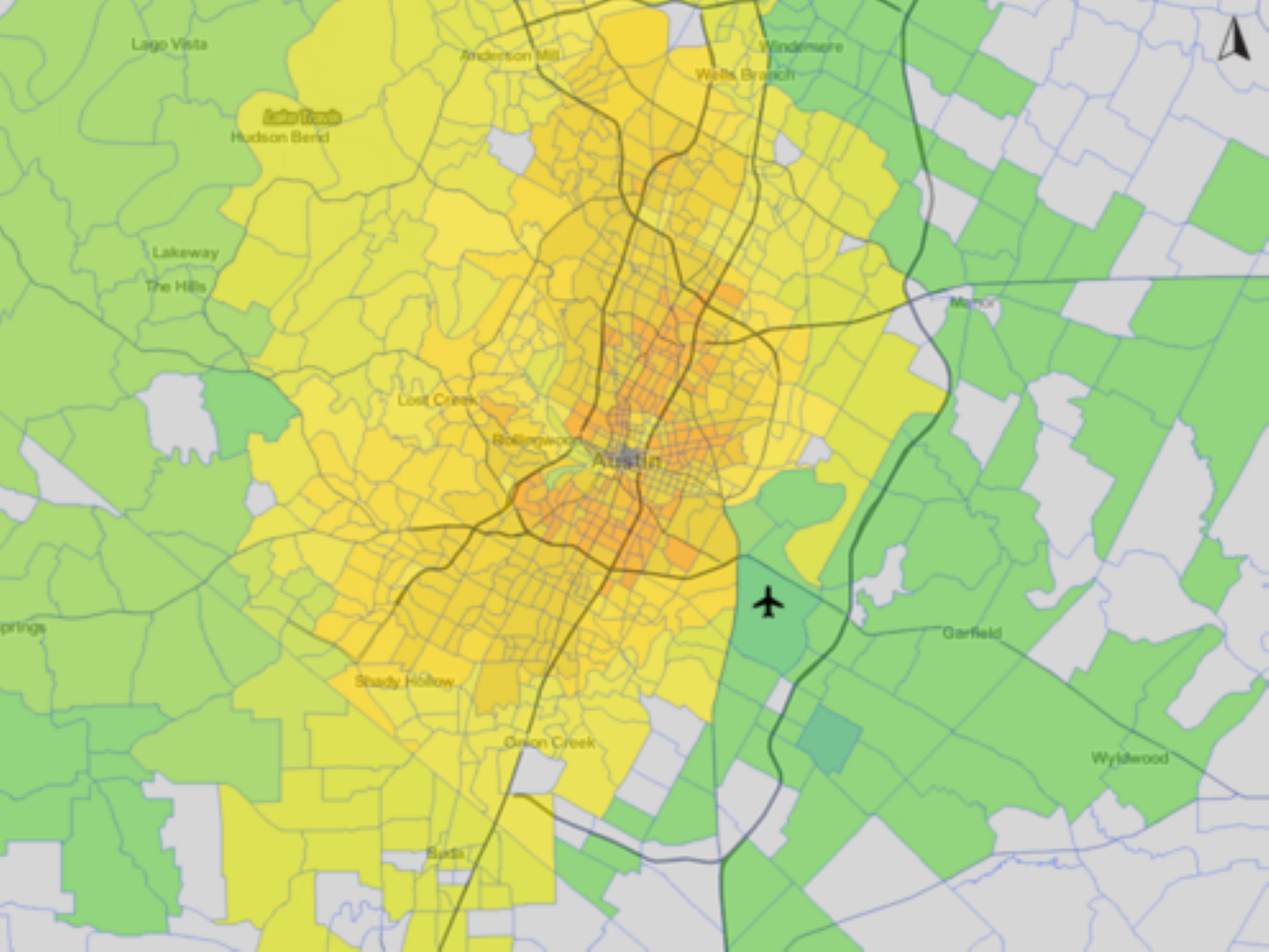}
				\put(0,0){\includegraphics[width=.09\linewidth]{fig/prodDs/Simb.pdf}}
			\end{overpic}
			\caption{Weekend}
		\end{subfigure}
		\begin{subfigure}[h]{0.325\linewidth}
			\begin{overpic}[width=.95\linewidth]{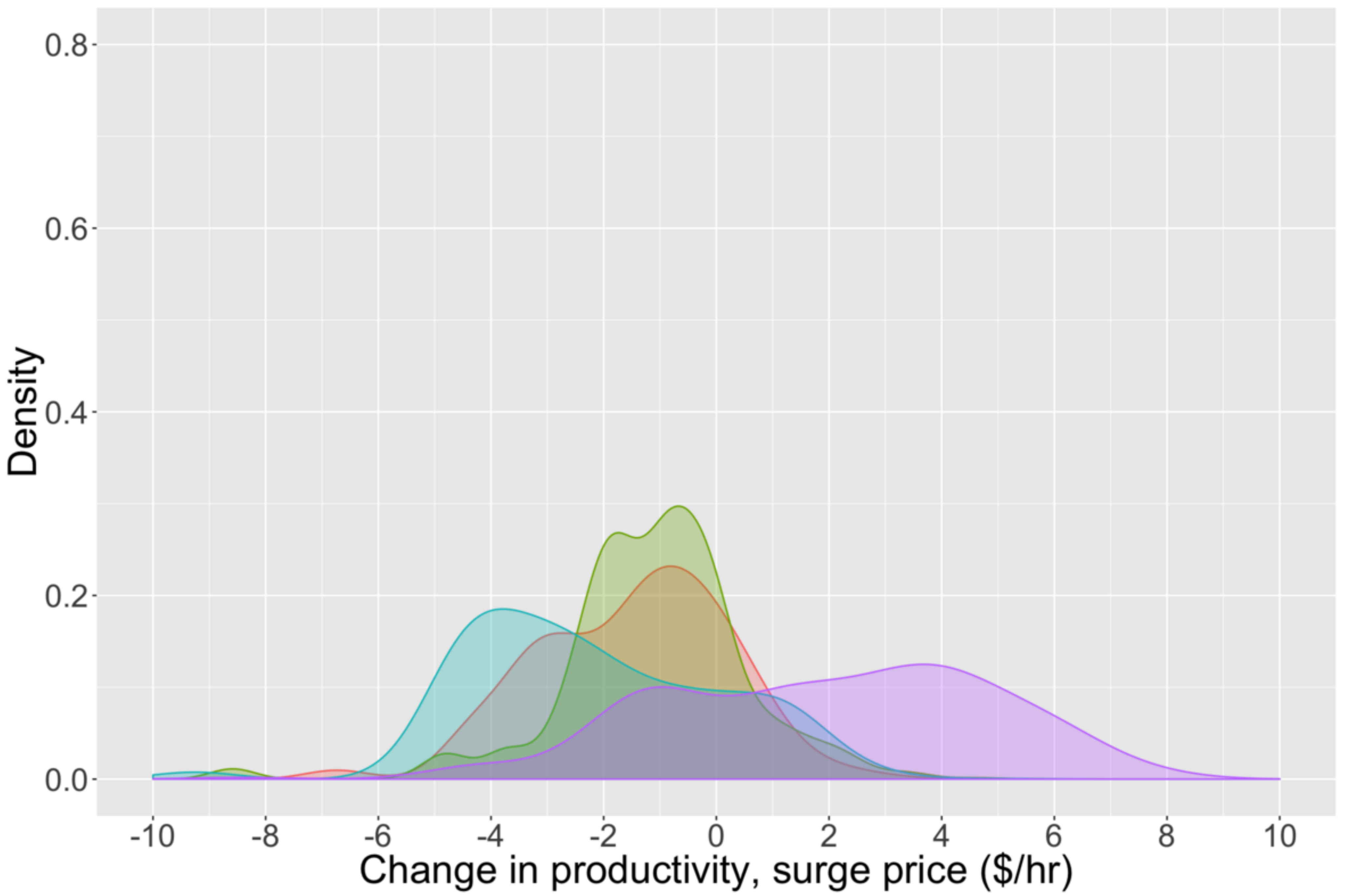}
				\put(40,60){\includegraphics[width=0.55\linewidth]{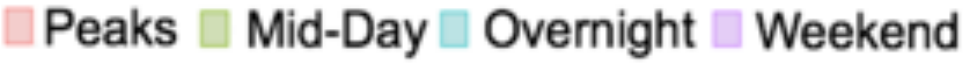}}
			\end{overpic}
			\caption{Density}
		\end{subfigure}
		\caption{Change in continuation payoff with respect to the average for trips with surge price, by destination}
		\label{fig:prods}
	\end{center}
\end{figure}

Table \ref{tab:res3s} presents the average productivity values for system-wide location as well as, CBD TAZs, and ABIA airport TAZ, for the four analysis periods. To highlight the differences, we provide the Austin-area maps for each period using the changes in productivity estimated by normalizing productivity metrics using the mean values.
The results are shown in Figures \ref{fig:prods} and \ref{fig:prodCs}. Figures \ref{fig:prods}(e) and \ref{fig:prodCs}(e) provide a representation of the distribution of the data using the kernel density, which allows for an easier interpretation of the maps.

The results for the estimations using the flat fare values are shown in \ref{app:S}. Table \ref{tab:res3} summarizes the average productivity results, while Figures \ref{fig:prod} and \ref{fig:prodC} present the Austin-area maps.


\begin{table}[H]
	\centering
	\caption{Summary of productivity results by location and period, surge price}
	\label{tab:res3s}
	\begin{tabular}{lllrrr}
		\hline
		\multicolumn{1}{c}{\multirow{2}{*}{Variable}}     & \multicolumn{2}{c}{\multirow{2}{*}{Period}} & \multicolumn{3}{c}{Destination Location}                                                \\ \cline{4-6} 
		\multicolumn{1}{c}{}                              & \multicolumn{2}{c}{}                        & \multicolumn{1}{c}{System-wide} & \multicolumn{1}{c}{CBD} & \multicolumn{1}{c}{Airport} \\ \hline
		\multirow{5}{3 cm}{Average continuation payoff, surge price (\$/hr)} & \multirow{3}{*}{Weekday}    & Peak Hours    & 19.2                            & 21.7                    & 16.5                        \\ 
		&                             & Mid-Day       & 19.6                            & 22.3                    & 15.3                        \\  
		&                             & Overnight     & 20.6                            & 24.5                    & 20.0                        \\ 
		& \multicolumn{2}{l}{Weekend}                 & 22.7                            & 25.4                    & 17.7                        \\ \cline{2-6} 
		& \multicolumn{2}{l}{\textit{Total}}          & \textbf{\textit{20.6}}                   & \textit{23.5}           & \textit{17.4}               \\ \hline
		\multirow{5}{3 cm}{Average driver prod., surge price (\$/hr)}  & \multirow{3}{*}{Weekday}    & Peak Hours    & 29.2                            & 26.8                    & 29.1                       \\  
		&                             & Mid-Day       & 28.3                           & 26.8                    & 26.8                        \\  
		&                             & Overnight     & 35.2                            & 31.9                   & 34.1                       \\ 
		& \multicolumn{2}{l}{Weekend}                 & 40.7                           & 33.9                    & 31.5                        \\ \cline{2-6} 
		& \multicolumn{2}{l}{\textit{Total}}          & \textbf{\textit{34.0}}                   & \textit{29.8}           & \textit{30.4}               \\ \hline
	\end{tabular}
\end{table}


\subsubsection{Continuation payoff}
\noindent
Destination productivity or continuation payoff is a measure of the expected future driver profit given that a trip ended in a specific location. 
Results for surge price values (Figure \ref{fig:prods}) show a change in productivity between the range of -\$8/hr and \$8/hr, a difference of \$16/hr. Trips that ended in the CBD and central area resulted in highest performance values, especially for trips during overnight hours and weekends. 
The spatial distribution during mid-days, the period with the lowest demand, varies approximately \$4/hr, while for weekends, the period with the highest demand, there is a \$14/hr variation, showing the most significant contrast. 
Airport trips present results below the system-wide mean, which can be related to the high idle time values discussed in the previous section.  
Results show that for flat fares (Figure \ref{fig:prod}), the change in productivity varies between -\$5/hr and \$5/hr, a global difference of \$10/hr. 
There is a high concentration of values that do not change drastically with respect to the mean, as shown in the density plot in \ref{fig:prod}(e).

Productivity values using surge prices suggest more significant spatial and temporal differences when compared to flat prices. For example, during weekends and overnight hours, the productivity contrasts increased significantly. Weekends and weekday overnight periods present the highest spatio-temporal heterogeneity for expected driver profit given a destination location. The differences are mainly due to the surge price, which is high during these periods (Table \ref{tab:surge} in \ref{app:S} describes the percentage of trips with surge price by period); thus, it generates more contrast if the trip destination, or nearby areas, had surge price. Also, airport trips were less productive, due to the high idle time, as discussed previously, and also due to the lack of surge price rates (refer to Figure \ref{fig:surge}).

The proposed productivity metric accounts for the idle time of the destination of a first trip. We evaluate the overall effect of this value on the measure of performance. Figure \ref{fig:prodidles}(a) presents the relationship between productivity and idle time (refer to Figure \ref{fig:prodidle}(a) for flat fare values). Results show high R-squared values, the lowest is 0.656 (mid-day hours) and the highest is 0.862 (weekends). 
Results indicate that locations with low idle time (e.g., areas with high ridership, population, and employment density) are related to a higher expected driver continuation payoff. Locations such as the central area, that contain universities, parks, and active nightlife represent higher productivity values, while suburban areas presented the lowest values.

\subsubsection{Driver Productivity}
\noindent
We now investigate the expectation of the performance of drivers considering not only the destination conditions but also the effect of trip characteristics. For this, we use the proposed driver productivity variable and combine this measure with a natural experiment, where we select only trips with origin in the CBD. Results show an almost inverted scenario with respect to the continuation payoff. In this case, the highest productivity areas are located in the suburbs and the lowest in the central area. For flat fares (Figure \ref{fig:prodC}), we observe that the changes in productivity vary from -\$8/hr to \$8/hr (\$16/hr differences). However, when taking the surge price into account (Figure \ref{fig:prodCs}) the variation increases to more than \$20/hr. 
The trips that started and ended in the CBD, present productivity results below the system-wide average, with the highest productive periods occurring during weekends and overnight hours. The airport area shows values similar to the CBD. 

Results indicate that the idle time from the drop-off location had a small effect in determining driver productivity for two subsequent trips. The fare of the first trip was more determinant. Figure \ref{fig:prodidles}(b) presents the relationship between driver productivity and destination idle time (refer to Figure \ref{fig:prodidle}(b) for trips with flat fare values). This figure suggests that productivity increased as the idle time of the destination increased, which is not consistent with results from the previous section.  Figure \ref{fig:prodidles}(c) presents the relationship between driver productivity and the length of the first trip (refer to Figure \ref{fig:prodidle}(c) for trips with flat fare values). 
This figure shows that the trip distance, used to estimate the fare, also has a positive relationship with productivity. Thus, the evaluation of this driver productivity measure suggests that drivers with long trips presented greater long-term earnings than drivers with short trips that ended in a high demand area.  

These results can help to explain drivers' destination preferences and strategies used for increasing earnings. The longer and more profitable trips seem to more than compensate for the idle time that drivers suffer from driving from the suburb back to the CBD. Longer trips being more profitable is a known issue. For example, Uber recently increased by 31 percent the time rate and decreased by 8 percent the distance rate \cite{uberlong}.


\begin{figure}[H]
	\centering
	\captionsetup{justification=centering}
	\begin{center}
		\begin{subfigure}[h]{0.325\linewidth}
			\begin{overpic}[width=0.95\linewidth]{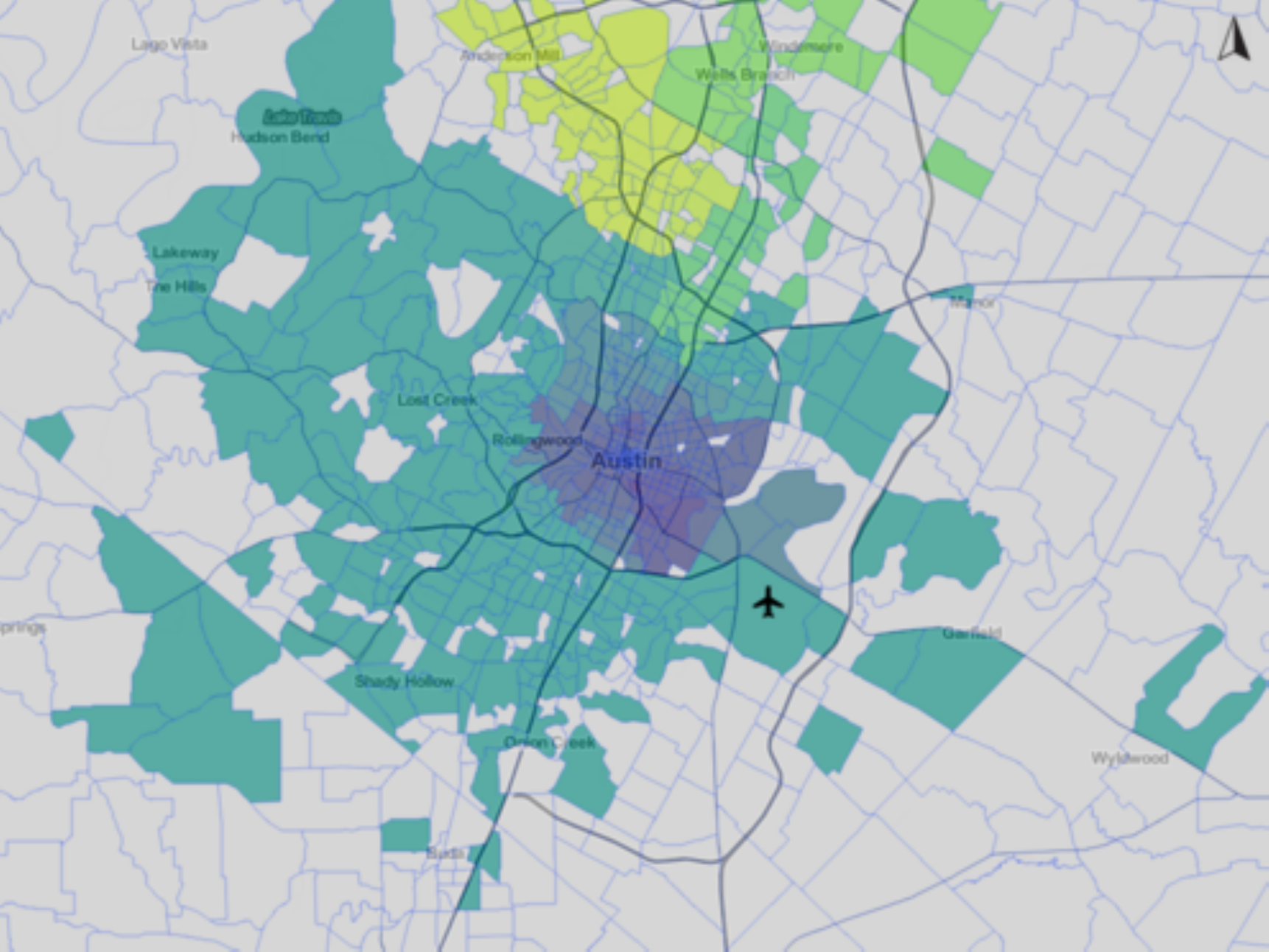}
				\put(0,0){\includegraphics[width=.09\linewidth]{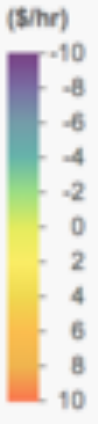}}
			\end{overpic}
			\caption{Peak hours}
		\end{subfigure}
		\begin{subfigure}[h]{0.325\linewidth}
			\begin{overpic}[width=0.95\linewidth]{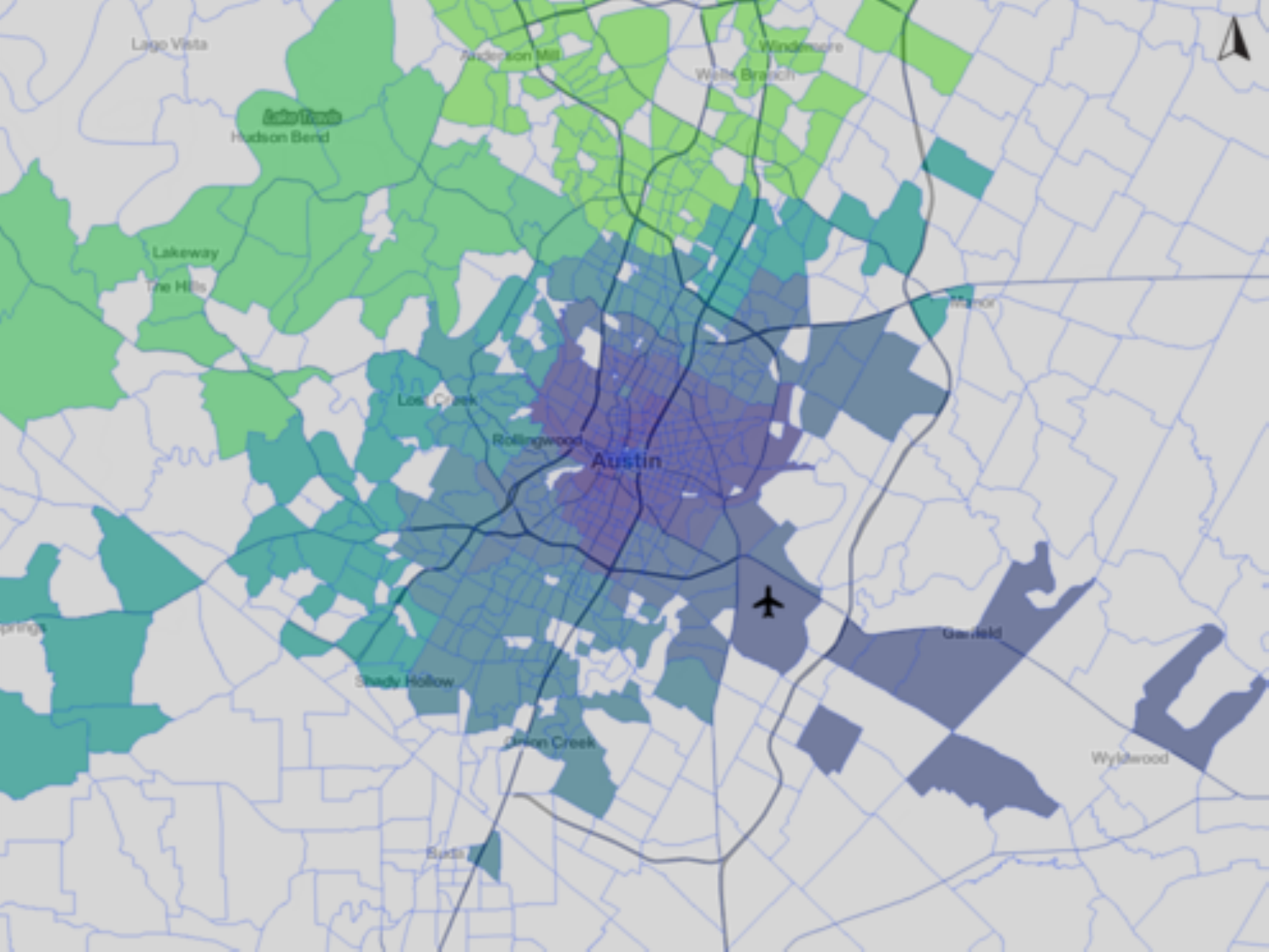}
				\put(0,0){\includegraphics[width=.09\linewidth]{fig/prodCs/Simb.pdf}}
			\end{overpic}
			\caption{Mid-day}
		\end{subfigure}
		
		\begin{subfigure}[h]{0.325\linewidth}
			\begin{overpic}[width=0.95\linewidth]{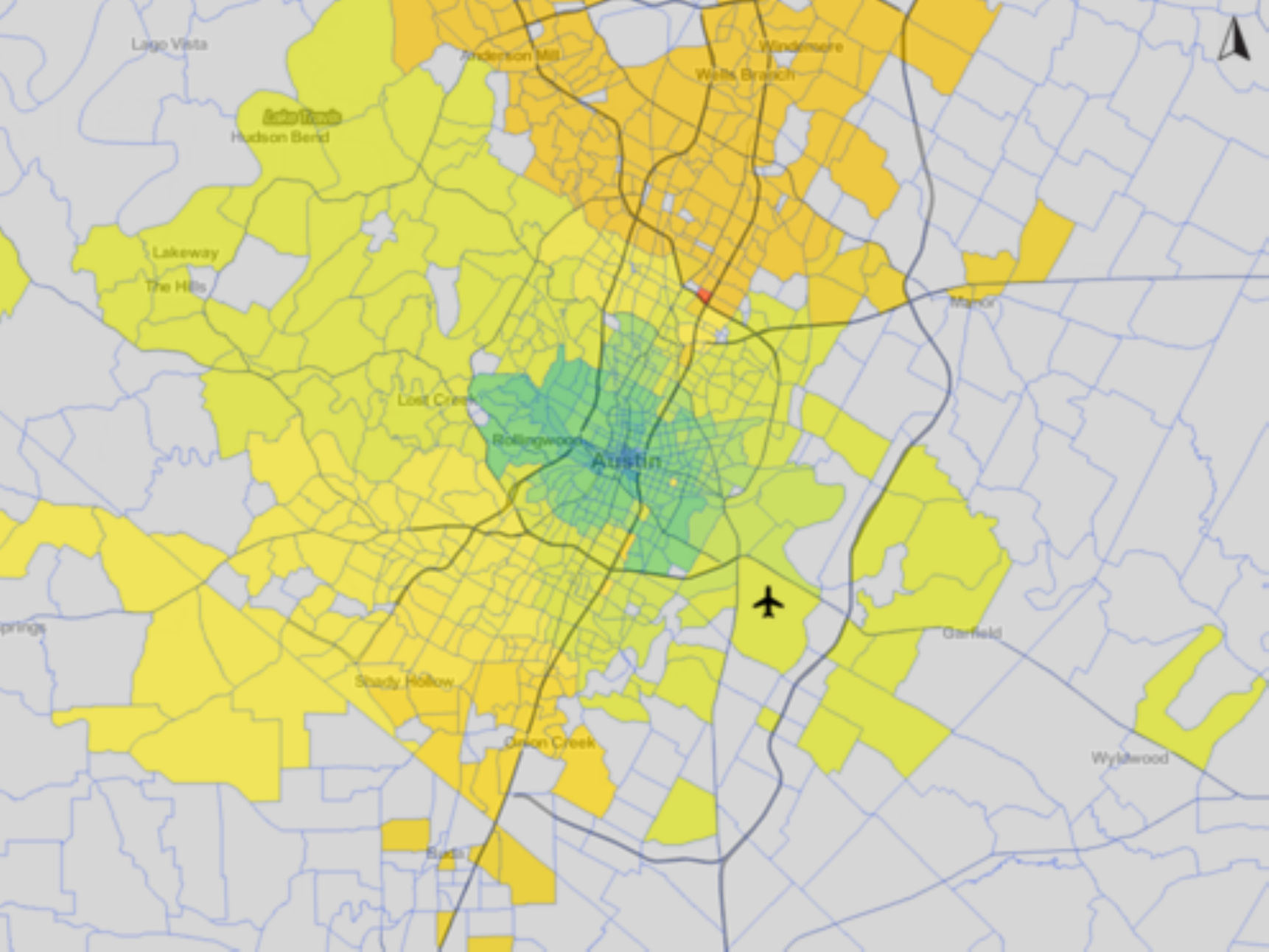}
				\put(0,0){\includegraphics[width=.09\linewidth]{fig/prodCs/Simb.pdf}}
			\end{overpic}
			\caption{Overnight}
		\end{subfigure}
		\begin{subfigure}[h]{0.325\linewidth}
			\begin{overpic}[width=0.95\linewidth]{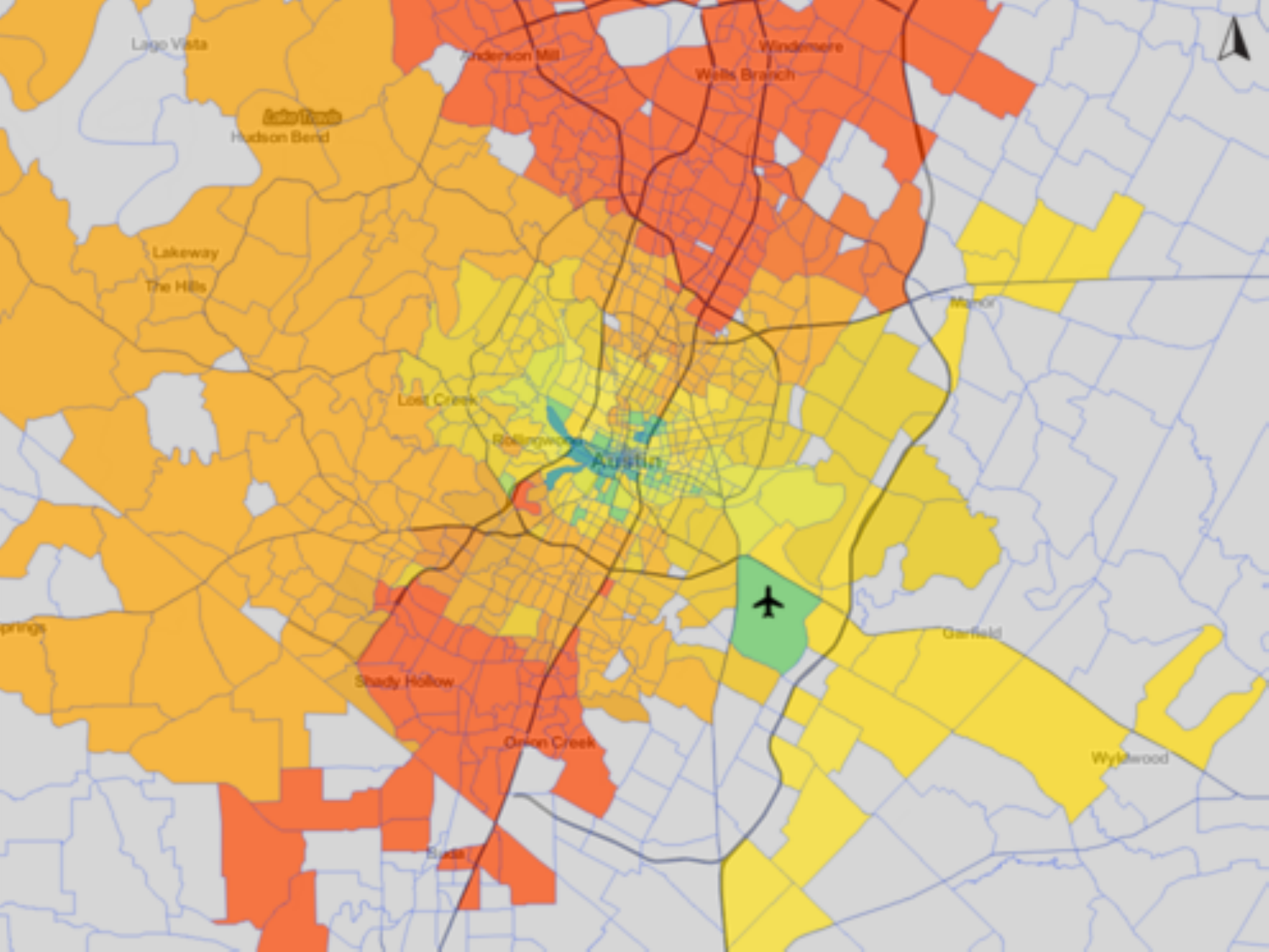}
				\put(0,0){\includegraphics[width=.09\linewidth]{fig/prodCs/Simb.pdf}}
			\end{overpic}
			\caption{Weekend}
		\end{subfigure}
		\begin{subfigure}[h]{0.325\linewidth}
			\begin{overpic}[width=.95\linewidth]{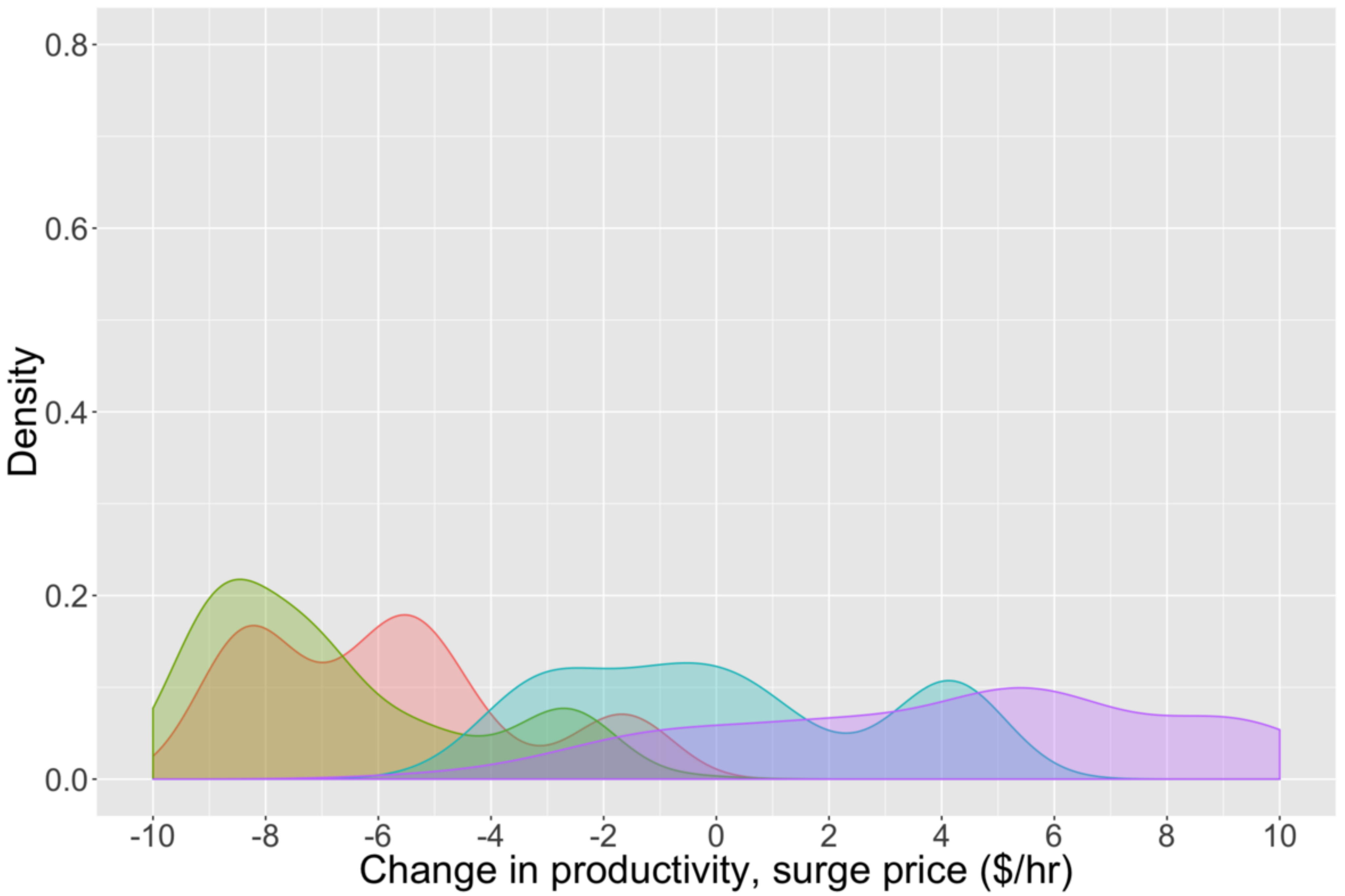}
				\put(40,60){\includegraphics[width=0.55\linewidth]{fig/res/Simb2.pdf}}
			\end{overpic}
			\caption{Density}
		\end{subfigure}
		\caption{Change in driver productivity with respect to the average for trips with surge price, by destination}
		\label{fig:prodCs}
	\end{center}
\end{figure}



\begin{figure}[H]
	\centering
	\captionsetup{justification=centering}
	\begin{center}
		\begin{subfigure}[h]{0.325\linewidth}
			\begin{overpic}[width=1\linewidth]{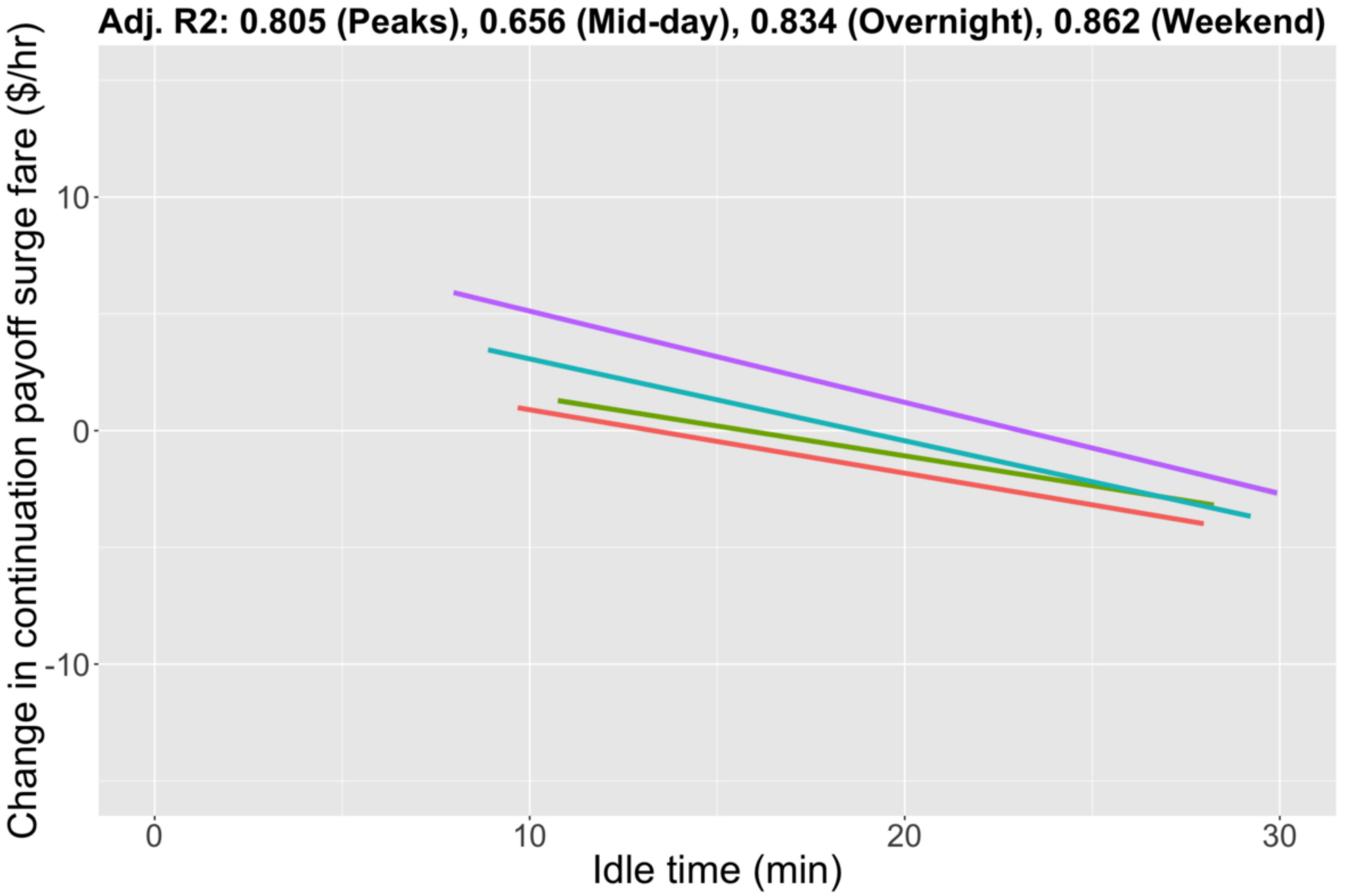}
				\put(40,57){\includegraphics[scale=.15]{fig/analysis/Simb2.pdf}}
			\end{overpic}
			\caption{Continuation payoff and idle time}
		\end{subfigure}
		\begin{subfigure}[h]{0.325\linewidth}
			\begin{overpic}[width=1\linewidth]{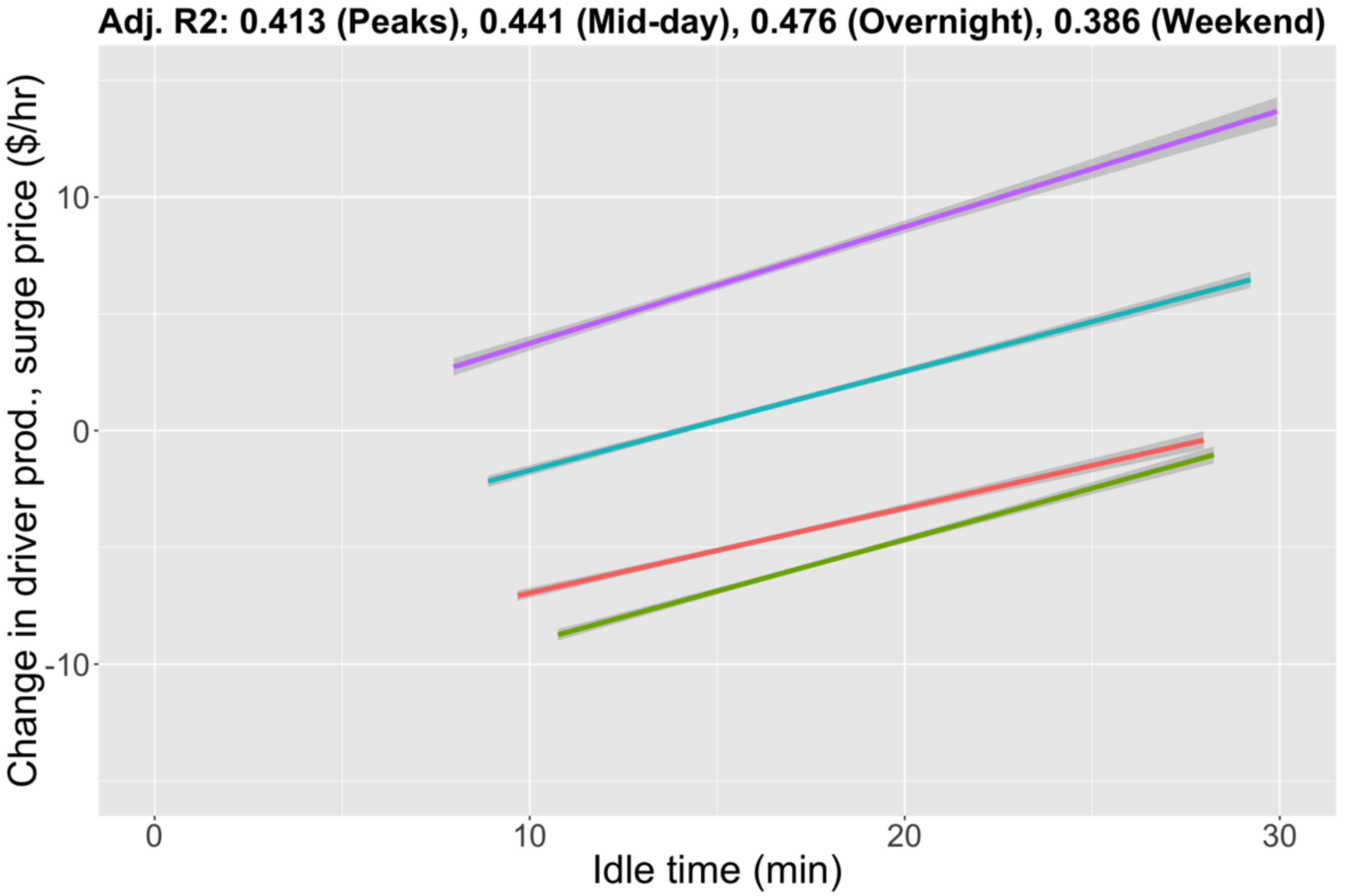}
				\put(40,57){\includegraphics[scale=.15]{fig/analysis/Simb2.pdf}}
			\end{overpic}
			\caption{Driver productivity and idle time}
		\end{subfigure}
		\begin{subfigure}[h]{0.325\linewidth}
			\begin{overpic}[width=1\linewidth]{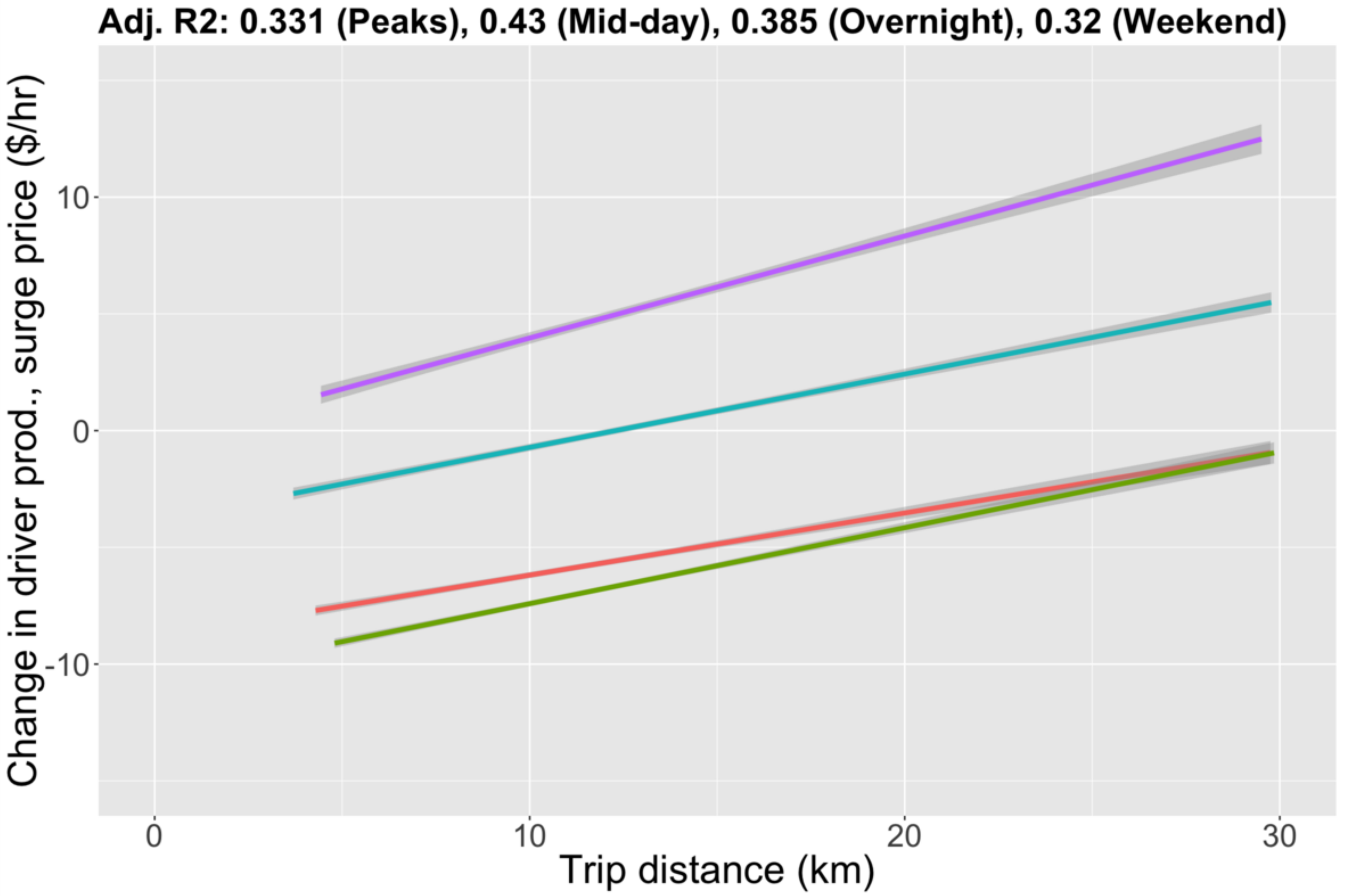}	
				\put(40,57){\includegraphics[scale=.15]{fig/analysis/Simb2.pdf}}
			\end{overpic}
			\caption{Driver productivity and trip distance}
		\end{subfigure}
		\caption{Relationship between change in productivity, idle time, and trip distance, for trips with surge price}
		\label{fig:prodidles}
	\end{center}
\end{figure}

\subsubsection{Summary}
\noindent
Our analysis shows that, given the current dispatching and pricing scheme employed by RideAustin, drivers have different continuation payoffs after getting dispatched on trips to different locations. The results suggest that there are significant spatio-temporal heterogeneities in driver productivity. 
However, given the optimal dispatching and pricing models discussed in \citet{spatial1}, this productivity will be the same for all locations, even when there exists significant spatial imbalance of trip flows.  

A limitation of our analysis is that it is hard to draw conclusions about true rider demand from our data, since only realized trips are recorded, and we cannot observe rider demands that go unfulfilled. Note, however, a major factor that determines whether destination­ based pricing is useful is whether there exists significant spatial and temporal imbalance in rider demand for trip flows (e.g. trips into versus out of airports), see \citet{spatial2}. In our data, however, this problem is mitigated, as we would expect the actual imbalance in rider trip flows to be more significant than the realized trip flows. 

Results from this analysis suggest that existing dispatching and pricing methods do not focus on the driver. 
There is a need for more efficient dispatching methods and new pricing strategies and policies that warrant driver equity. Examples include methods that take into account spatio-temporal variations, as in \citet{spatial1}, and pricing schemes that have a non-linear relation with trip distance, as in  \citet{friction9}.

\section{Conclusions and Future Work}
\label{sec:sec7}
\noindent
This study explored the spatial structure of operational and driver performance variables in ride-sourcing systems using empirical data. We used information from more than 1.4 million rides in the Austin area, provided by a local TNC, during a period in which the leading companies were not operating within the city. 
We proposed performance metrics using two consecutive trips to capture the effects of market conditions at drop-off locations. Further, we developed a natural experimental framework by analyzing trips with a common origin and varying destinations, isolating spatial effects on productivity.
The principal findings suggest that drivers presented different productivity after being dispatched to trips with different destinations; moreover, the origin-based surge price scheme increased the drivers' earning disparities.
Results point out that trip distance is the dominant factor in driver productivity and short-distance trips showed lower productivity, even when ending in areas with high demand. 
Based on the developed methods, results suggest that current dispatching and pricing schemes do not focus on the driver and that there is a need to provide policies that warrant more equitable driver earnings in the ride-sourcing market.

The results and methods presented in this study can serve multiple purposes. First, from a driver and operator point of view, we identified the spatial and temporal distribution of the principal operational and performance variables that can lead to a more efficient driver supply method. Second, from the planners' and engineers' perspective, we provided insights on ride-sourcing travel patterns in the Austin area that can help to understand the characteristics of the ride-sourcing service. Third, we provide empirical evidence of driver performance inequality due to spatial and temporal factors. This evaluation can lead to pricing strategies and policies that warrant fair conditions in driver compensation.
Finally, our results have relevance to transportation research in that we provide an application of spatial smoothing to a transportation problem. This method can provide a more appropriate high-definition spatial evaluation, reduce noisy measures and enhance interpretability. 

The present study suggests performance metrics that account for the trip's profit, duration, and market frictions of the destination area, in term of driver idle and reach time. However, we did not account for the cost that the driver incurs while waiting for the next trip and the cost of empty miles. It is recommended that future work explores such costs and provide a sensitivity analysis of the effects on driver earnings under the conditions studied in this paper.


\appendix 

\section{Ride-Sourcing Data Description}
\label{app:A}
\setcounter{figure}{0}
\setcounter{table}{0}
\begin{figure}[H]
	\centering
	\captionsetup{justification=centering}
	\begin{center}
		\begin{subfigure}[h]{0.495\linewidth}
			\includegraphics[width=1\linewidth]{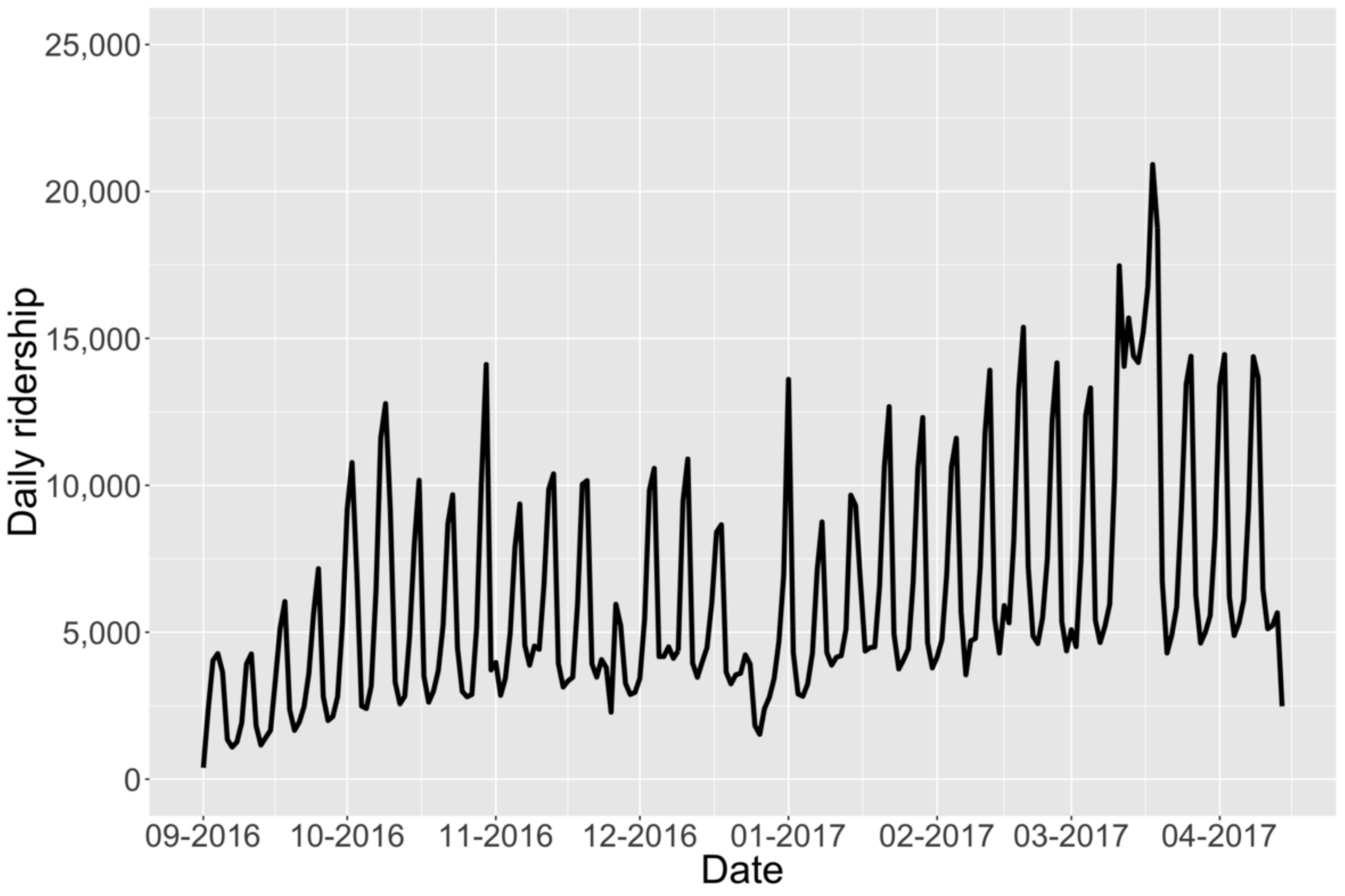}
			\caption{Daily ridership}
		\end{subfigure}
		\begin{subfigure}[h]{0.495\linewidth}
			\includegraphics[width=1\linewidth]{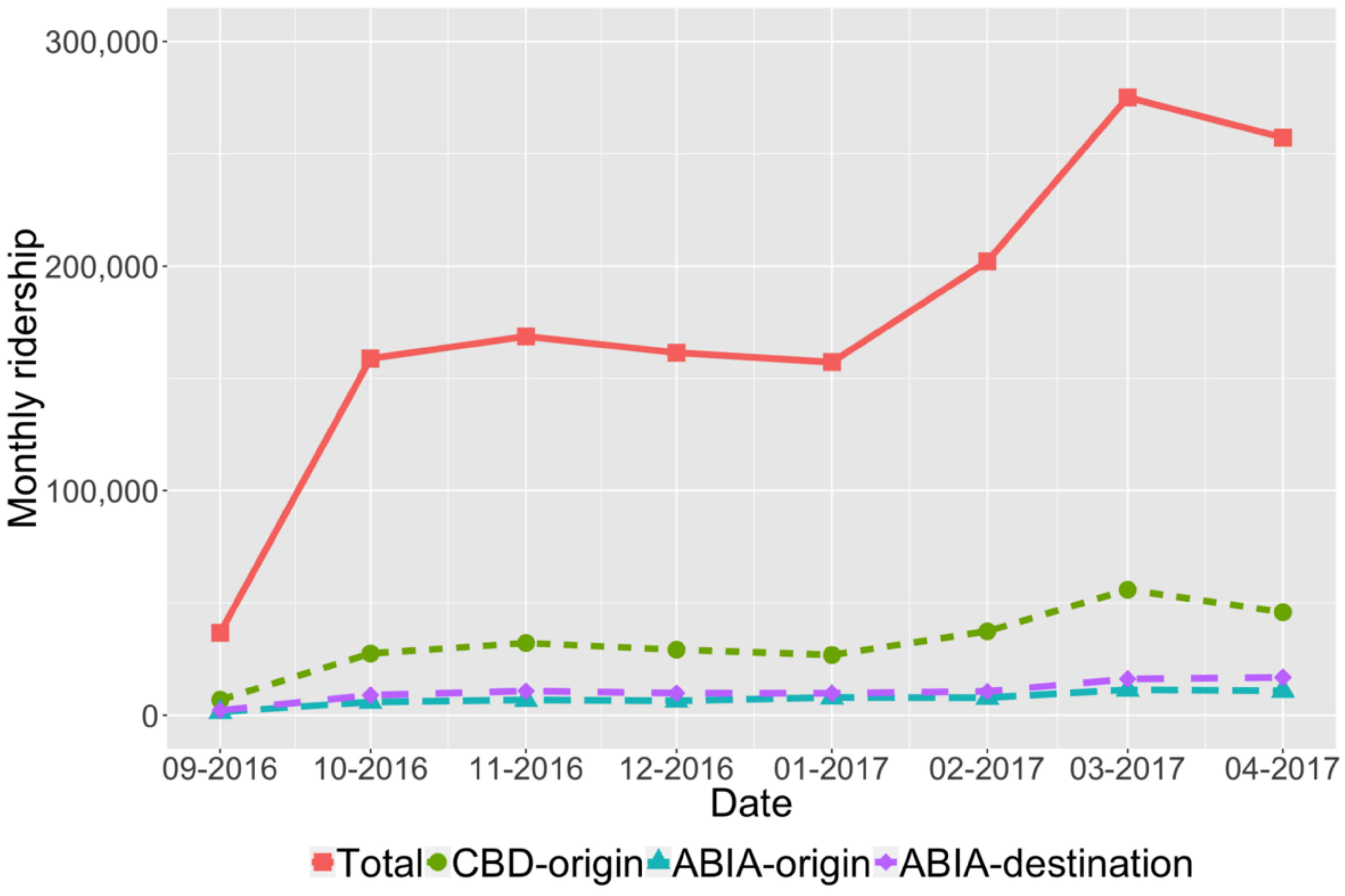}
			\caption{Monthly ridership}
		\end{subfigure}
		\begin{subfigure}[h]{0.495\linewidth}
			\includegraphics[width=1\linewidth]{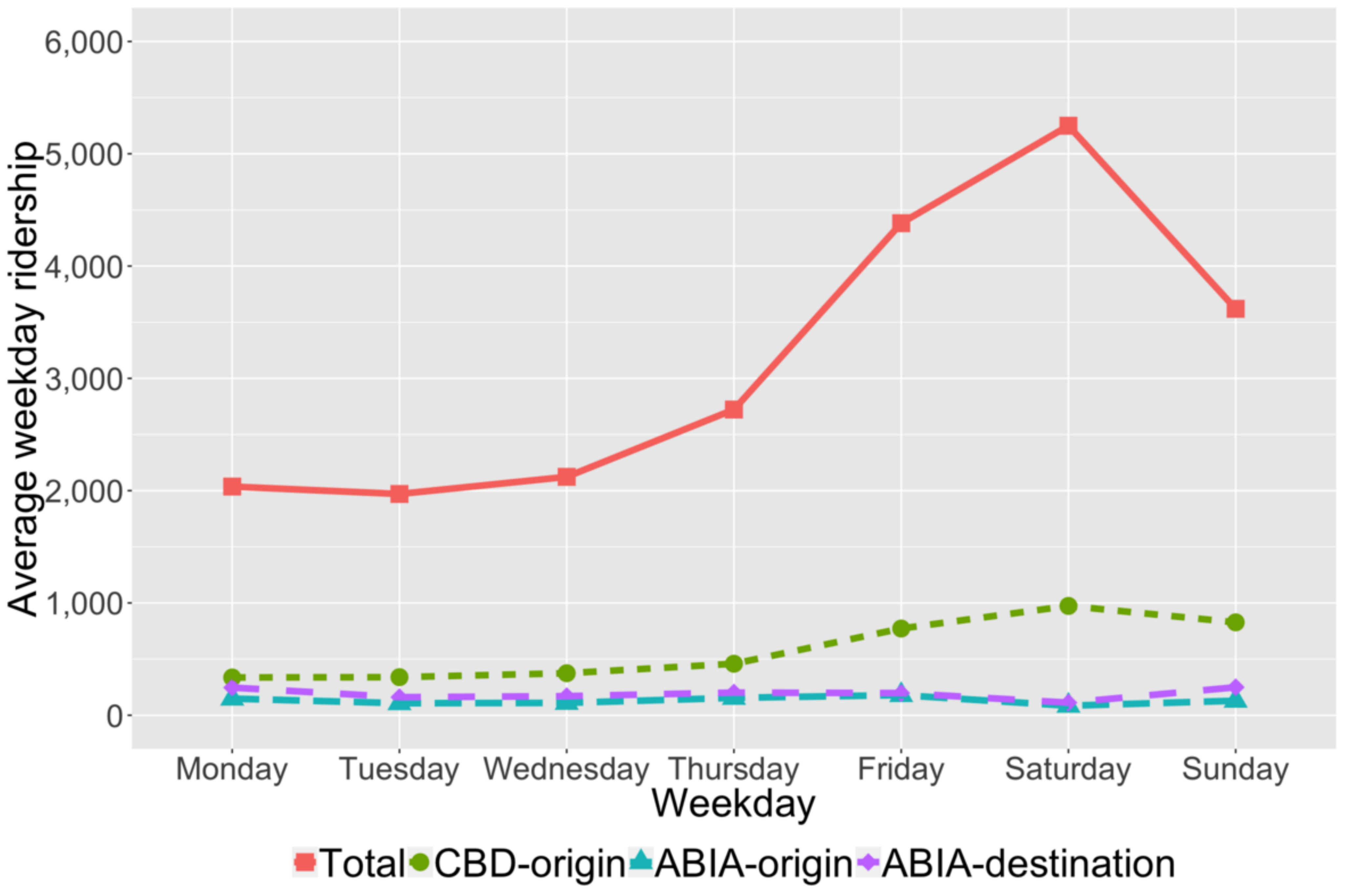}
			\caption{Average weekday ridership}
		\end{subfigure}
		\begin{subfigure}[h]{0.495\linewidth}
			\includegraphics[width=1\linewidth]{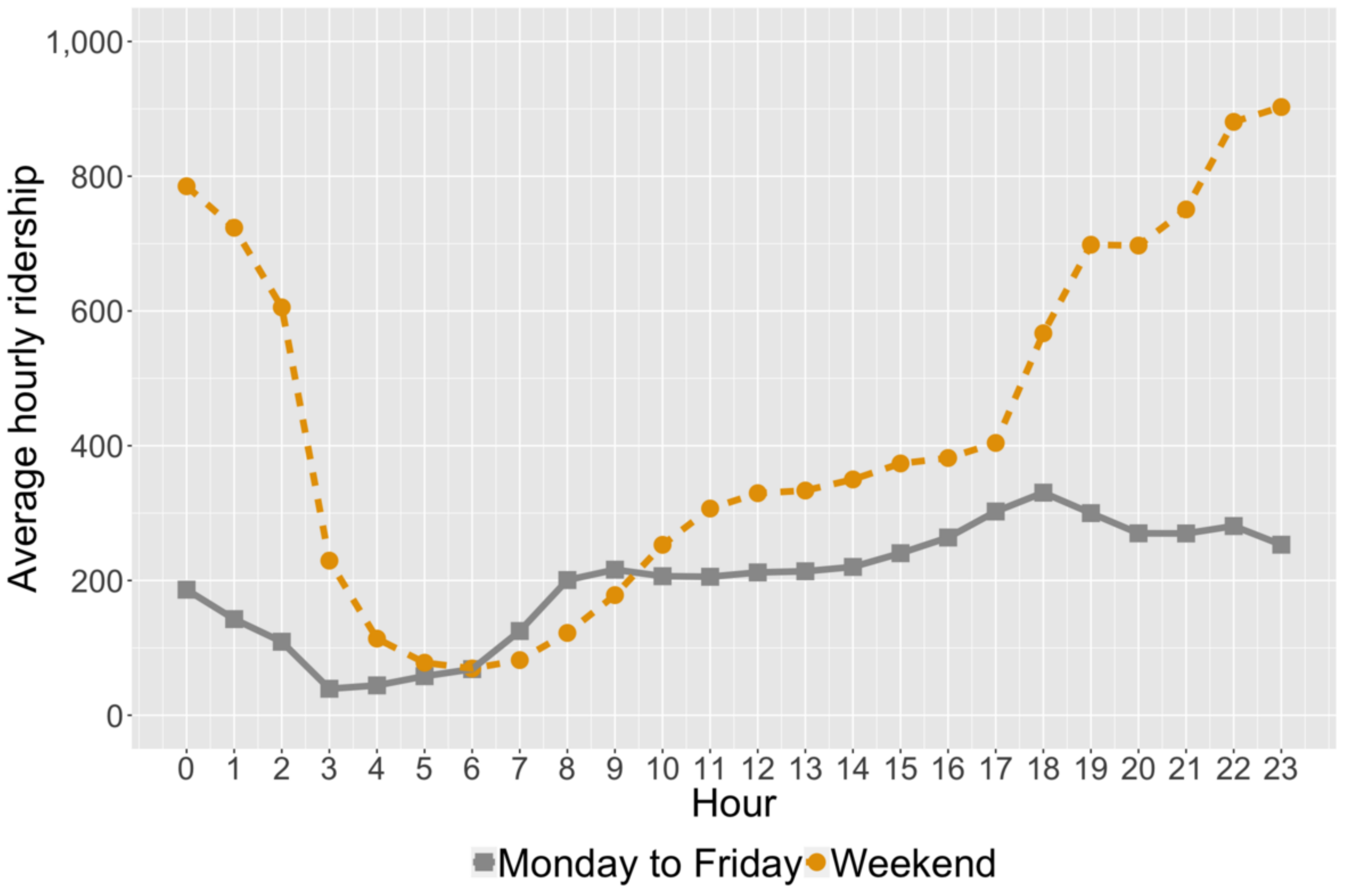}
			\caption{Average hourly ridership}
		\end{subfigure}
		\caption{Ride-sourcing data description}
		\label{fig:des}
	\end{center}
\end{figure}

\begin{figure}[H]
	\begin{center}
		\begin{subfigure}[h]{0.495\linewidth}
			\begin{overpic}[width=1\linewidth]{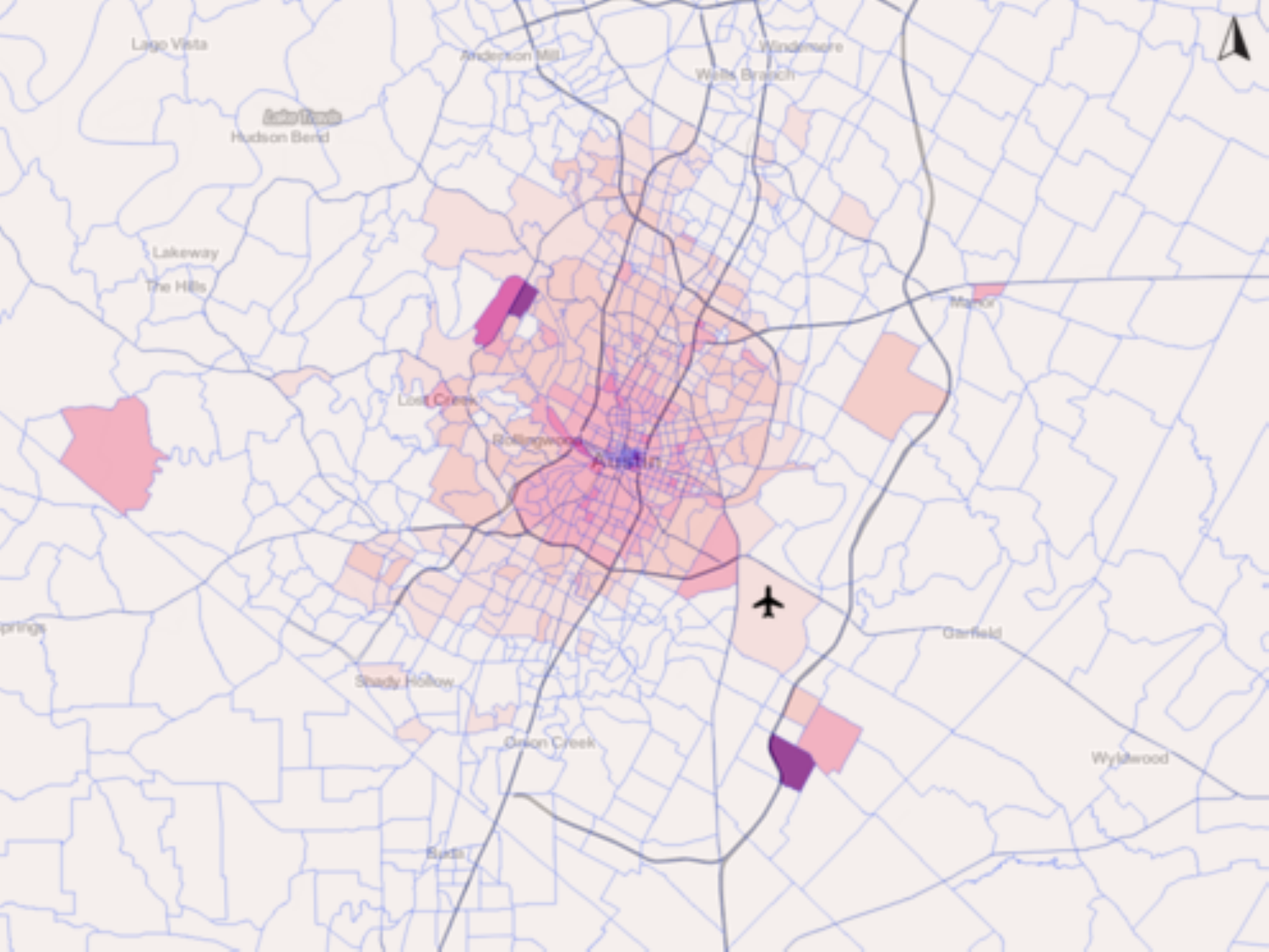}
				\put(0,0){\includegraphics[width=.14\linewidth]{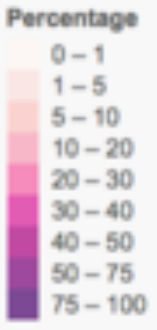}}
			\end{overpic}
			\caption{Trips with surge price during Weekdays}
		\end{subfigure}
		\begin{subfigure}[h]{0.495\linewidth}
			\begin{overpic}[width=1\linewidth]{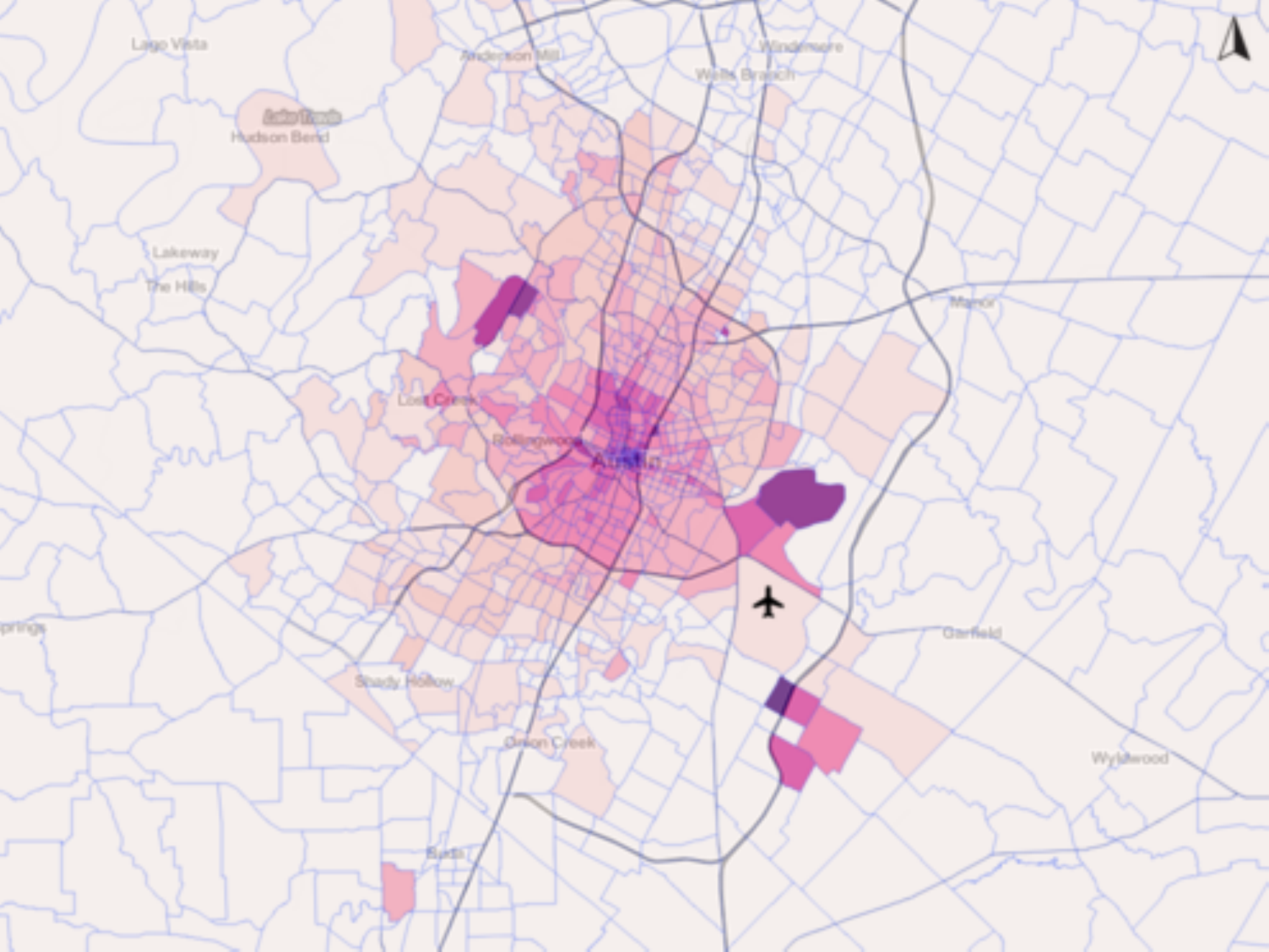}
				\put(0,0){\includegraphics[width=.14\linewidth]{fig/area/surge/Simb.pdf}}
			\end{overpic}
			\caption{Trips with surge price during Weekends}
		\end{subfigure}
		\caption{Location of the trips with surge price}
		\label{fig:surge}
	\end{center}
\end{figure}

\begin{figure}[H]
	\centering
	\captionsetup{justification=centering}
	\begin{center}
		\begin{overpic}[width=.495\linewidth]{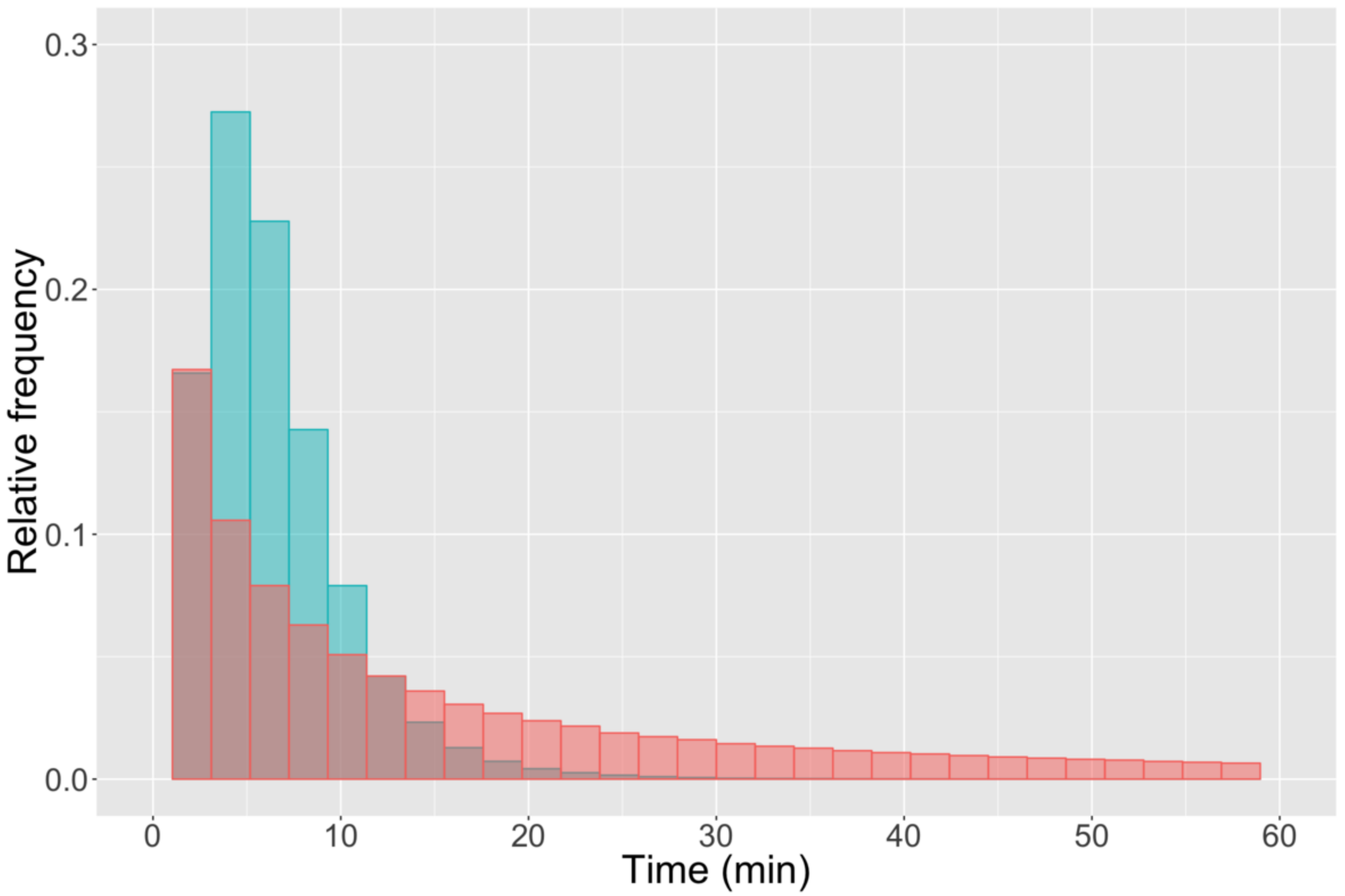}				
			\put(60,60){\includegraphics[width=.18\linewidth]{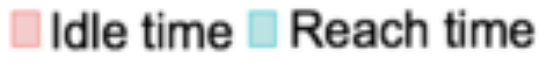}}
		\end{overpic}
		\caption{Histogram of idle and reach time}
		\label{fig:fric}
	\end{center}
\end{figure}


\section{Background on Spatial Smoothing} \label{app:B}
\noindent
Space can be modeled as continuous or discrete. Smoothing techniques for the continuous case include Gaussian processes, Gaussian kernel smoothing, and continuous random fields, while for the discrete case some methods include Graph kernel smoothing and Graph Laplacian smoothing. 
Spatial-smoothing techniques can be classified into local and global approaches \cite{tansey2017}, where local approaches smooth only a local window around each point, such as neighboring pixels in an image, while global methods typically define an objective function over the entire graph and simultaneously optimize the whole set of points. 
The most simplistic local approaches simply replace each point with the average or median of the points in its window  \cite{tansey2017}. 

An important aspect of using spatial data is specification of the spatial covariance function. Data can be isotropic, meaning that the spatial dependence does not depend on the direction of the spatial separation between sampling locations, only on the distance \cite{weller2016}. Methods such as the Gaussian kernel assume isotropy. However, this assumption is often violated by real-world data, where arbitrary discontinuities may be present. In some cases, it is more appropriate to rely on anisotropic smoothing techniques, which can smooth differently in distinct directions and locations. Anisotropic local methods for images include bilateral filter and guided filter used to preserve edges. Global techniques for anisotropy include discrete Markov random fields (MRFs). This method defines a joint distribution over a graph via a product of exponentiated potential functions over cliques, or as a conditional autoregressive (CAR) model where each node's unnormalized likelihood is written conditioned on all other nodes in the graph \cite{tansey2017}. 

An alternative to MRFs for global smoothing is the graph-based trend filtering (GTF)  \cite{wang2015}, which is a special use of the generalized lasso \cite{tibshirani2011} that applies an  $\ell_1$ penalty\footnote{Formally, the $\ell_p$-norm of $x$ is defined as:  $\Vert x \Vert_p =  \sqrt[p]{    \sum_{i}{|x_i|^p}} $, where $p \in \R$. } to the vector of $(k+1)^{st}$-order differences, where the integer $k \geq 0$ is a hyperparameter. While global approaches like MRFs and GTF typically yield better results, they often fail to scale to large graphs because every node being dependent on the rest of the graph  \cite{tansey2017}. One exception is a particular case of the GTF with $k=0$, known as graph-based total variation denoising, also called graph-fused lasso (GFL).


\section{A Fast and Flexible Algorithm for the GFL} \label{app:C}
\noindent
\citet{tansey2015} proposed an ADMM approach to solving the GFL, where the key insight is to decompose the graph into a set of trails that can each be solved efficiently using techniques for the ordinary (1D) fused lasso. The resulting technique is both faster than previous GFL methods and more flexible in the choice of loss function and graph structure \cite{tansey2015}. This section provides a summary of the method.

The core idea of the algorithm is to decompose a graph  $\mathcal{G}= (\mathcal{V}, \mathcal{E})$ with node set  $\mathcal{V}$ of size $2k$ and edge set $\mathcal{E}$ into a set of non-overlapping trails $\mathcal{T} = \{ t_1, t_2, ..., t_k \} $, on which the optimization algorithm can operate, and allows one to rewrite the penalty function in Equation \ref{eq:37} as: 

\begin{equation}\label{eq:39}
\sum_{(r,s) \in \mathcal{E}} |x_r - x_s | = \sum_{t \in \mathcal{T}} \sum_{(r,s) \in t} |x_r - x_s |
\end{equation}

The updated penalty function allows proposing an efficient ADMM algorithm. The next sections present details of the updated optimization method and the trial decomposition approaches suggested by the authors. 

\subsection{Optimization via the ADMM}
\noindent
The objective function (Equation \ref{eq:37}) can be rewritten using the updated penalty function as shown in Equation  \ref{eq:310}. For each trail $t$ (where $|t| = m$), we introduce $m + 1$ slack variables\footnote{In an optimization problem, a slack variable is a variable that is added to an inequality constraint to transform it into an equality.}, one for each vertex along the trail. Multiple slack variables are introduced if a vertex is visited more than once in a trail.  

\begin{equation}\label{eq:310}
\begin{aligned}
& \underset{\mathbf{x} \in \R^n}{\text{minimize}}
& & 
\ell ( \mathbf{y, x} )  + \lambda \sum_{t \in \mathcal{T}} \sum_{(r,s) \in t} |z_r - z_s | \\
& \text{subject to}
&&  x_r = z_r \\
&&&  x_s = z_s 
\end{aligned}
\end{equation}

This problem can be solved using the ADMM algortihm \cite{boyd2011} based on the following updates:
\begin{equation}\label{eq:311}
\mathbf{x}^{k+1} = \underset{\mathbf{x}}{\text{argmin}}
\bigg( 
\ell ( \mathbf{y, x} )  + \frac{\alpha}{2} \Vert \mathbf{Ax} - \mathbf{z}^{k} +  \mathbf{u}^k \Vert ^2
\bigg) 
\end{equation}
\begin{equation}\label{eq:312}
\mathbf{z}_t^{k+1} = \underset{\mathbf{z}}{\text{argmin}}
\Bigg( 
w \sum_{(r \in t} (\tilde{y}_r - z_r)^2 + \sum_{(r,s) \in t} |z_r - z_s|
\Bigg), t \in \mathcal{T}
\end{equation}
\begin{equation}\label{eq:313}
\mathbf{u}^{k+1} = \mathbf{u}^k + \mathbf{Ax}^{k+1} - \mathbf{z}^{k+1}
\end{equation}
where $u$ is the scaled dual variable, $\alpha$ is the scalar penalty parameter, $w = \frac{\alpha}{2}$, $\tilde{y}_r = x_r - u_r$ and $A$ is a sparse binary matrix used to encode the appropriate $x_i$ for each $z_j$ . Here $t$ is used to denote both the vertices and edges along trail $t$.

For the squared-error loss function $\ell(  \textbf{y},  \textbf{x})  = \sum_{i=1}^{n} \frac{1}{2} (y_i-x_i)^2$, the $x$ updates have the simple closed-form solution:
\begin{equation}\label{eq:314}
x_i^{k+1} = \frac{2y_i + \alpha \sum_{j \in \mathcal{J}} (z_j - u_j)}{2 + \alpha |\mathcal{J}| },
\end{equation}
where $\mathcal{J}$ is the set of dual variable indices that map to $x_j$. Crucially, the trail decomposition approach means that each trail's $z$ update in Equation \ref{eq:312} is a one-dimensional fused lasso problem which can be solved in linear time via an efficient dynamic programming routine.

\subsection{Trail decomposition}
\noindent
The two approaches for the trail decomposition are summarized as follows\footnote{ For a broader explanation see  \citet{tansey2015} or \citet{zuniga2018}, \cite{zuniga2019}.}:

\begin{enumerate}
	\item Create $k$ ``pseudoedges" connecting the $2k$ odd-degree vertices and then find an Eulerian tour on the surgically altered graph. To decompose the graph into trails, we then walk along the tour (which by construction enumerates every edge in the original graph exactly once). Every time a pseudo-edge is encountered, we mark the start of a new trail. 
	\item Iteratively choose a pair of odd-degree vertices and select a shortest path connecting them based on an heuristic (e.g., a trail with median length). Any component that is disconnected from the graph then has an Eulerian tour and can be appended onto the trail at the point of disconnection.
\end{enumerate}


\section{Results Using Flat Fare Values}
\setcounter{figure}{0}
\setcounter{table}{0}
\label{app:S}

\begin{table}[H]
	\centering
	\caption{Summary of productivity results by location and period, flat fare}
	\label{tab:res3}
	\begin{tabular}{lllrrr}
		\hline
		\multicolumn{1}{c}{\multirow{2}{*}{Variable}}     & \multicolumn{2}{c}{\multirow{2}{*}{Period}} & \multicolumn{3}{c}{Destination Location}                                                \\ \cline{4-6} 
		\multicolumn{1}{c}{}                              & \multicolumn{2}{c}{}                        & \multicolumn{1}{c}{System-wide} & \multicolumn{1}{c}{CBD} & \multicolumn{1}{c}{Airport} \\ \hline
		\multirow{5}{3 cm}{Average continuation payoff, flat fare (\$/hr)}  & \multirow{3}{*}{Weekday}    & Peak Hours    & 18.4                            & 19.6                    & 15.7                        \\  
		&                             & Mid-Day       & 19.2                            & 20.6                    & 14.9                        \\  
		&                             & Overnight     & 19.1                            & 22.3                    & 19.3                       \\ 
		& \multicolumn{2}{l}{Weekend}                 & 19.9                            & 21.6                    & 16.9                        \\ \cline{2-6} 
		& \multicolumn{2}{l}{\textit{Total}}          & \textbf{\textit{19.2}}                   & \textit{21.0}           & \textit{16.7}               \\ \hline
		\multirow{5}{3 cm}{Average driver prod., flat fare (\$/hr)}  & \multirow{3}{*}{Weekday}    & Peak Hours    & 26.8                            & 24.7                    & 26.4                       \\  
		&                             & Mid-Day       & 27.2                           & 25.4                    & 25.9                        \\  
		&                             & Overnight     & 30.1                            & 28.6                   & 31.5                       \\ 
		& \multicolumn{2}{l}{Weekend}                 & 31.0                            & 27.4                    & 29.1                        \\ \cline{2-6} 
		& \multicolumn{2}{l}{\textit{Total}}          & \textbf{\textit{29.0}}                   & \textit{26.5}           & \textit{28.2}               \\ \hline
	\end{tabular}
\end{table}

\begin{figure}[H]
	\centering
	\captionsetup{justification=centering}
	\begin{center}
		\begin{subfigure}[h]{0.325\linewidth}
			\begin{overpic}[width=0.95\linewidth]{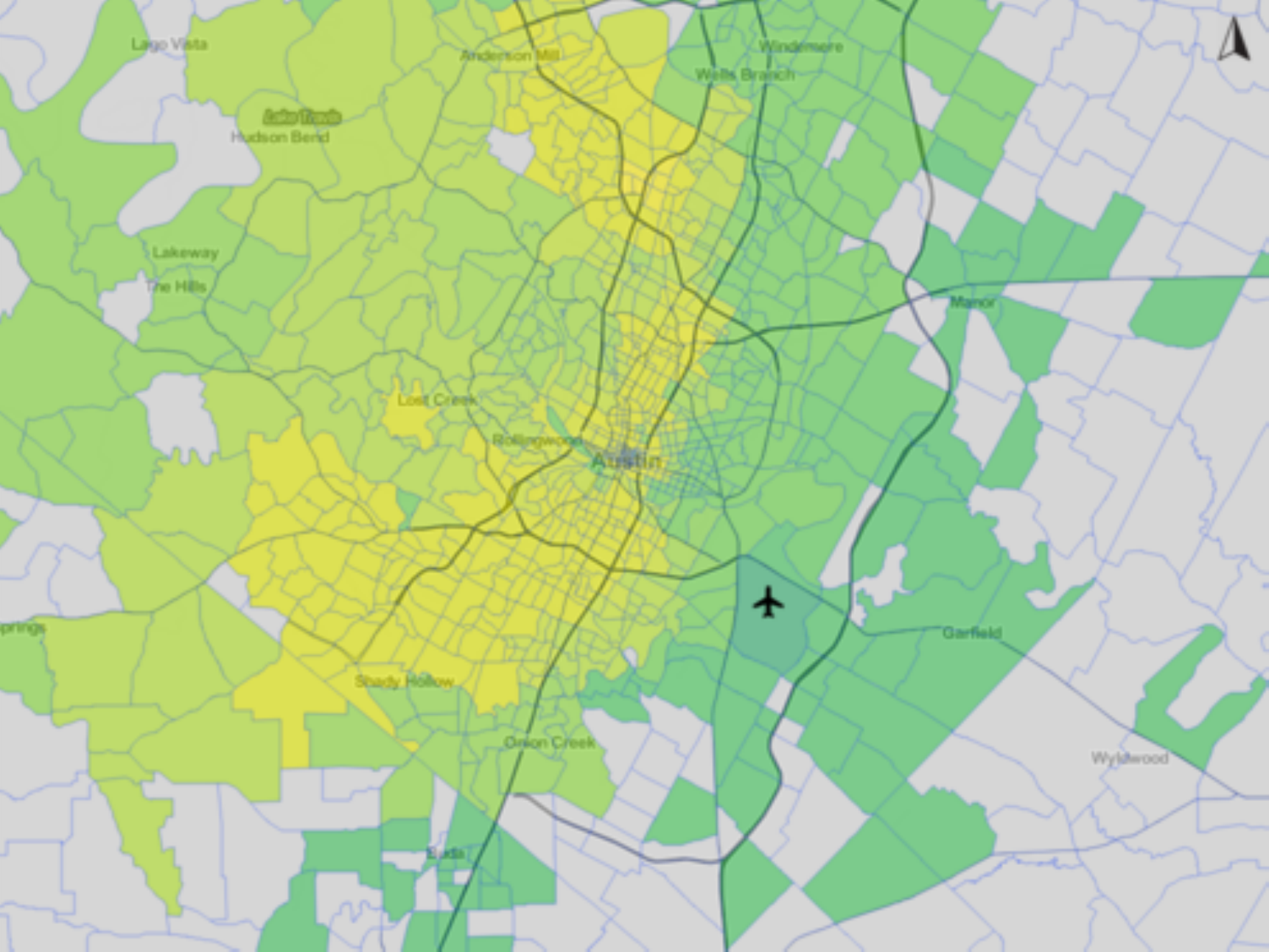}
				\put(0,0){\includegraphics[width=.09\linewidth]{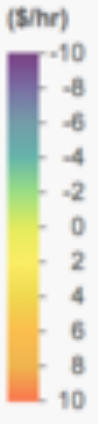}}
			\end{overpic}
			\caption{Peak hours}
		\end{subfigure}
		\begin{subfigure}[h]{0.325\linewidth}
			\begin{overpic}[width=0.95\linewidth]{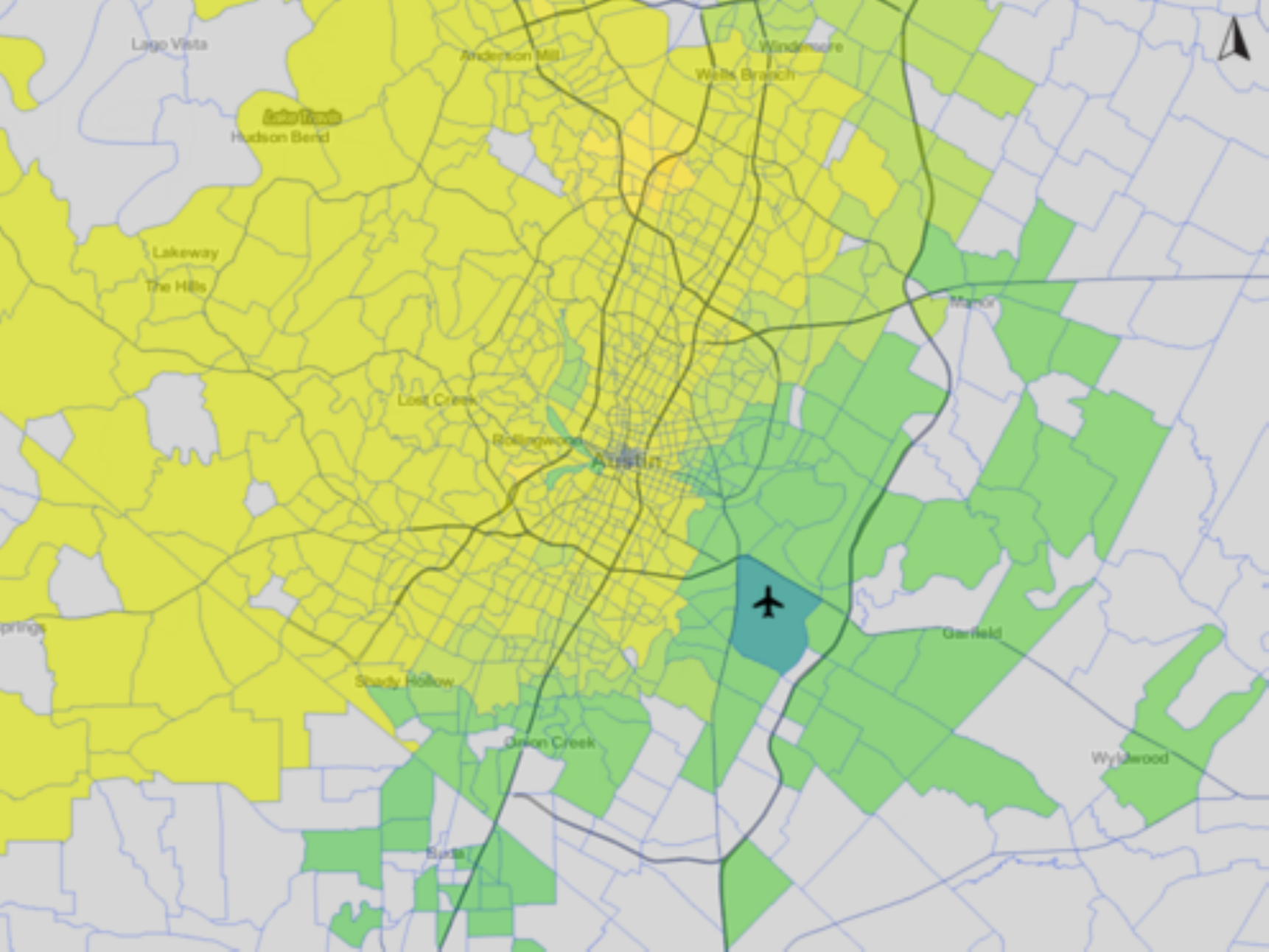}
				\put(0,0){\includegraphics[width=.09\linewidth]{fig/prodD/Simb.pdf}}
			\end{overpic}
			\caption{Mid-day}
		\end{subfigure}
		
		\begin{subfigure}[h]{0.325\linewidth}
			\begin{overpic}[width=0.95\linewidth]{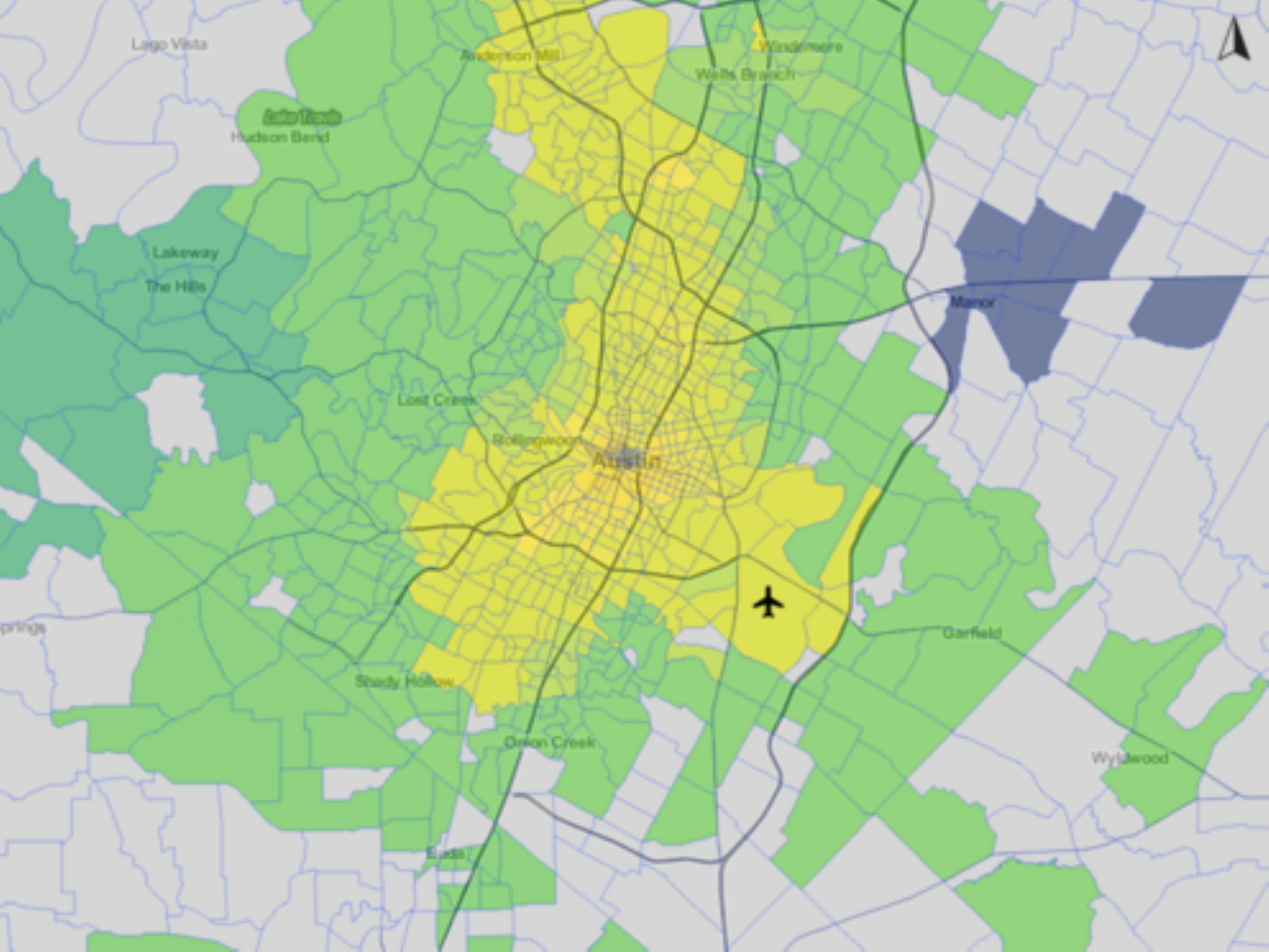}
				\put(0,0){\includegraphics[width=.09\linewidth]{fig/prodD/Simb.pdf}}
			\end{overpic}
			\caption{Overnight}
		\end{subfigure}
		\begin{subfigure}[h]{0.325\linewidth}
			\begin{overpic}[width=0.95\linewidth]{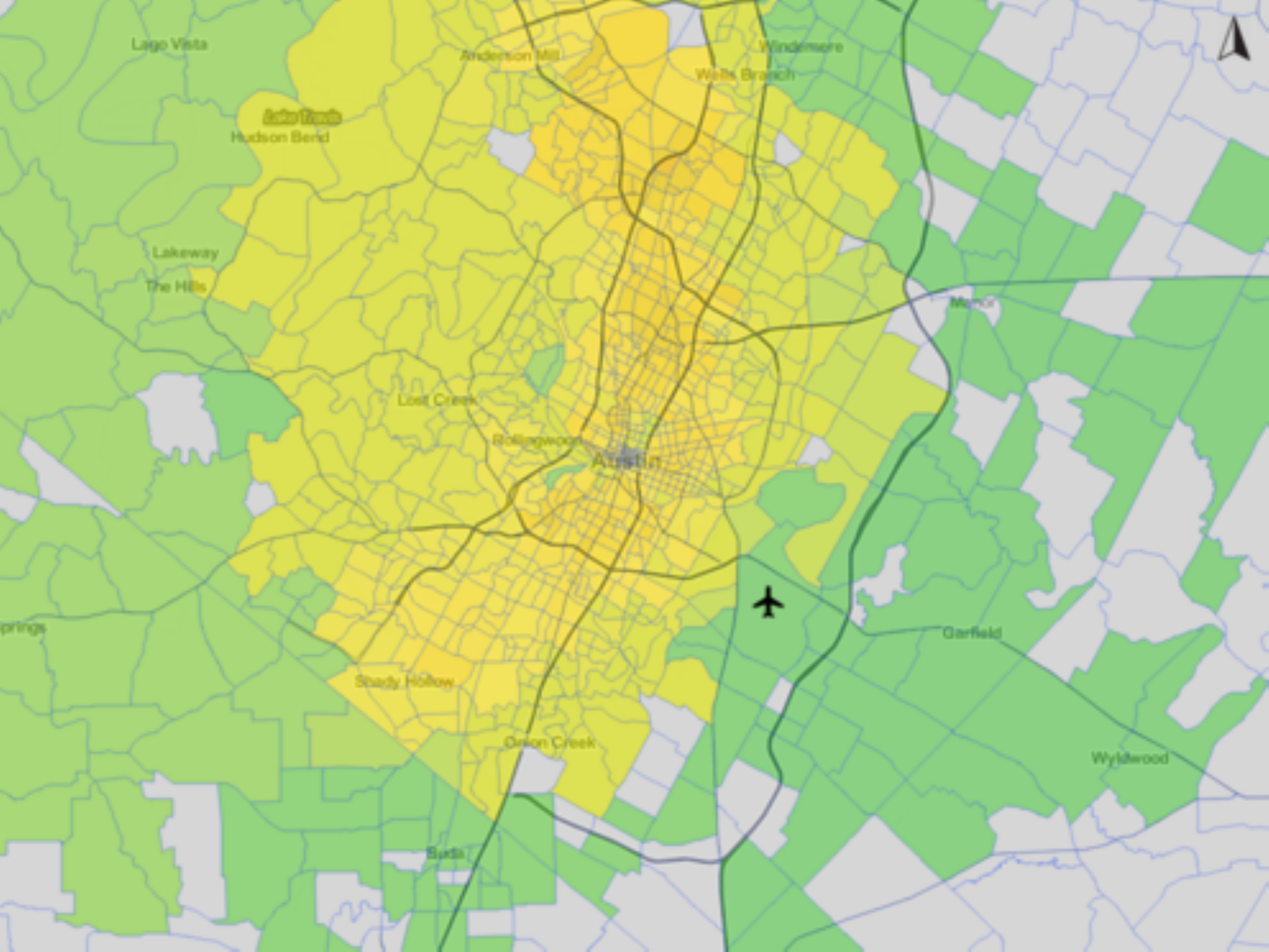}
				\put(0,0){\includegraphics[width=.09\linewidth]{fig/prodD/Simb.pdf}}
			\end{overpic}
			\caption{Weekend}
		\end{subfigure}
		\begin{subfigure}[h]{0.325\linewidth}
			\begin{overpic}[width=.95\linewidth]{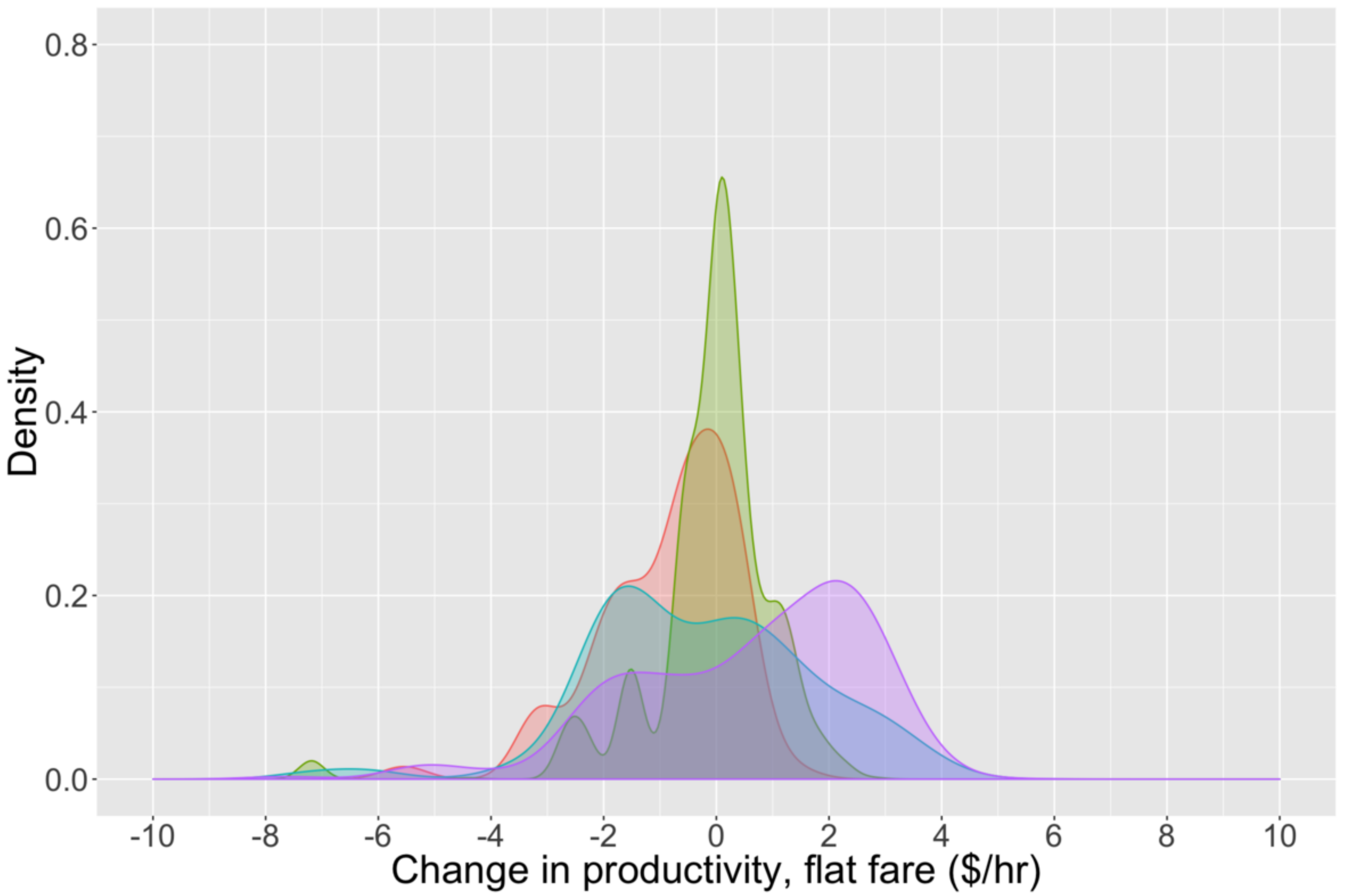}
				\put(40,60){\includegraphics[width=.55\linewidth]{fig/res/Simb2.pdf}}
			\end{overpic}
			\caption{Density}
		\end{subfigure}
		\caption{Change in continuation payoff with respect to the average for trips with flat fare, by destination}
		\label{fig:prod}
	\end{center}
\end{figure}

\begin{figure}[H]
	\centering
	\captionsetup{justification=centering}
	\begin{center}
		\begin{subfigure}[h]{0.325\linewidth}
			\begin{overpic}[width=0.95\linewidth]{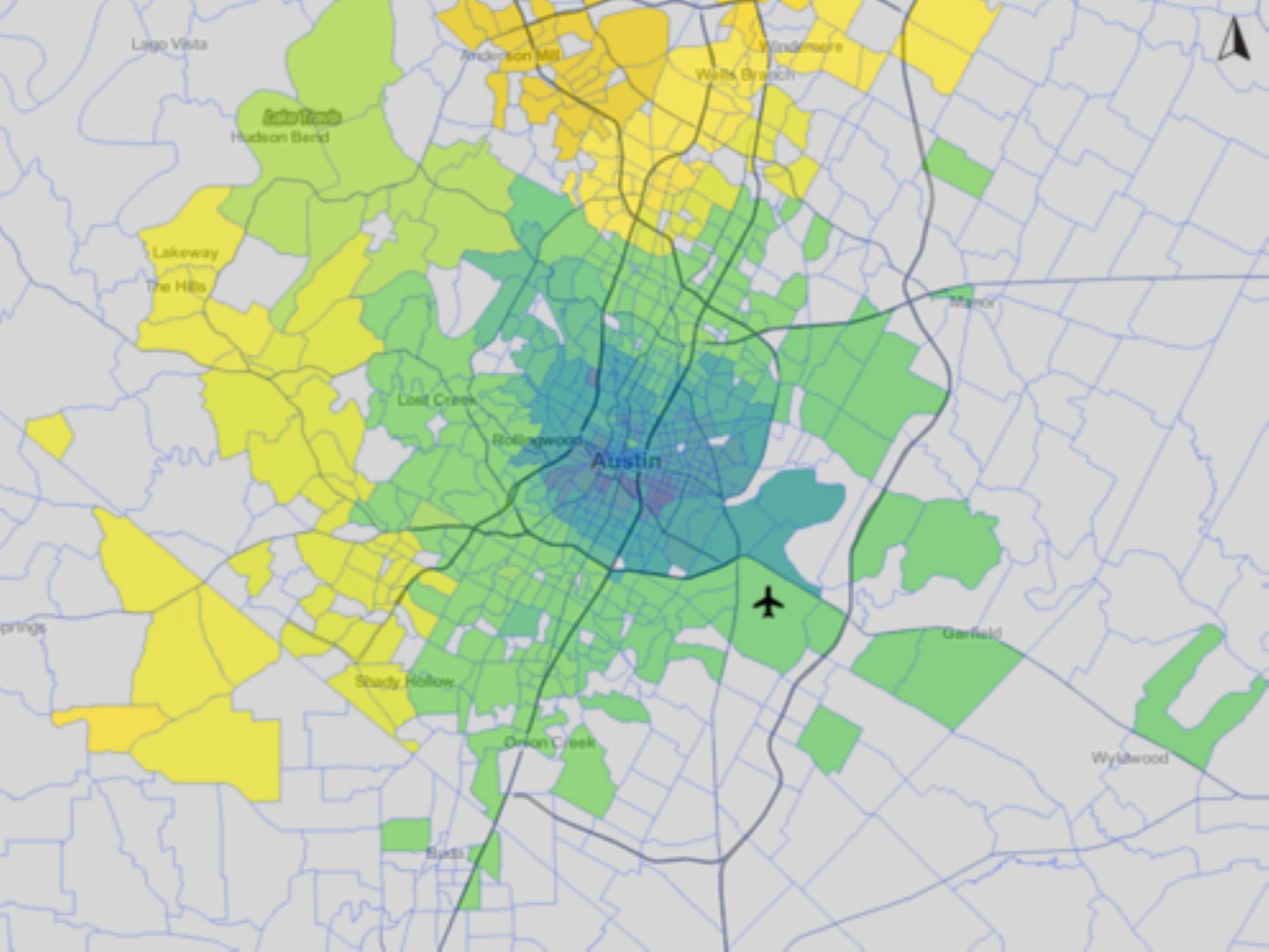}
				\put(0,0){\includegraphics[width=.09\linewidth]{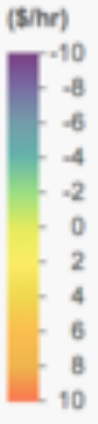}}
			\end{overpic}
			\caption{Peak hours}
		\end{subfigure}
		\begin{subfigure}[h]{0.325\linewidth}
			\begin{overpic}[width=0.95\linewidth]{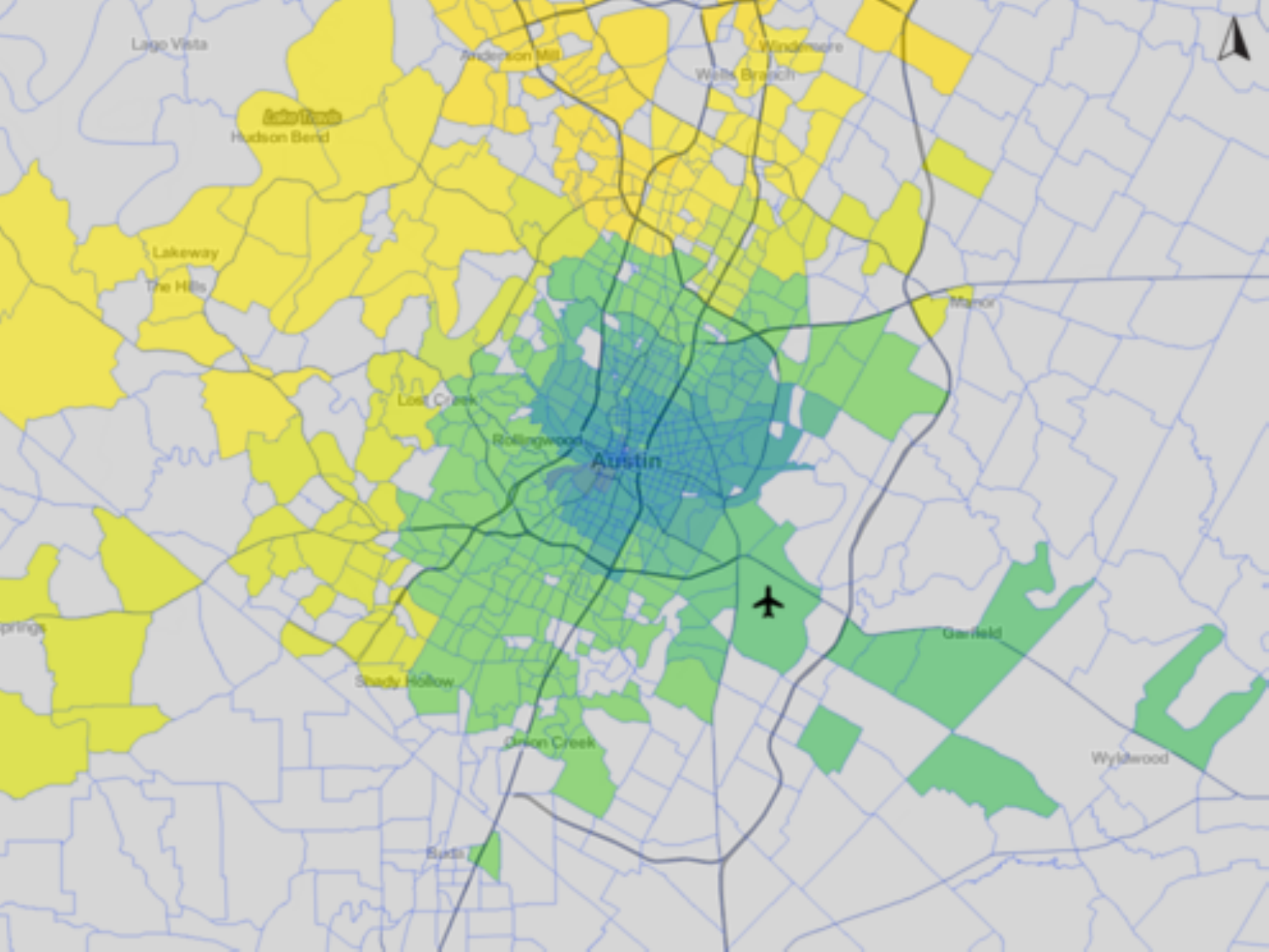}
				\put(0,0){\includegraphics[width=.09\linewidth]{fig/prodC/Simb.pdf}}
			\end{overpic}
			\caption{Mid-day}
		\end{subfigure}
		
		\begin{subfigure}[h]{0.325\linewidth}
			\begin{overpic}[width=0.95\linewidth]{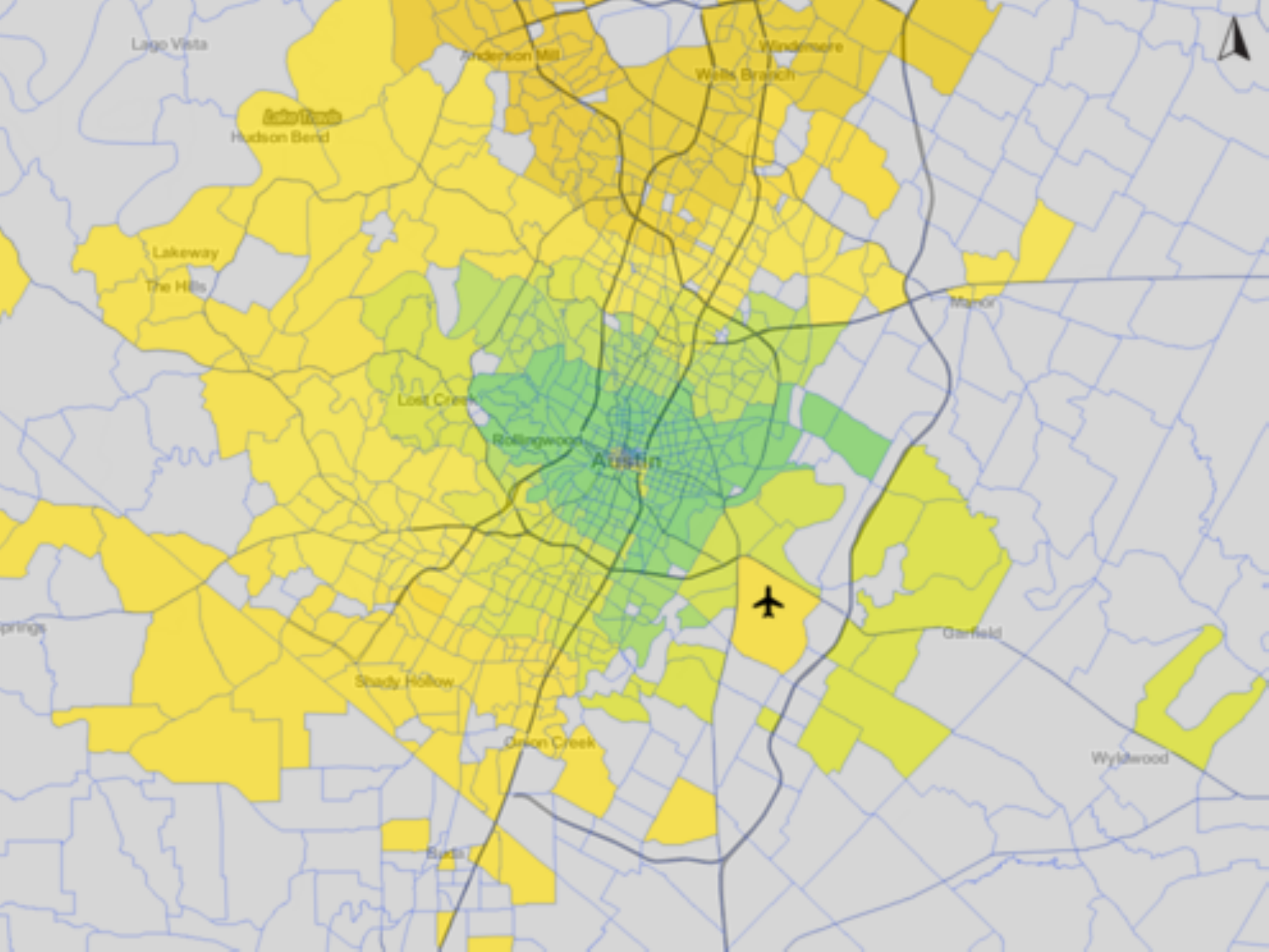}
				\put(0,0){\includegraphics[width=.09\linewidth]{fig/prodC/Simb.pdf}}
			\end{overpic}
			\caption{Overnight}
		\end{subfigure}
		\begin{subfigure}[h]{0.325\linewidth}
			\begin{overpic}[width=0.95\linewidth]{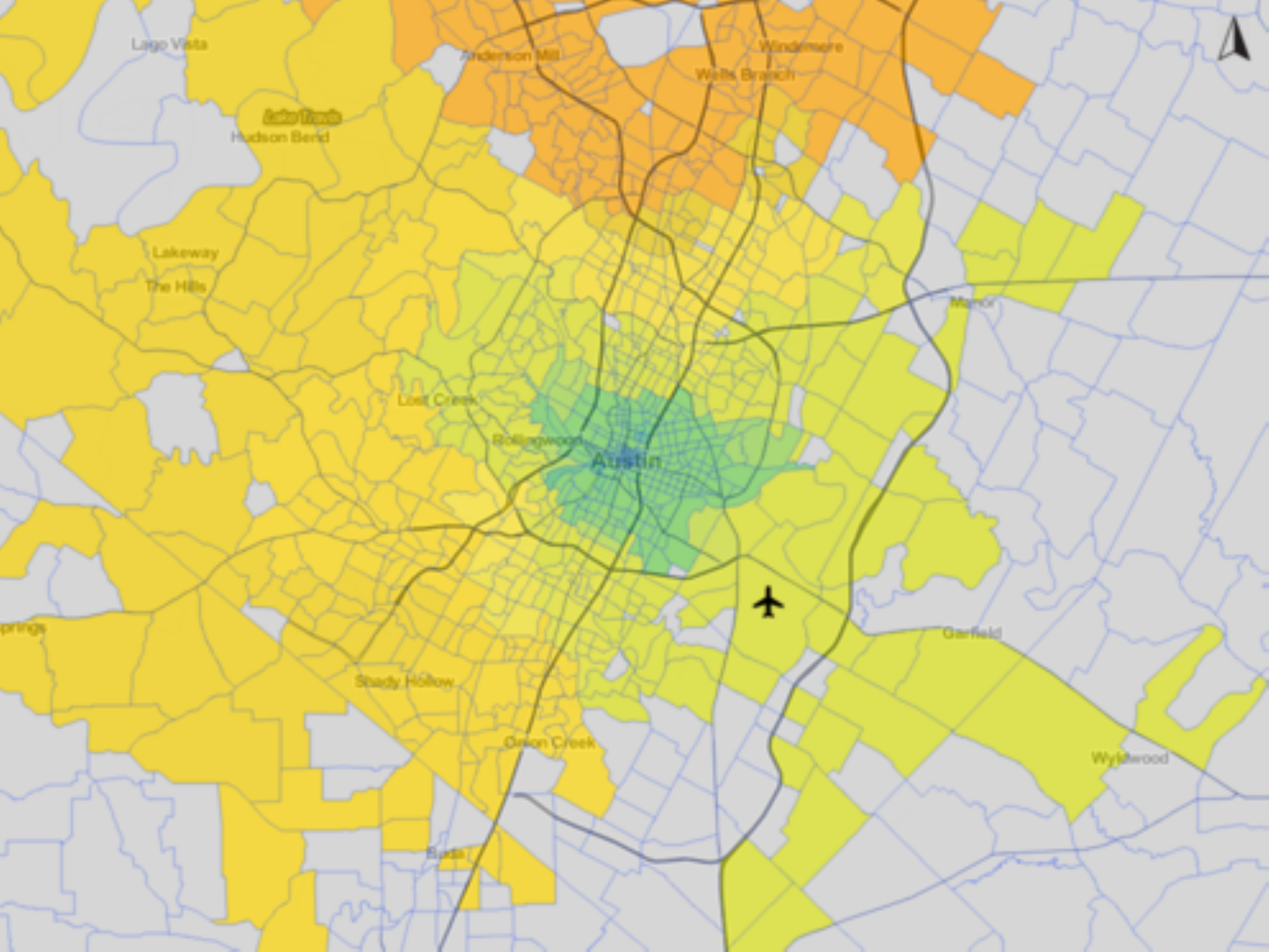}
				\put(0,0){\includegraphics[width=.09\linewidth]{fig/prodC/Simb.pdf}}
			\end{overpic}
			\caption{Weekend}
		\end{subfigure}
		\begin{subfigure}[h]{0.325\linewidth}
			\begin{overpic}[width=.95\linewidth]{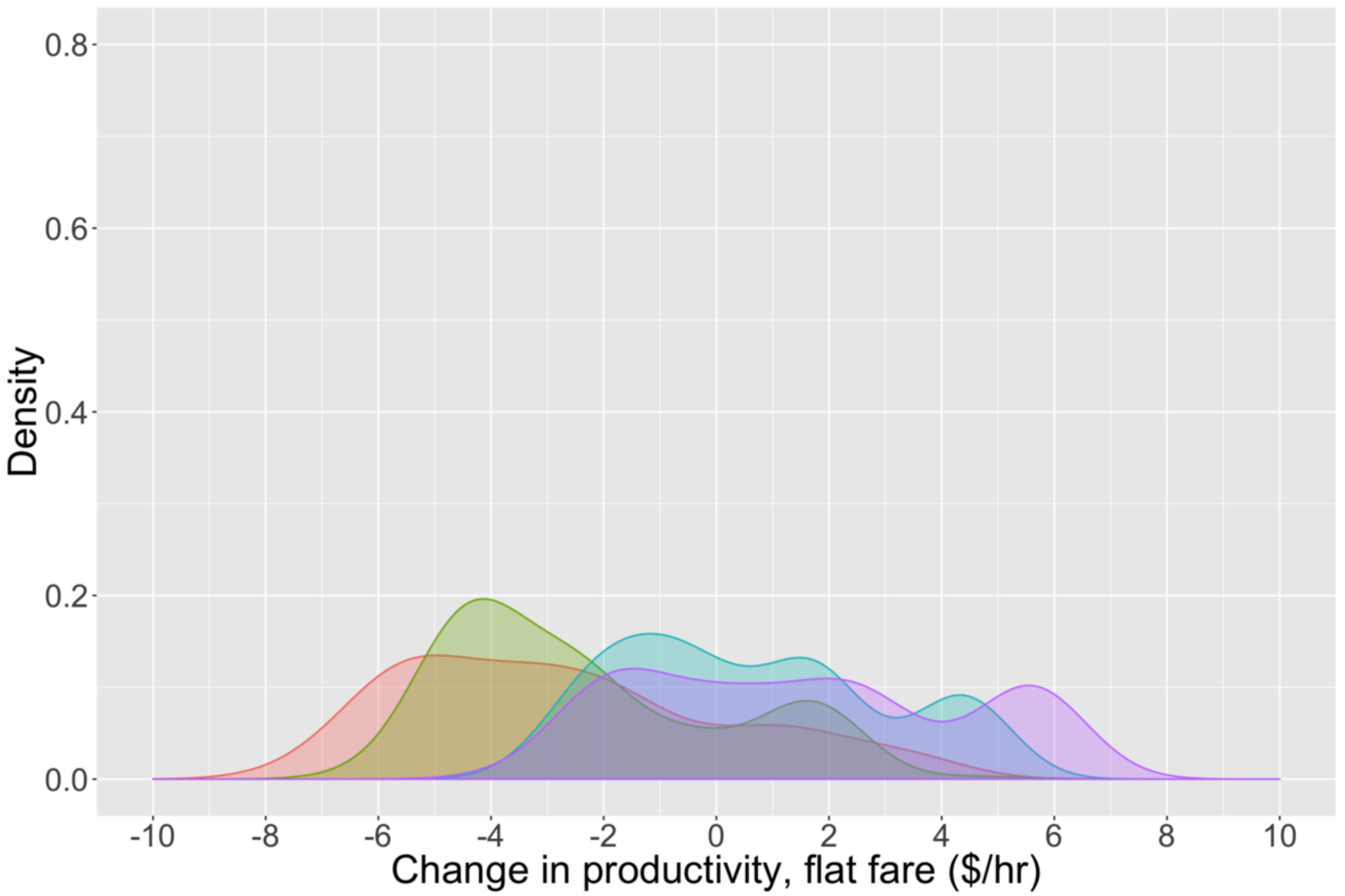}
				\put(40,60){\includegraphics[width=.55\linewidth]{fig/res/Simb2.pdf}}
			\end{overpic}
			\caption{Density}
		\end{subfigure}
		\caption{Change in driver productivity with respect to the average for trips with flat fare, by destination}
		\label{fig:prodC}
	\end{center}
\end{figure}


\begin{table}[H]
	\centering
	\caption{Description of trips with surge price by period}
	\begin{tabular}{llrrrr}
		\hline
		\multicolumn{2}{c}{\multirow{2}{*}{Period}} & \multicolumn{2}{c}{System-wide}                            & \multicolumn{2}{c}{CBD-origin}                             \\ \cline{3-6} 
		\multicolumn{2}{c}{}                        & \multicolumn{1}{c}{Total} & \multicolumn{1}{c}{Percentage} & \multicolumn{1}{c}{Total} & \multicolumn{1}{c}{Percentage} \\ \hline
		\multirow{3}{*}{Weekday}    & Peak Hours    &  38,528                     & 2.7\%                          &  7,793                     & 3.0\%                          \\ 
		& Mid-Day       &  15,490                     & 1.1\%                          &  2,600                     & 1.0\%                          \\ 
		& Overnight     &  35,214                     & 2.5\%                          &  17,307                     & 6.6\%                          \\ 
		\multicolumn{2}{l}{Weekend}                 &  158,701                    & 11.2\%                         &  46,972                     & 18.0\%                         \\ \hline
		\multicolumn{2}{l}{\textit{Total}}          & \textit{247,933}          & \textit{17.5\%}                & \textit{74,672}           & \textit{28.5\%}                \\ \hline
	\end{tabular}
	\label{tab:surge}
\end{table}

\begin{figure}[H]
	\centering
	\captionsetup{justification=centering}
	\begin{center}
		\begin{subfigure}[h]{0.325\linewidth}
			\begin{overpic}[width=1\linewidth]{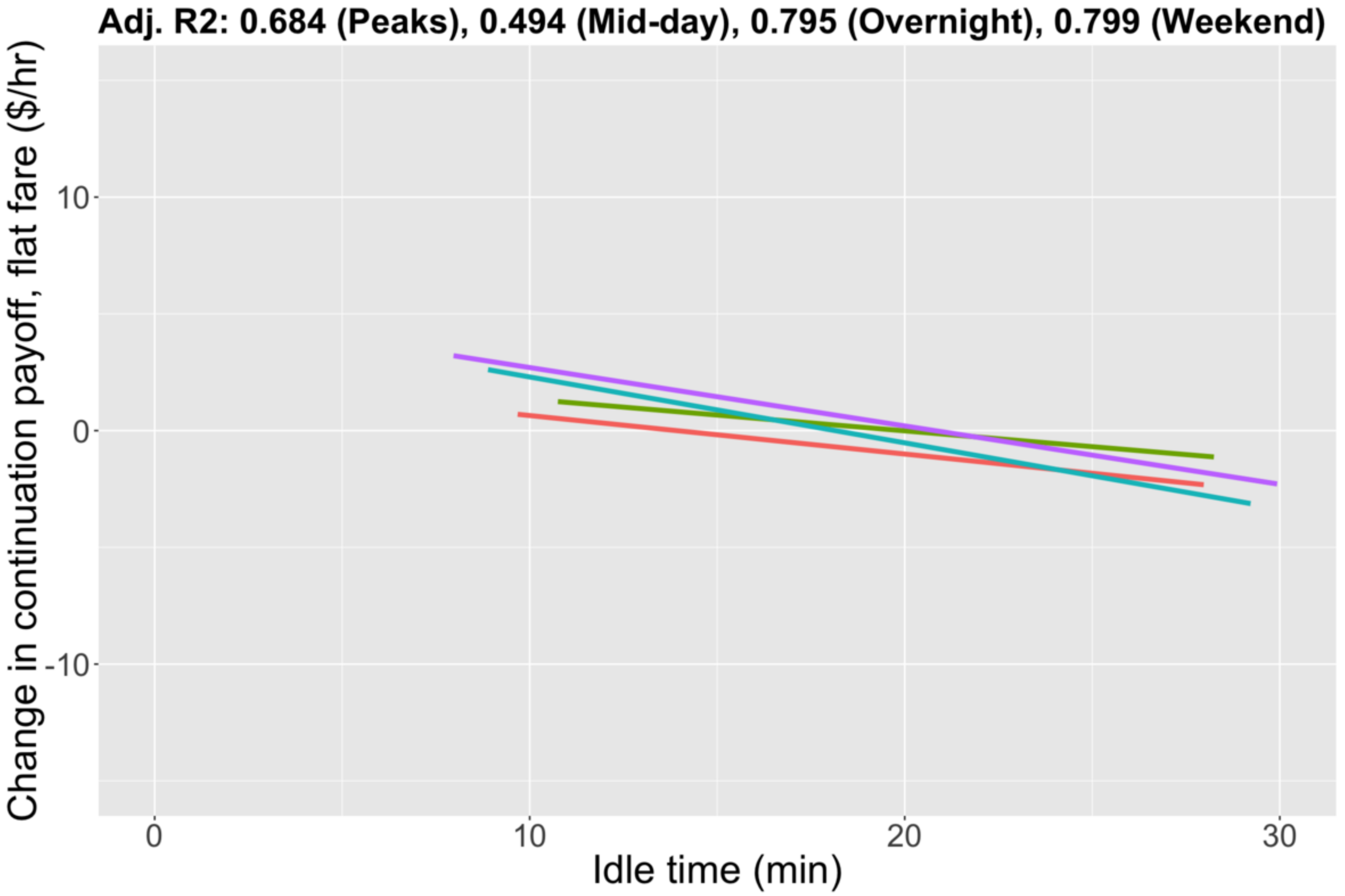}
				\put(40,57){\includegraphics[width=0.55\linewidth]{fig/analysis/Simb2.pdf}}
			\end{overpic}
			\caption{Continuation payoff and idle time}
		\end{subfigure}
		\begin{subfigure}[h]{0.325\linewidth}
			\begin{overpic}[width=1\linewidth]{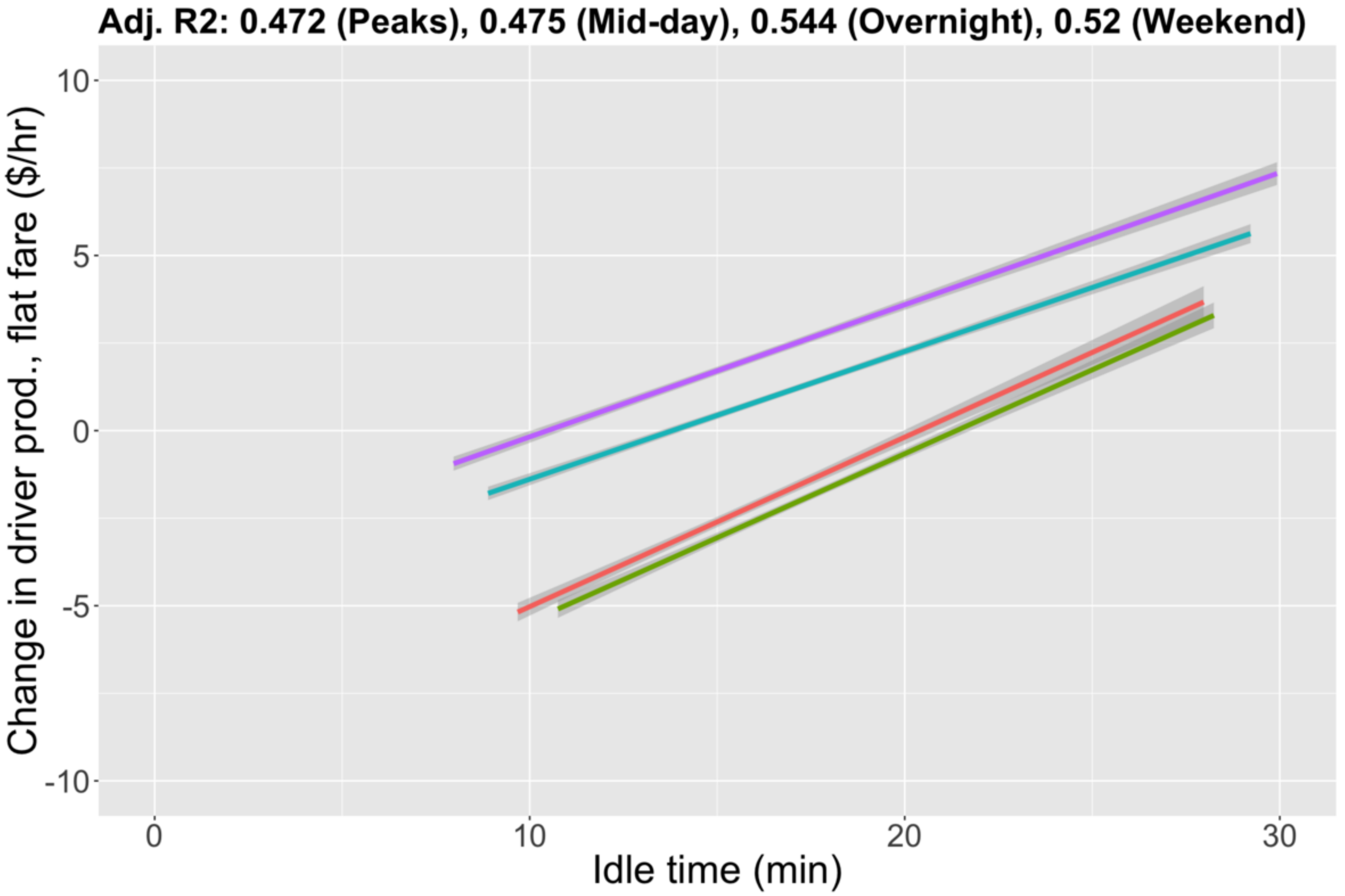}
				\put(40,57){\includegraphics[width=0.55\linewidth]{fig/analysis/Simb2.pdf}}
			\end{overpic}
			\caption{Driver productivity and idle time}
		\end{subfigure}
		\begin{subfigure}[h]{0.325\linewidth}
			\begin{overpic}[width=1\linewidth]{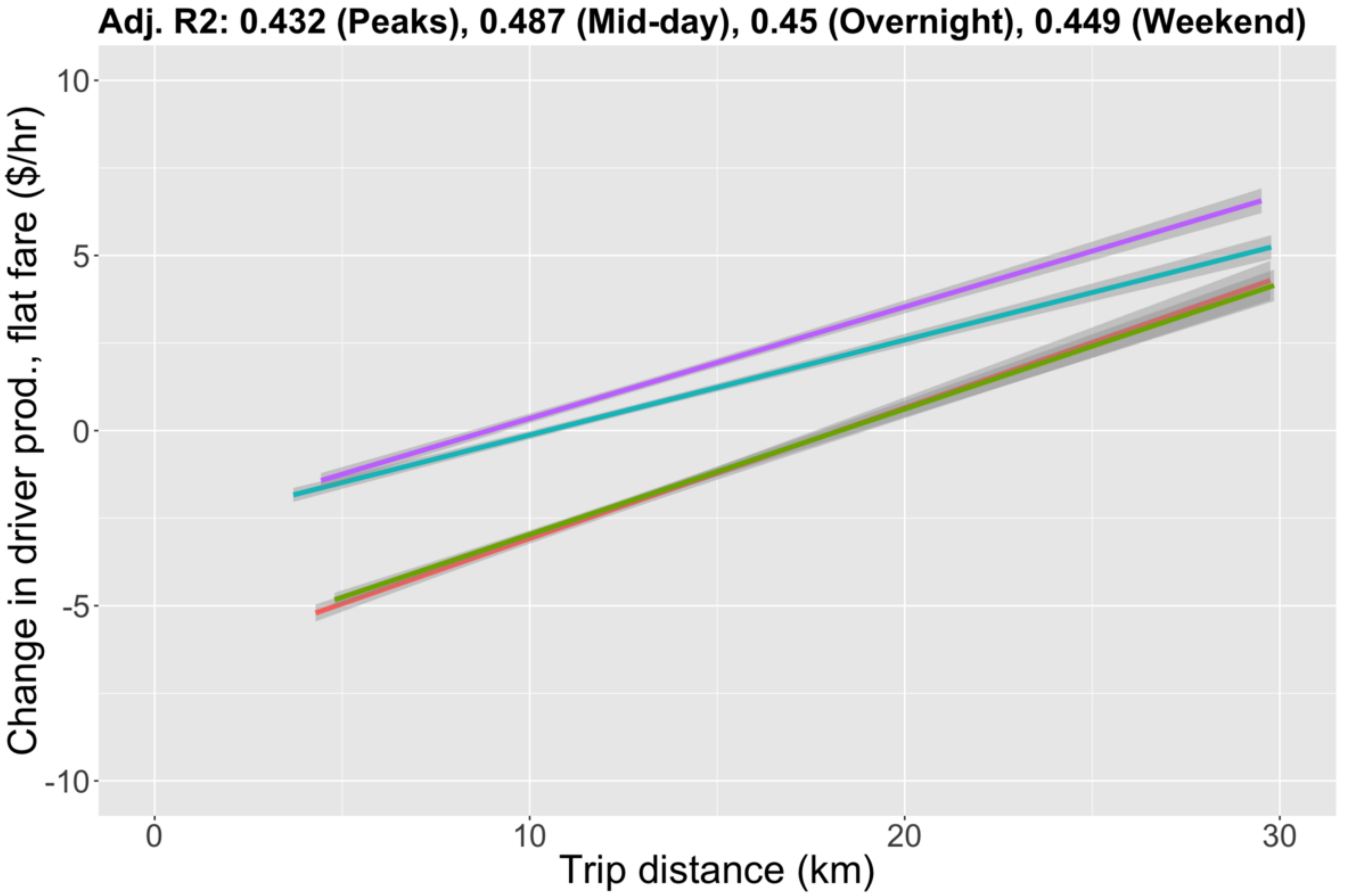}	
				\put(40,57){\includegraphics[width=0.55\linewidth]{fig/analysis/Simb2.pdf}}
			\end{overpic}
			\caption{Driver productivity and trip distance}
		\end{subfigure}
		\caption{Relationship between change in productivity, idle time, and trip distance}
		\label{fig:prodidle}
	\end{center}
\end{figure}

\bibliographystyle{model1-num-names}
\bibliography{partc_ridesourcing}

\end{document}